\documentclass[a4paper,11pt]{article}
\usepackage{microtype}
\usepackage{epsfig}
\usepackage{microtype}
\usepackage{graphicx}
\usepackage{enumitem}
\usepackage{epsfig,enumerate,amsmath,amsfonts,amssymb,amsthm,mathrsfs,ifpdf}
\usepackage{indentfirst,relsize}
\usepackage[numbers]{natbib}
\usepackage{setspace}
\usepackage{enumerate}
\usepackage{latexsym}
\usepackage{stackrel}
\usepackage[all]{xy}
\usepackage[usenames,dvipsnames]{pstricks}
\usepackage{pst-grad} 
\usepackage{pst-plot} 
\usepackage{xspace}

\newcommand{\size}[1]{\left| #1 \right|}

\newcommand{\remove}[1]{}

\newcommand{\bu}{{\bf u}}
\newcommand{\bv}{{\bf v}}
\newcommand{\bw}{{\bf w}}
\newcommand{\br}{{\bf r}}
\newcommand{\bl}{{ \bf \ell}}
\newcommand{\bt}{{\bf t}}
\newcommand{\bz}{{\bf z}}

\newcommand{\md}{\mbox{mid}}

\newcommand{\cS}{\mathcal{S}}
\newcommand{\cT}{\mathcal{T}}
\newcommand{\cC}{\mathcal{C}}
\newcommand{\cL}{\mathcal{L}}
\newcommand{\cA}{\mathcal{A}}
\newcommand{\cB}{\mathcal{B}}
\newcommand{\cD}{\mathcal{D}}
\newcommand{\cF}{\mathcal{F}}
\newcommand{\cP}{\mathcal{P}}
\newcommand{\cN}{\mathcal{N}}

\newcommand{\cG}{\mathcal{G}}

\newcommand{\Oh}{{O}}

\newcommand{\cX}{\mathcal{X}}
\newcommand{\cY}{\mathcal{Y}}
\newcommand{\eps}{\epsilon}

\newcommand{\cbound}{N \cdot n \cdot (4d(2d+2))^{2d+2}}

\newcommand{\ubound}{n^\alpha \cdot ( \alpha \cdot \Delta ^{d+1})^{\Delta^{\Oh(\alpha \cdot d^2)}}}

\newcommand{\ctwconsbound}{4(\alpha + {\alpha \choose 2}(\ell_g(\alpha-2)-1))}

\newcommand{\bijbound}{\Oh( \alpha^{2}n^{\alpha +3})\cdot\Delta^{d+1}\cdot(\alpha \Delta^{d+1})^{\Delta^{O(\alpha d^{2})}}}

\newcommand{\genbijbound}{\Oh( \alpha^{2}n^{\alpha +11})\cdot\Delta^{d+1}\cdot(\alpha \Delta^{d+1})^{\Delta^{O(\alpha d^{2})}}}

\newcommand{\fuvtype}{(4(\Gamma+d))^{\mu^2  \cdot \Delta^{d+1}}}

\newcommand{\ufbound}{n\cdot \left( \mu \cdot \Delta ^{d+1} \right)^{\Delta^{\Oh(\Gamma\cdot d+d^2)}}}

\newcommand{\ustate}{\ufbound  \cdot 2^{\Oh ( \fuvtype)}}

\newcommand{\ctwbound}{\Oh(n^2\cdot N)\cdot( \mu \cdot \Delta ^{d+1} )^{\Delta^{\Oh(\mu \cdot d+d^2)}} \cdot 2^{\Oh ((4(\mu+d))^{\mu^2  \cdot \Delta^{d+1}})}+n^{\Oh(1)}}

\newcommand{\gdash}{G\setminus ({\sf Dom}_s^F \cup {\sf Dom}_t^F)}

\newcommand{\gdashdash}{G\setminus ({\sf Dom}_s^{F'} \cup {\sf Dom}_t^{F'})}

\newcommand{\domst}{{\sf Dom}_s^F \cup {\sf Dom}_t^F}
\newcommand{\domstd}{{\sf Dom}_{s'}^F \cup {\sf Dom}_{t'}^F}
\newcommand{\domstdash}{{\sf Dom}_s^{F'} \cup {\sf Dom}_t^{F'}}
\newcommand{\domstddash}{{\sf Dom}_{s'}^{F'} \cup {\sf Dom}_{t'}^{F'}}
\newcommand{\tst}{{\sc st }}

\newcommand{\embed}{{\sc Embed}}

\newcommand{\linethetabound}{n^2\cdot d^{2d+3}\cdot (4d(2d+2))^{2d+2}}

\newcommand{\FPT}{\text{\normalfont FPT}}

\theoremstyle{plain}
\newtheorem{theo}{Theorem}[section]
\newtheorem{lem}[theo]{Lemma}
\newtheorem{pre}[theo]{Proposition}
\newtheorem{coro}[theo]{Corollary}

\newtheorem{cl}[theo]{Claim}
\theoremstyle{definition}
\newtheorem{defi}[theo]{Definition}
\newtheorem{rem}{Remark}
\newtheorem{obs}[theo]{Observation}

\newcommand{\defproblem}[3]{
  \vspace{1mm}
\noindent\fbox{
  \begin{minipage}{0.96\textwidth}
  \begin{tabular*}{\textwidth}{@{\extracolsep{\fill}}lr} #1 \\ \end{tabular*}
  {\bf{Input:}} #2  \\
  {\bf{Question:}} #3
  \end{minipage}
  }
  \vspace{1mm}
}

\usepackage[margin = 2.50cm]{geometry}

\title{
FPT algorithms for embedding into low complexity graphic metrics
}

\author{
Arijit Ghosh
\footnote{Indian Statistical Institute, Kolkata, India.}
\and
Sudeshna Kolay
\footnote{Eindhoven University of Technology, Netherlands.}
\and
Gopinath Mishra
\footnotemark[1]
}

\begin{document}

\maketitle

\begin{abstract}
Given metric spaces $(X, D_{X})$ and $(Y, D_{Y})$, an {\em embedding} $F: X \rightarrow Y$ is an injective mapping
from $X$ to $Y$. {\em Expansion} $e_{F}$ and {\em contraction} $c_{F}$ of an embedding $F: X \rightarrow Y$ are defined 
as 
$$
e_{F} = \max_{x_{1}, x_{2} (\neq x_{1}) \in X} \frac{D_{Y}(F(x_{1}), F(x_{2}))}{D_{X}(x_{1}, x_{2})} \; \; \mbox{and} \; \; 
c_{F} = \max_{x_{1}, x_{2} (\neq x_{1}) \in X} \frac{D_{X}(x_{1}, x_{2})}{D_{Y}(F(x_{1}), F(x_{2}))},
$$
respectively and {\em distortion} $d_{F}$ is defined as $d_{F}= e_{F} \cdot c_{F}$.
Observe that $d_{F} \geq 1$. An embedding $F : X \rightarrow Y$ is {\em non-contracting} if $c_{F} \leq 1$. When 
$d=1$, then $F$ is \emph{isometry}.

 The {\sc Metric Embedding} problem takes as input two metric spaces $(X,D_X)$ and $(Y,D_Y)$, and a positive integer $d$. The objective is to determine whether there is an embedding $F:X \rightarrow Y$ such that $d_{F} \leq d$. Such an embedding is called a \emph{distortion $d$ embedding}. The bijective {\sc Metric Embedding} problem is a special case of the {\sc Metric Embedding} problem where $|X| = |Y|$. In parameterized complexity, the {\sc Metric Embedding} problem, in full generality, is known to be W-hard and therefore, not expected to have an FPT algorithm. In this paper, we consider the {\sc Gen-Graph Metric Embedding} problem, where the two metric spaces are graph metrics. We explore the extent of tractability of the problem in the parameterized complexity setting. We determine whether an unweighted graph metric $(G,D_G)$ can be embedded, or bijectively embedded, into another unweighted graph metric $(H,D_H)$, where the graph $H$ has low structural complexity. For example, $H$ is a cycle, or $H$ has bounded treewidth or bounded connected treewidth. The parameters for the algorithms are chosen from the upper bound $d$ on distortion, bound $\Delta$ on the maximum degree of $H$, treewidth $\alpha$ of $H$, and the connected treewidth $\alpha_{c}$ of $H$. 

Our general approach to these problems can be summarized as trying to understand the behavior of the shortest paths in $G$ under a low distortion embedding into $H$, and the structural relation the mapping of these paths has to shortest paths in $H$.

\end{abstract}

\section{Introduction}
\label{sec:intro}

\remove{
\textcolor{red}{
Given metric spaces $(X, d_{X})$ and $(Y, d_{Y})$, an {\em embedding}
of $f: X \rightarrow Y$ is an injective mapping 
from $X$ to $Y$. {\em Expansion} $e_{f}$ and {\em contraction} $c_{f}$
of an embedding $f: X \rightarrow Y$ are defined as
$e_{f} = \max_{x_{1}, x_{2} (\neq x_{1}) \in X} \frac{d_{Y}(f(x_{1}),
  f(x_{2}))}{d_{X}(x_{1}, x_{2})}$ and $c_{f} = \max_{x_{1}, x_{2} (\neq x_{1}) \in X} \frac{d_{X}(x_{1},
  x_{2})}{d_{Y}(f(x_{1}), f(x_{2}))}$ respectively, and {\em
  distortion} $\delta_{f}$ is defined as $d_{f}= e_{f} \cdot
c_{f}$.
Observe that $\delta_{f} \geq 1$. An embedding $f : X \rightarrow Y$
is {\em non-contracting} if $c_{f} \leq 1$ and {\em isometry} if 
$d_{f} = 1$.} \textcolor{blue}{This paragraph not required anymore.}
}

Let $(X,D_X)$ and $(Y,D_Y)$ be two metric spaces. A small distortion
embedding $F: X \rightarrow Y$ of a metric space $(X, D_{X})$ into a metric space $(Y,
D_{Y})$ essentially means that distances in the metric space $(F(X), D_{Y})$ 
have similar lengths as the corresponding distances in $(X, D_{X})$. Therefore, an
algorithmic problem in $(X, D_{X})$ could, in principle, be
solved (or approximated) by solving a similar problem in $(F(X), D_{Y})$. Therefore, there is an interest in finding an efficient way to compute
a low distortion embedding of $(X, D_{X})$ into $(Y, D_{Y})$.  
This simple insight has been used extensively to
design fast algorithms for many classical problems in computer
science. The central idea is, given an input metric space $(X,
D_{X})$,  we find an efficient small distortion embedding of $(X, D_{X})$ into a simpler metric space whose
structure is well understood. For example, the metric space into which we intend to embed $(X,D_X)$ could be a simple graphic metric like a path or a tree, or well-studied metric spaces like a Euclidean space etc. Examples of problems where this strategy has been successfully
implemented include problems like {\sc Sparsest Cut}, {\sc Approximate Nearest
Neighbor Search}, {\sc Clustering}, {\sc Multicommodity Flow}, {\sc Multicut}, {\sc
Low Diameter Decomposition}, {\sc Small Balanced Separators} etc.
(See~\cite{LinialLR95,Indyk01,GuptaNRS04, ABRAHAM20113026,Matousek-book-2013}).    

The problem of finding a low distortion embedding of a metric space into
a simple metric space has been extensively
studied both in Mathematics and Computer Science. For
example, any metric space with $n$ points can be isometrically embedded into
$\ell^{n}_{\infty}$ \cite{LinialLR95}. In~\cite{LinialLR95}, Linial et al. 
showed that any
unweighted $n$ vertex tree with $l$ leaves can be isometrically
embedded into $\ell^{O(\log l)}_{\infty}$. Bourgain \cite{BourgainLEFM85} showed that any metric space on $n$ points can be
embedded into $\ell_{p}$ with distortion $\Oh(\log n)$. Johnson and Lindenstrauss \cite{johnson84extensionslipschitz} showed that any $n$ points can be embedded into $\ell_{2}^{\Oh\left(\log^{2} n/\eps^{2}\right)}$ with distortion at most $(1+\epsilon)$.  Bartal \cite{BartalFOCS96} and Fakcharoenphol et al. \cite{FakcharoenpholRT03} showed that any metric space on $n$ points can be embedded into a distribution of trees with expected distortion $\Oh(\log n)$. 

\subsection{Embedding into low complexity graphs}
The need for getting small distortion embeddings into simpler
spaces naturally led to the question of finding minimum distortion embedding 
of $(X, D_{X})$ into $(Y,D_{Y})$ when both the metric spaces come from shortest 
path metrics on graphs with positive weights, and $(Y,D_{Y})$ has a simple topology 
like paths, cycles, trees etc. 
Kenyon et al. \cite{KenyonRS09} showed that this problem is APX-hard even
when both the graphs are unweighted, have the same number of vertices, 
and one of the graphs 
is a simple wheel graph. 
Kenyon et al.'s \cite{KenyonRS09} result implies that the problem is APX-hard even
when we are looking for a bijective embedding and both the graphs are unweighted.
Badoiu et al. \cite{BadoiuDGRRRS05} also proved
APX-hardness when both the graphs are unweighted and $(Y, D_{Y})$ is the metric space of a path. Badoiu et al. \cite{BadoiuCIS05} showed that computing the minimum distortion is hard to approximate up to a factor 
polynomial in $\vert X \vert$, even when $(X,D_{X})$ is a weighted tree
with polynomial spread and $(Y,D_{Y})$ is a path.  Fellows et al. \cite{FellowsFLLRS13} showed that the problem of embedding a weighted graph metric into a path with distortion at most $d > 2$ is NP-complete.

Badoiu et al. \cite{BadoiuDGRRRS05} gave the first algorithm for deciding if an unweighted graph metric has a non-contracting embedding into a path with distortion $d$. The running time of their algorithm was $n^{4d+2} \cdot d^{\Oh\left( 1\right)}$, where $n$ denotes the number of vertices in the graph. 
The study of small distortion embeddings through the lens of Parameterized Complexity \cite{saketbook15} started with the work of Fellows et al. \cite{FellowsFLLRS13}. 
They gave the first fixed parameter tractable(FPT) algorithm with
running time $\Oh\left(n\, d^{4}(2d+1)^{2d}\right)$ for
finding a non-contracting embedding of an $n$ vertex unweighted graph metric into a path with
distortion at most $d$ ($d$ is the parameter of the
algorithm). Fellows et al. \cite{FellowsFLLRS13} also showed that
their FPT algorithm can be extended to get an FPT algorithm for the
case of non-contracting embeddings for weighted graphs into paths, where the parameters to the
algorithm are both the distortion and the maximum weight of an edge in
the graph. Nayyeri et al. \cite{NayyeriR15} gave improved exact
algorithms for embedding weighted path metrics into weighted paths. 

Results for embedding into paths have been extended to the case of trees. 
Kenyon et al. \cite{KenyonRS09} gave the first FPT algorithm for finding a bijective 
embedding $f$ of an unweighted graph metric on $n$ vertices into a tree with maximum degree bounded by $\Delta$ in $\Oh\left( n^{2}\cdot 2^{\Delta^{\mu^{3}}}\right)$ time, where 
$\mu = \max \left\{ e_{f}, \, c_{f}\right\}$.
Fellows et al. \cite{FellowsFLLRS13} extended this result to give an
algorithm for the problem of finding a non-contracting embedding of unweighted graphs into bounded degree trees with distortion at most $d$ in $\Oh(n^{2}\cdot |V(T)|)\cdot 2^{\Oh( (5d)^{\Delta^{d+1}}\cdot d)}$ time, where $V(T)$ denotes the vertex set of the tree and where the maximum degree in $T$ is bounded by $\Delta$. In a follow-up paper, Nayyeri et al. \cite{NayyeriR17} gave the first $(1+\epsilon)$-approximation algorithm to embed weighted graphs with spread $\Sigma$ into graphs on $m$ vertices with bounded treewidth $\alpha$ and doubling dimension $\lambda$ in $m^{O(1)}\cdot n^{O(\alpha) \cdot (d_{opt}\Sigma)^{\alpha\cdot (1/\epsilon)^{\lambda+2} \cdot \lambda \cdot \left(O(d_{opt})\right)^{2\lambda}}}$ time, where $d_{opt}$ denotes the minimum distortion. 

\subsection{Our contributions}

In this paper, we further investigate the problem of embedding a general graph metric $(G,D_{G})$ into
a low complexity graph metric $(H, D_{H})$ with distortion at most
$d$. We will denote by $n$ and $N$ the number of vertices in graphs
$G$ and $H$, respectively. Also, we denote distortion by $d$, the maximum degree of $H$ by $\Delta$, and the length of a longest geodesic (or induced) cycle of $H$ by $\ell_g$. We approach the metric embedding problem by trying to understand the behavior of the shortest paths in $G$ under a low distortion embedding into $H$, and what relation the mapping of these paths has to shortest paths in $H$. Careful analysis of this connection helps us solve a number of problems in this area, in the parameterized setting. {\bf All the algorithmic results mentioned below are regarding non-contracting bounded distortion embeddings. However, all these results can be extended to find bounded distortion embeddings, without the assumption on non-contraction. For all the results, if the running time of the stated algorithm is $T$, then the running time of finding a bounded distortion embedding will be $(nN)^{\Oh(1)}\cdot T$.} For more details, refer to Proposition~\ref{prop:scaling} in Section~\ref{sec:prelim}.

\begin{itemize}
\item[(a)] {\bf In Section~\ref{sec:cycle}, we find a non-contracting embedding of distortion $d$ of $G$ into $H$, when $H$ is a cycle. We present an FPT algorithm, parameterized by $d$ and with time complexity $2^{\Oh\left( d \log d \right)}
\cdot n^{\Oh(1)}$.} The techniques used for obtaining this algorithm involve a few different ideas from the ones used in the previous papers where $H$ is either a path or a bounded degree tree. This is due to the existence of the large geodesic cycle that is the graph $H$. The technique of pushing embeddings, introduced in \cite{FellowsFLLRS13} for embedding into paths, does not work and some new ideas are required to solve this problem. This resolves a question left open in \cite{FellowsFLLRS13}. Moreover, our FPT algorithm can be extended to the weighted setting, where the input graph $G$ has edge weights and we parameterize by the distortion $d$ as well as the maximum edge weight in $G$. On the other hand, we show that when we do not take the maximum edge weight as a parameter, then the problem becomes NP-Complete for any distortion $d >2$.

\item[(b)] Observe that the \emph{treewidth} of a cycle is $2$, but the \emph{connected
treewidth} of a cycle is $\Omega(n)$ (see the definitions of treewidth
and connected treewidth in Section \ref{sec:prelim}). These two parameters (treewidth
and connected treewidth of graphs) play important roles in this
paper. 
\begin{itemize}
\item {\bf First, in Section~\ref{sec:bij}, we find a non-contracting bijection of distortion $d$ of $G$ into $H$, which has constant treewidth $\alpha$. Parameterized by $d$ and $\Delta$, the FPT algorithm has time complexity $\Oh( \alpha^{2}n^{\alpha +3})\cdot\Delta^{d+1}\cdot(\alpha  \Delta^{d+1})^{\Delta^{O(\alpha d^{2})}}$.} This is an extension of the result of Kenyon et al. \cite{KenyonRS09} for bijection into
bounded degree trees. Even though our FPT algorithm is different from the algorithm in \cite{KenyonRS09},
the fact that we are looking for bijective embeddings plays an important
role in the analysis and the algorithm in \cite{KenyonRS09} also uses
this fact crucially.  

\item {\bf Next, in Section~\ref{sec:ctw}, we find a non-contracting embedding of distortion $d$ of $G$ into $H$, which has treewidth $\alpha$ and length of longest geodesic cycle $\ell_g$. Parameterized by $d, \alpha, \ell_g$ and $\Delta$, the FPT algorithm has time complexity $\Oh(n^2\cdot N)\cdot ( \mu \cdot \Delta ^{d+1} )^{\Delta^{\Oh(\mu \cdot d+d^2)}} \cdot 2^{\Oh ( (4(\mu+d))^{\mu^2  \cdot \Delta^{d+1}}} + n^{\Oh(1)}$, where $\mu = \ctwconsbound$.} This result crucially uses the result in~\cite{Diestel2017} that a graph has bounded connected treewidth if and only if the graph has bounded treewidth and no long geodesic cycle. It is to be noted that a wheel graph has constant connected treewidth, and by a result in~\cite{KenyonRS09}, embedding into wheel graphs is NP-hard even when the distortion $d=2$. However, when the wheel graph has bounded degree, then the number of vertices in the wheel graph becomes bounded, and we obtain a trivial FPT algorithm parameterized by the degree and the distortion $d$. This motivated us to consider the above variant of metric embedding. Our FPT algorithm extends the result of Fellows et al. \cite{FellowsFLLRS13} for embedding into trees with bounded degree. Controlling the behavior of shortest paths in the graph $G$  under a low distortion embedding into the class of graphs with bounded degree and bounded connected treewidth is algorithmically considerably harder than the case of bounded degree trees. The algorithm given in \cite{FellowsFLLRS13} remembers the information of one shortest path between two vertices in $G$ but in our case we have to efficiently maintain information of a set of useful paths between two vertices. 
\end{itemize}
\item[(c)] {\bf In Section~\ref{sec:theta}, we consider generalized theta graphs: defined by the union of $k$ paths all of which have common endpoints $s$ and $t$. We find a non-contracting embedding of distortion $d$ of $G$ into $H$, which is a generalized theta graph. Parameterized by $d$ and $k$, our FPT algorithm has running time $\Oh(N)+ n^5\cdot k^{2k+1}\cdot (kd+1)^{(2d)^{\Oh(kd)}}\cdot d^{\Oh(d^2)}$.} As mentioned earlier, it was shown in~\cite{Diestel2017} that a graph has bounded connected treewidth if and only if the graph has bounded treewidth and no long geodesic cycle. In general, embedding into graphs with large geodesic cycles is not amenable to known algorithmic techniques in the parameterized settings. Intuitively, all known techniques for designing FPT algorithms in this area used the fact that if a low distortion embedding $F$ exists, then the embedding of a shortest path between two vertices $u,v \in V(G)$ and the shortest path in $H$ between $F(u)$ and $F(v)$ are somewhat structurally related. With the presence of large geodesic cycles this structural relation may completely break down: although the two paths have similar lengths, structurally they could be completely different. This poses a problem for designing dynamic programming algorithms, a staple for FPT algorithms in this area. The class of generalized theta graphs has treewidth $2$, but may have large geodesic cycles. Hence, these graphs are more general than cycles and have constant treewidth, but they do not have bounded connected treewidth. Even for this very structured graph class, by virtue of the graphs having long geodesic cycles, we needed to develop completely new ideas in order to find low distortion embeddings into generalized theta graphs via FPT algorithms. The problem arises from the fact that any two geodesic cycles of a generalized theta graph intersect at at least two vertices, and there are many pairs of geodesic cycles with large intersections. Our algorithm is still a dynamic programming algorithm, but a more involved one. The way to work around the apparent barriers is to investigate more closely the structural properties of an input graph $G$ that can be
embedded with small distortion into a generalized theta graph. Independently a generalization of this result was obtained in~\cite{Carpenter17}.
\end{itemize}

\section{Preliminaries}
\label{sec:prelim}

\subsection{General Notations}
\label{ssec:general-notations}

We denote $\{1,\ldots,t\}$ as $[t]$. For a set $S$, $\size{S}$ denotes the number of elements present in $S$. Given a function $f:U'\rightarrow D'$ and a function $F: U\rightarrow D$, where $U'\subseteq U$ and $D'\subseteq D$, we say that $F$ \emph{extends} $f$ if for all $x\in U'$, $F(x) = f(x)$. A graph is denoted by $G$ while its vertex set and edge set are denoted by $V(G)$ and $E(G)$, respectively. We denote the set of neighbours of a vertex $v \in V(G)$ as $N_G(v)$. The degree of a vertex $v \in V(G)$ is denoted as $deg_G(v)$. We also define $\Delta(G)=\max\limits_{v \in V(G)} deg_G(v)$. We also define the set $B(v,r) = \{u\in V(H)~\vert~ D_H(u,v) \leq r\}$, and refer to is as an $r$-ball around $v$. For a subgraph $G'$ of $G$, $v\in V(G)\setminus V(S)$ is said to be a neighbour of $G'$ if there is a vertex $u \in V(G')$ such that $(u,v)\in E(G)$. A subgraph $G'$ of $G$ is said to be an \emph{induced subgraph} if $E(G') = \{(u,v)\in E(G) \vert u,v \in V(G')\}$. An induced cycle in a graph is also called a geodesic cycle.

A \emph{generalized theta graph} is the union of $k$ paths $\mathcal{P} = \{P_1,P_2,\ldots P_k\}$ such that the endpoints of all the paths are two vertices $s$ and $t$, while every pair of paths are internally vertex and edge disjoint. Such a graph will also be referred to as a generalized theta graph defined at $s$,$t$, and the family $\mathcal{P}$ is said to define the generalized theta graph.

\subsection{Tree decompositions and treewidth} 
\label{ssec:tree-decompositions-treewidth}

We define treewidth and tree decompositions.  
\begin{defi}[Tree Decomposition ~\cite{saketbook15}] 
A tree decomposition of a (undirected or directed) graph $G$ is a tuple $\mathcal{T} = (T,\{X_\bu\}_{\bu \in V(T)})$, where $T$ is a 
tree in which each vertex $\bu \in V(T)$ has an assigned set of vertices $X_\bu \subseteq V(G)$ (called a bag) such 
that the following properties hold:
\begin{itemize}
\item $\bigcup_{\bu \in V(T)} X_\bu = V(G)$
 \item For any $(x,y) \in E(G)$, there exists a $\bu \in V(T)$ such that $x, y \in X_\bu$.
\item If $x \in X_\bu$ and $x \in X_\bv$, then $x \in X_\bw$ for all $\bw$ on the path from $\bu$ to $\bv$ in $T$.
\end{itemize}
In short, we denote $\mathcal{T} = (T,\{X_\bu\}_{\bu \in V(T)})$ as $T$.
\end{defi}

The \emph{treewidth} $tw_{\mathcal{T}}$ of a tree decomposition $\mathcal{T}$ is the size of the largest bag of $\mathcal{T}$ minus one. A graph may have several distinct tree decompositions. The treewidth $tw(G)$ of
a graph $G$ is defined as the minimum of treewidths over all possible tree decompositions of $G$. Note that for the tree $T$ of a  tree decomposition, we denote a vertex of $V(T)$ in bold font.

A tree decomposition  ${\mathcal{T}}=(T,\{X_\bu\}_{\bu \in V(T)}))$ is called a {\em nice tree decomposition} if $T$ is a tree rooted at some node $\br$ where $ X_{\br}=\emptyset$, each node of $T$ has at most two children, and each node is of one of the following kinds:
\begin{itemize}
\item {\bf Introduce node}: a node $\bu$ that has only one child $\bu'$ where $X_{\bu}\supset X_{\bu'}$ and  $|X_{\bu}|=|X_{\bu'}|+1$.

\item {\bf  Forget vertex node}: a node $\bu$ that has only one child $\bu'$  where $X_{\bu}\subset X_{\bu'}$ and  $|X_{\bu}|=|X_{\bu'}|-1$.

\item {\bf Join node}:  a node  $\bu$ with two children $\bu_{1}$ and $\bu_{2}$ such that $X_{\bu}=X_{\bu_{1}}=X_{\bu_{2}}$.

\item {\bf Leaf node}: a node $\bu$ that is a leaf of $T$, and $X_{\bu}=\emptyset$. 
\end{itemize}
One can  show that  a tree decomposition of width $w$ can be transformed into a nice tree decomposition of the same width $w$ and  with $\Oh(w |V(G)|)$ nodes, see~e.g.~\cite{saketbook15}. 
 
A \emph{connected tree decomposition} is a tree decomposition where the vertices in every bag induce a connected subgraph of $G$~\cite{Diestel2017}. The \emph{connected treewidth} $ctw(G)$ of a graph $G$ is defined as the minimum of treewidths over all possible connected tree decompositions of $G$.

\subsection{Parameterized Complexity}
\label{ssec:parameterized-complexity}

The instance of a parameterized problem/language is a pair containing the problem instance of size $n$ and a positive integer $k$, which is called a parameter. The problem is said to be in \FPT{} if there exists an algorithm that solves the problem in $f(k) n^{\Oh(1)}$ time, where $f$ is a computable function. Such an algorithm is called an \FPT\ algorithm and the running time of such an algorithm is called \FPT\ running time. There is also an accompanying theory of parameterized intractability using which one determines that a parameterized problem is unlikely to admit \FPT{} algorithms. A parameterized problem is said to be in the class para-NP if it has a nondeterministic algorithm with FPT running time. To show that a problem is para-NP-hard we need to show that the problem is NP-hard for some constant value of the parameter.
Readers are requested to refer~\cite{saketbook15} for more details on Parameterized Complexity.

\subsection{Metric Embedding}
\label{ssec:metric-embedding}

Given a graph $G$, the function $D_G:V(G)\times V(G)\rightarrow \mathbb{R}$ is the shortest distance function defined on $G$; for any pair $u,v \in V(G)$, $D_G(u,v)$ is the length of the shortest path between $u$ and $v$ in the graph $G$. When the graph is unweighted, then note that the range of $D_G$ is $\mathbb{Z}^{\geq 0}$. When we talk of a graph metric, then we denote it as the tuple $(G,D_G)$. In this paper, unless otherwise mentioned, a graph metric is that of an {\bf unweighted undirected graph}.

A metric embedding of a graph metric $(G,D_G)$ into a graph metric $(H,D_H)$ is a function $F:V(G) \rightarrow V(H)$. When the graph metrics are clear, we also use the terminology that the metric embedding is that of $G$ into $H$, or that $G$ is embedded into $H$. We also denote $(G,D_G)$ as the \emph{input metric space} and $(H,D_H)$ as the \emph{output metric space}. The \emph{expansion} of $F$ is defined as 
$$
e_{F} = \max\limits_{u,v \in V(G)} \frac{D_H(F(u),F(v))}{D_G(u,v)}. 
$$
Similarly, the \emph{contraction} of $F$ is defined as 
$$
c_{F} = \max\limits_{u,v \in V(G)} \frac{D_G(u,v)}{D_H(F(u),F(v))}. 
$$
The metric embedding $F$ is said to be of distortion $d$ if $d = e_{F}\cdot c_{F}$. When $F$ has the property that $c_{F} <1$, then the metric embedding is called a \emph{non-contracting embedding}. In this case, a non-contracting distortion $d$ metric embedding implies that the expansion is at most $d$. Therefore, for any pair $u,v \in V(G)$, 
$$
D_G(u,v) \leq D_H(F(u),F(v)) \leq d\cdot D_G(u,v). 
$$
When $d=1$, then the metric embedding is called an \emph{isometry}.

The {\sc Gen-Graph Metric Embedding} problem is defined below:

\defproblem{\sc Gen-Graph Metric Embedding}{Two graph metrics $(G,D_G)$ and $(H,D_H)$, where $G$ is a connected graph, and a positive integer $d$}{Is there a distortion $d$ metric embedding of $(G,D_G)$ into $(H,D_H)$?}

A related problem is the {\sc Graph Metric Embedding} problem, where we are only interested in finding a non-contracting distortion $d$ metric embedding.

\defproblem{\sc Graph Metric Embedding}{Two graph metrics $(G,D_G)$ and $(H,D_H)$, where $G$ is a connected graph, and a positive integer $d$}{Is there a non-contracting distortion $d$ metric embedding of $(G,D_G)$ into $(H,D_H)$?}

We also define the following restricted version of the problem.

\defproblem{\sc Red-Blue Graph Metric Embedding}{Two graph metrics $(G,D_G)$ and $(H,D_H)$, where $G$ is a connected graph and $V(H)= R \uplus B$, and a positive integer $d$}{Is there a metric embedding $F: V(G) \rightarrow R$ such that it is a non-contracting distortion $d$ metric embedding of $(G,D_G)$ into $(H,D_H)$?}

In {\sc Red-Blue Graph Metric Embedding}, the objective is to find a non-contracting distortion $d$ metric embedding that maps the vertices of $V(G)$ only to the vertices in $R \subseteq V(H)$. Now, we will explain the necessity of this restricted problem in the graph metric setting.

When $(G,D_G)$ and $(H,D_H)$ are graph metrics, then the contraction $c_{F}$ is a rational number $\frac{p}{q}$. By a result of Nayyeri et al.~\cite{NayyeriR15}, we can assume that $F$ is a non-contracting distortion $d$ metric embedding. We state a modified version of the result, tailored to our need.

\begin{pre}\label{prop:scaling}
Let $(G, D_G)$ and $(H, D_H)$ be graph metrics with $n$ and $N$ vertices, respectively. Then the {\sc Gen-Graph Metric Embedding} problem reduces to solving $\Oh(N^4n^4)$ instances of the {\sc  Red-Blue Graph Metric Embedding} problem.
\end{pre}

\begin{proof}
Let $F$ be any metric embedding with distortion $d$. Let $x,x' \in V(G)$ and $y,y' \in V(H)$ realize the expansion $e_{F}$ and the contraction $c_{F}$, respectively. Since we consider graph metrics, the contraction $c_{F}$ is a rational $\frac{p}{q}$. We subdivide all the edges of $H$ $p$ times. This results in a new graph $H'$, where we mark all the original vertices of $H$ with red and all the newly introduced subdivision vertices with blue. We define a bipartition of $V(H')$ into the set $R$ of red vertices and the set $B$ of blue vertices. Now, the mapping $F$ still is a distortion $d$ embedding of $G$ into $H'$. The extra property is that $F$ is a non-contracting distortion $d$ embedding of $G$ into $H'$. Since the original vertices of $H$ are remembered in the set $R$, we can derive a distortion $d$ metric embedding of $G$ into $H$ from a non-contracting distortion $d$ metric embedding of $G$ into $H'$, where $V(G)$ is mapped to $R$.
 Since we do not know the metric embedding $F$, we guess the vertices $x,x', F(x), F(x'), y, y', F^{-1}(y)$ and $F^{-1}(y')$. There are $\Oh(N^4n^4)$ choices for these values. Thus, if we solve $\Oh(N^4n^4)$ instances of {\sc  Red-Blue Graph Metric Embedding}, we can solve {\sc  Gen-Graph Metric Embedding}.
\end{proof}

The {\sc Gen-Graph Metric Embedding} problem or the {\sc Graph Metric Embedding} problem for a graph class $\mathcal{G}$ is a variant where the output metric space $(H,D_H)$ is such that $H\in \mathcal{G}$. In this paper, the graph classes we consider are the class of cycles, the class of generalized theta graphs and the class of graphs with treewidth $\alpha$, for any constant $\alpha$. All these classes are closed under the subdivision operation. Therefore, by Proposition~\ref{prop:scaling} solving the {\sc Gen-Graph Metric Embedding} for such a graph class is same as solving polynomially many instances of the {\sc Red-Blue Graph Metric Embedding} problem for the graph class. In this paper, we present detailed FPT algorithms for the {\sc Graph Metric Embedding} problem. All these algorithms are easily modified to solve the {\sc Red Blue Graph Metric Embedding} problem, and therefore by Proposition~\ref{prop:scaling}, FPT algorithms for the {\sc Gen-Graph Metric Embedding} problem can be designed.

The bijective {\sc Gen-Graph Metric Embedding} problem takes the same input but aims to determine whether the distortion $d$ embedding is a bijective function. The following result in~\cite{Gupta03} is useful for us to reduce the problem to that of finding a non-contracting distortion $d$ metric embedding.

\begin{pre}\label{prop:bijection}
Let $(G,D_G)$ and $(H,D_H)$ be two graph metrics. Let $F:V(G) \rightarrow V(H)$ be a bijection. Then the expansion of $F$ is achieved by an adjacent pair $(x,x') \in E(G)$ and the contraction of $F$ is achieved by an adjacent pair $(y,y') \in E(H)$. 
\end{pre} 

From the above proposition, we obtain the following result.

\begin{lem}\label{lem:bij-prelim}
Let $(G, D_G)$ and $(H, D_H)$ be graph metrics with $n$ and $m$ vertices, respectively. Then the bijective {\sc Gen-Graph Metric Embedding} problem reduces to solving $(mn)^{\Oh(1)}$ instances of the {\sc  Red-Blue Graph Metric Embedding} problem with the following property: consider a reduced instance of the {\sc  Red-Blue Graph Metric Embedding} problem where the input metric is $(G,D_{G})$ and the output metric is $(H',D_{H'})$ with $V(H') = R' \uplus B'$. (i) Any path in $H'$ of length at least $d+1$ has at least one internal vertex in $R'$; (ii) The solution metric embedding should be a bijection between $V(G)$ and $R'$ 
\end{lem}

\begin{proof}
By Proposition~\ref{prop:bijection}, for a bijective distortion $d$ metric embedding $F$, an edge in $H$ realizes the contraction $c_{F}$. Therefore,  $c_{F}\in \mathbb{N}$ and $c_{F}\leq d$. From Proposition~\ref{prop:scaling}, it is enough to solve the {\sc Red-Blue Graph Metric Embedding} problem where the output graph metric of each instance has had each of its edges subdivided at most $d$ times. Therefore, in the new output graph metric any path of length $d+1$ contains at least one internal red vertex. Also, a bijective mapping of the {\sc Gen-Graph Metric Embedding} problem implies that in a reduced instance of {\sc Red-Blue Graph Metric Embedding} we look for a bijection between the vertices of the input graph metric $(G,D_{G})$ and the red vertices $R'$ of the output graph metric $(H',D_{H'})$.
\end{proof}

In Section~\ref{sec:bij}, we describe an FPT algorithm for bijective {\sc Graph Metric Embedding} for constant treewidth graphs. Using Lemma~\ref{lem:bij-prelim}, it is possible to easily modify this algorithm to design an FPT algorithm for bijective {\sc Gen-Graph Metric Embedding}.

The following Observation relates $\Delta(H)$ to $\Delta(G)$ when $G$ can be embedded into $H$ by a non-contracting distortion $d$ embedding.

\begin{obs}\label{obs:degree-bd}
Given graph metrics $(G,D_G)$ and $(H,D_H)$, if $\Delta(G)> \Sigma_{0\leq i \leq d-1} \Delta(H)(\Delta(H)-1)^{i}$ then there is no non-contracting distortion $d$ metric embedding of $G$ into $H$.
\end{obs}

\begin{proof}
Consider any vertex $v \in V(H)$. Consider the set $B(v,d) = \{u\in V(H)~\vert~ D_H(u,v) \leq d\}$. Then $\vert B(v,d)\vert \leq \Sigma_{0\leq i \leq d-1} \Delta(H)(\Delta(H)-1)^{i}$. Let $x \in V(G)$ be such that $deg_G(x) = \Delta(G) > \Sigma_{0\leq i \leq d-1} \Delta(H)(\Delta(H)-1)^{i}$. If there is a non-contracting distortion $d$ embedding $F$ of $G$ into $H$, then without loss of generality we assume that $F(x) = v$. However, by non-contracting, $F$ is an injective function. Since $F$ has distortion $d$, every neighbour of $x$ in $G$ must be mapped to a vertex of $B(v,d)$. However, since  $\Delta(G) > \Sigma_{0\leq i \leq d-1} \Delta(H)(\Delta(H)-1)^{i}$, this is not possible. Therefore, there cannot exist a non-contracting distortion $d$ embedding of $G$ into $H$.
\end{proof}

The following observation is about non-contracting distortion $d$ embeddings into lines or cycles.
\begin{obs}
\label{obs:unwt-length_linecycle}
Let $F$ be a non-contracting distortion $d$ embedding of a graph $G$, into $H$ such that $H$ is either a line or a cycle. Let $P'$ be the maximal subpath of $H$ such that $F(x) \in V(P')$ for all $x \in V(G)$. Then $\size{P'} \leq 2dn$, where $\size{V(G)}=n$.
\end{obs}
\begin{proof}
For any two vertices $x,y \in V(G)$, $D_G(x,y) \leq n-1$. This implies $D_H(F(x),F(y)) \leq d(n-1)$.
Let $u \in V(H)$ be such that some vertex of $G$ is mapped to it. Observe that any other vertex of $H$ that has a vertex of $G$ as its pre-image, must be within a distance of $d(n-1)$ from $u$ in $H$. As the degree of $u$ is at most 2, the result follows.
\end{proof}

\section{{\sc Graph Metric Embedding} for cycles}
\label{sec:cycle}

In this Section, we are going to look at the parameterized complexity of the {\sc Graph Metric Embedding} problem for cycles. Recall that this means that the output metric $(C,D_C)$ is such that $C$ is a cycle. First, we consider the problem of embedding unweighted graphs into cycles, parameterized by the distortion $d$. We show that the problem is in FPT. Next, we consider the input graph to have edge weights, while the parameters are the distortion $d$ and the largest edge weight $W$. This problem also has an FPT algorithm, very similar to the one for unweighted graphs. However, in general the weights on the edges need not be reasonably bounded, and therefore, we consider the weighted problem with only distortion $d$ as parameter. In this case, the problem becomes NP-hard for any distortion $d>2$, where $d$ is a rational number.

\subsection{Embedding an unweighted graph into a cycle}
\label{ssec:cycle_unwt}

In this part, we will present an FPT algorithm for embedding an input unweighted graph into a given cycle. Our algorithm is similar to the FPT algorithm for {\sc Graph Metric Embedding} for lines, given in~\cite{FellowsFLLRS13}. We too try to build the metric embedding function by stitching together partial embeddings that we find locally. Similar to the definition in~\cite{FellowsFLLRS13}, a partial embedding describes what a metric embedding could possibly look like when restricted to a small subpath of the given cycle. The main idea is to employ a dynamic programming strategy to build a solution metric embedding using a set of computed partial embeddings. We compute the set of partial embeddings by an FPT algorithm and the dynamic programming algorithm is also FPT. Thereby we find a metric embedding into the given cycle in FPT time. Now we formalize the ideas. In~\cite{FellowsFLLRS13}, the algorithm searched for a special kind of embedding, called a \emph{pushing embedding}, that always exists for embedding into lines. Such an embedding need not exist for embedding into cycles. Our algorithm does not require the assumption of a pushing embedding.

Let $(G,D_G)$ be the input connected graph metric and $(C,D_C)$ be the output graph metric such that $C$ is a cycle. We denote the vertices of the cycle $C$ as $0,\ldots, N-1$ in the clockwise direction. Whenever we use $u$ to denote a vertex of $C$ in this section, consider the vertex to be numbered $u~\mbox{mod}~N$.

First, we make an observation about a property of a non-contracting distortion $d$ embedding into a cycle.
\begin{obs}
\label{obs:specialtype_cycle1}
Let $F$ be a non-contracting distortion $d$ embedding of $G$ into $C$. Then there exists at most one maximal connected subgraph $S$ of $C$ such that the pre-image of $V(S)$ is $\emptyset$
and $\size{V(S)}\geq 2d+3$.
\end{obs}
\begin{proof}
We prove the observation by contradiction. The input graph $G$ is connected. Let $S_1$ and $S_2$ be two maximal connected subgraph of $C$ such that the pre-images of both $V(S_1)$ and $V(S_2)$ are $\emptyset$ and $\size{V(S_1)},\size{V(S_2)} \geq 2d+3$. Without loss of generality, assume that the vertex set of $S_1$ is $\{1,\ldots,i\}$ and that of $S_2$ is $\{j,\ldots,k\}$, where $2d+3 \leq i < j-1 $ and $j+2d+2 < k +1 \leq N$. By definition of maximality, $0, i+1,~j-1$ and $k+1$ have pre-images. Consider the sets $\hat{S_1} = \{v\in V(G) \vert F(v) \in \{i+1,\ldots,j-1\}\}$ and $\hat{S_2} = \{v\in V(G) \vert F(v) \in \{k+1,\ldots,N=0~\mbox{mod}~N \}\}$. These subsets are nonempty and form a partition of $V(G)$.

Now,  $D_C(x,y) > d$, for all $x \in \{i+1,\ldots,j-1\}$ and $y \in \{k+1,\ldots,N\}$. Consider any pair of vertices $u ,v \in V(G) $ such that $u \in \hat{S_1}$ and $v \in \hat{S_2}$. Since $D_C(F(u),F(v)) > d$ and $F$ is a non-contracting distortion $d$ embedding of $G$, it is true that $(u,v) \notin E(G) $. This implies the graph $G$ is not connected, which is a contradiction.
\end{proof}

From now on, we assume that if there is a non-contracting distortion $d$ embedding of $G$ into $C$, the pre-image of the vertex $0 \in V(C)$ is nonempty. Also, let us assume that if there is a subgraph $S$ of $C$ such that $\size{V(S)} \geq 2d+3$ and the pre-image of $V(S)$ is $\emptyset$, then $V(S) = \{k,\ldots, N-1\}$ for some $k\leq N-(2d+3)$. This can be ensured by making a guess for the vertex of $G$ that will be mapped to $0\in V(C)$; there are polynomially many such guesses. For any $\ell \in \{0,\ldots,N-1\}$, we define $S_{\ell}=\{\ell-(d+1), \ldots, \ell+(d+1) \}$. We fix a function $\Psi : W \rightarrow S_0$, where $\Psi$ is a non-contracting distortion $d$ embedding of a vertex set $W \in V(G)$. Since we are under the assumption that the preimage of the vertex $0 \in V(C)$ is nonempty, we also make sure that $W \neq \emptyset$. All of the following definitions are with respect to $\Psi$. First, we define the notion of a partial embedding at a vertex $a \in V(C)$.

\begin{defi}
\label{defi:partial_cycle}
\begin{itemize}
\item[(i)] For an $a \in V(C)$, a \emph{partial embedding} of $U \subseteq V(G)$, with respect to $\Psi$, is a function $f_a : U \rightarrow S_a$. Here, $U$ is referred to as ${\sf Dom}_{f_a}$ and $a$ is referred to as $\md(f_a)$.
\item[(ii)] ${\sf Dom}_{f_a}^{[p,q]}$ is the set of vertices of $U$ that are mapped into
 $\{a+p,\ldots a+q\}$, where $-(d+1) \leq p \leq q \leq (d+1)$. ${\sf Dom}_{f_a}^x $ is the set of vertices of $U$ that are mapped into $\{a+x\}$.
 \item[(iii)] ${\sf Dom}_{f_a}^L= Dom_{f_a}^{[-(d+1), -1]}$ and ${\sf Dom}_{f_a}^R= Dom_{f_a}^{[1, (d+1)]}$. $L(f_a)$ ($R(f_a))$ is the union of the vertex sets of connected components of $V(G) \setminus (W \cup U)$ that have neighbours in ${\sf Dom}_{f_a}^L$ ($Dom_{f_a}^R)$.
\end{itemize}   
\end{defi}

It is important to define the feasibility of a partial embedding. Since we ultimately want to use the partial embeddings to build a metric embedding of $G$ into $C$, it is necessary that a feasible partial embedding behaves like a metric embedding locally.

\begin{defi}
\label{defi:feasible_cycle}
Let $f_a:U \rightarrow S_a$ be a \emph{partial embedding} with respect to $\Psi$, where $U \subseteq V(G) $. Then $f$ is said to be \emph{feasible with respect to $\Psi$}, if it satisfies the following conditions. 
\begin{itemize}
\item[(i)] $U \neq \emptyset$, $U \cap W = \emptyset$ and $S_a \cap S_0 = \emptyset$.
\item[(ii)] $f_a$ is a non-contracting distortion $d$ embedding of $U$.
\item[(iii)] For every pair $u \in U$ and $w \in W$, $D_C(f_a(u),\Psi(w))$ is non-contracting and has expansion at most $d$, 
\item[(iv)] If $u$ is a neighbour of a vertex in ${\sf Dom}_{f_a}^0$, then $f_a(u) \in S_a$.
\item[(v)] $L(f_a) \cap R(f_a) = \emptyset$.
\end{itemize}
\end{defi}

\remove{
\begin{defi}

\label{defi:lambda_cycle}
Let $f_a:U \rightarrow S_a$ be a \emph{feasible partial embedding} with respect to $\Psi$. Then
  $\lambda\left(f_a \right)$ is the maximum $m < a-(d+1)$ satisfying one of the the following condition. 
  \begin{itemize}
  \item  There exists a $v \in V(G) \setminus  U$ and a feasible partial embedding $f_{a'}$ such that $f_{a'}(v)=m $, where $a' < a$. 
  \item  There exists a $w \in W $ such that $\Psi(w)=m $.
  
\end{itemize}   
\end{defi}
}

Next, we define the notion of succession from one partial embedding to another. This definition will help us put together many partial embeddings to get one metric embedding of $G$ into $C$.

\begin{defi}
\label{defi:succeed_cycle}
 Let $f_a:U \rightarrow S_a$ be a feasible partial embedding of $U$ with respect to $\Psi$. Let $f_b:U' \rightarrow S_b$ be a feasible partial embedding of $U'$ with respect to $\Psi$. Then $f_b$ \emph{succeeds} $f_a$ with respect to $\Psi$
  if the following conditions are satisfied.
\begin{itemize}

\item[(i)] ${\sf Dom}_{f_a}^{[-d,d+1]} = {\sf Dom}_{f_b}^{[-(d+1),d]}$.
\item[(ii)] $\forall u \in {\sf Dom}_{f_a} \cap {\sf Dom}_{f_b} $, $f_a(u)=f_b(u)$.
\remove{
\item[(iii)] If ${\sf Dom}_f^{-(d+1)}=\emptyset$, then $\lambda(f_b)=\lambda(f_a)$. Otherwise, $\lambda(f_b)$
             is the vertex of $G$ that is mapped to $a-(d+1)$.
 \item[(iv)] Let $u_{\lambda}$ is the vertex that is mapped to $\lambda(f_b)$ by some feasible partial embedding $f_{a'}$. The path between $f_{a'}(u_{\lambda})$ and $f_b(u)$ is neither contracting nor expanding by a factor of more than $d$ for any $u \in U_b$.
 }
\item[(iii)]  ${\sf Dom}_{f_a}^{-(d+1)} \subseteq  L(f_b)$.
\item[(iv)] ${\sf Dom}_{f_b}^{(d+1)} \subseteq  R(f_a)$.
\end{itemize}
\end{defi}

While we are considering a partial embedding $f$, the sets $L(f)$ and $R(f)$ help us to have a view of the mapping of the rest of the graph $G$ in terms of this partial embedding $f$. The next Lemma proves that our definition of succession implies that we have a consistent view of the mapping of the rest of $G$ when we move from one partial embedding to another.

\begin{lem}
\label{lem:succeed}
Let $f_a:U \rightarrow S_a$ be a feasible partial embedding of $U$ with respect to $\Psi$, and $f_b:U' \rightarrow S_b$ be a feasible partial embedding of $U'$ with respect to $\Psi$. If $f_b$ \emph{succeeds} $f_a$ with respect to $\Psi$, then the following properties hold.
\begin{itemize}
\item[(i)] $R(f_a)  = R(f_b) \cup {\sf Dom}_{f_b}^{d+1}$.
\item[(ii)] $L(f_b) = L(f_a) \cup {\sf Dom}_{f_a}^{-(d+1)}$.
\end{itemize}
\end{lem}

\begin{proof}
We prove property (i), and property (ii) can be proved similarly. Note that $f_b$ succeeds $f_a$. 
In the forward direction, we prove that $R(f_a)  \subseteq  R(f_b) \cup {\sf Dom}_{f_b}^{d+1}$.
  Let us consider a connected component $Y$ such that $V(Y) \subseteq R(f_a)$. Recall the definition of $R(f_a)$. Since the two embeddings are feasible, $L(f_a)\cap R(f_a)=\emptyset$. This implies that $Y$ does not contain any neighbour of ${\sf Dom}_{f_a}^{-(d+1)}$. Hence, $Y$ is a connected component of $ G \setminus  \left( {\sf Dom}_{f_a}^{[-d,d+1]} \cup W \right) = G \setminus  \left( {\sf Dom}_{f_b}^{[-(d+1),d]} \cup W \right)$.

If $Y$ does not contain any vertex from ${\sf Dom}_{f_b}^{d+1}$, then $Y$ is a connected component of $G \setminus \left( {\sf Dom}_{f_b} \cup W \right)$. As $V(Y) \subseteq R(f_a)$, by definition $Y$ contains a neighbour of 
${\sf Dom}_{f_a}^R = {\sf Dom}_{f_a}^{[1,d+1]} = {\sf Dom}_{f_b}^{[0,d]} $. By definition of feasibility, if $Y$ had a neighbour in ${\sf Dom}_{f_b}^{[1,d]}$, then $Y$ would have to contain a vertex in ${\sf Dom}_{f_b}^{[1,d+1]}$. Since $Y$ is a connected component of $G \setminus \left( {\sf Dom}_{f_b} \cup W \right)$ with no vertex in ${\sf Dom}_{f_b}^{d+1}$, $Y$ cannot have a neighbour in ${\sf Dom}_{f_b}^{0}$. So, $Y$ has neighbours in ${\sf Dom}_{f_b}^{[1,d]} \subseteq {\sf Dom}_{f_b}^R$. Hence, $V(Y) \subseteq R(f_b)$.

On the other hand, if $Y$ contains a vertex from ${\sf Dom}_{f_b}^{d+1}$, then each connected 
component $Y'$ of $G[Y \setminus \left( {\sf Dom}_{f_b}^{d+1} \cup W \right)]$,  
is also a connected component of $ G \setminus \left({\sf Dom}_{f_b} \cup W \right)$. So, $V(Y') \subseteq R(f_b)$. This implies that $V(Y) \subseteq R(f_b) \cup {\sf Dom}_{f_b}^{d+1}$.

To complete the proof, in the backward direction, we prove that $R(f_b) \cup {\sf Dom}_{f_b}^{d+1} \subseteq R(f_a)$.
By condition (iv) of Definition~\ref{defi:succeed_cycle}, ${\sf Dom}_{f_b}^{d+1} \subseteq  R(f_a)$. Now consider a connected component $Y$ such that $V(Y) \subseteq R(f_b)$. By the definition of $R(f_b)$, $Y$ is a connected
 component of $G \setminus \left( {\sf Dom}_{f_b} \cup W \right)$. By condition (iii) of Definition~\ref{defi:embed_cycle}, ${\sf Dom}_{f_a}^{-(d+1)} \subseteq  L(f_b)$. Also, $L(f_b) \cap R(f_b) = \emptyset$ as
 $f_b$ is a feasible partial embedding. So, $Y$ does not contain any vertex of ${\sf Dom}_{f_a}^{-(d+1)}$. This
 implies $Y$ is a connected component of $G \setminus ({\sf Dom}_{f_a} \cup {\sf Dom}_{f_b} \cup W)$.
 
  If $Y$ does not have a neighbour in ${\sf Dom}_{f_b}^{d+1}$, then $Y$ has neighbours only in ${\sf Dom}_{f_b}^{[1,d]}={\sf Dom}_{f_a}^{[2,d+1]} \subseteq {\sf Dom}_{f_a}^R $. Thus, $V(Y) \subseteq R(f_a)$.
    If $Y$ has a neighbour in $Dom_{f_b}^{d+1}$, then in the graph $G \setminus ({\sf Dom}_{f_a} \cup W)$, $Y$ and at least one vertex from ${\sf Dom}_{f_b}^{d+1}$ are in the same 
    connected component. Since ${\sf Dom}_{f_b}^{d+1} \subseteq  R(f_a)$ and since $L(f_a) \cap R(f_a) = \emptyset$, $V(Y) \subseteq R(f_a)$.

Putting everything together, $R(f_b) \cup {\sf Dom}_{f_b}^{d+1} \subseteq R(f_a)$.
Hence, we have proved property (i).  
\end{proof}

Finally, the following definitions help to establish the meaning of building a metric embedding using a sequence of partial embeddings.

\begin{defi}
\label{defi:union_function}
Let $\Pi =\{f_i: A_i \rightarrow B_i, i \in [t]\}$ be a set of functions such that for any $i,j \in [t]$, $x \in A_i \cap A_j$ implies $f_i(x)=f_j(x)$. Then we define $\Phi_{\Pi}:\bigcup\limits_{i=1}^{t}A_i\rightarrow \bigcup\limits_{i=1}^{t} B_i$ such that $\Phi_{\Pi}(x)=f_i(x)$ for $i \in [t], x \in A_i$.
\end{defi}
\begin{defi}
\label{defi:embed_cycle}
A sequence of feasible partial embedding $\Phi_1,\ldots,\Phi_t$ with respect to $\Psi$ is \emph{embeddable with respect to $\Psi$} if it satisfies the following conditions. 
\begin{itemize}
  \item[(i)] $\Phi_i$ \emph{succeeds} $\Phi_{i-1}$ with respect to $\Psi$ for every $2 \leq i \leq t$.
  \item[(ii)] Let $F = \Phi_{\Pi}$, where $\Pi = \{\Phi_i : i \in [t]\} \cup {\Psi}$. Suppose $\Phi_1: A \rightarrow S_a$ and $ \Phi_t:B \rightarrow S_b$. Let us call the vertices $\{1,2,\ldots d\} \in V(C)$ as $C_b$ and the vertices $\{-(d+1),\ldots,-1\} \in V(C)$ as $C_e$. If $u\in V(G)$ is a neighbour of some vertex in ${\sf Dom}_{\Phi_1}^L$, then $F(u ) \in S_a \cup C_b$ and if $u\in V(G)$ is a neighbour of some vertex in ${\sf Dom}_{\Phi_t}^R$, 
   then $F(u) \in S_b \cup C_e$.
  \item[(iii)] $L(\Phi_1)=\emptyset, R(\Phi_1)=V(G)\setminus ({\sf Dom}_{\Phi_1} \cup W)$ and $L(\Phi_t)=V(G) \setminus ({\sf Dom}_{\Phi_t} \cup W), R(\Phi_t)=\emptyset$.
  \end{itemize}
\end{defi}

We prove the relation between a non-contracting distortion $d$ embedding of $G$ into $C$ and a sequence of feasible partial embeddings.

\begin{lem}
\label{lem:succeed_embed_cycle}
A graph $G$ has a non-contracting distortion $d$ embedding into a cycle $C$ 
 if and only if there exist a $W \subseteq V(G)$, a non-contracting distortion $d$ partial embedding $\Psi: W \rightarrow S_0$ and an \emph{embeddable} sequence of feasible partial embeddings $\Phi_1,\ldots,\Phi_t$ with respect to $\Psi$. 
 \end{lem}

\begin{proof}
First, we show that if $F$ is a non-contracting distortion $d$ embedding of $G$ into $C$ then there is an embeddable sequence of feasible partial embeddings $\Phi_1,\ldots,\Phi_t$ with respect to some $\Psi$. By Observation~\ref{obs:specialtype_cycle1}, there exists at most one connected subgraph $S$ of $C$ such that the pre-image of $S$ under $F$ is $\emptyset$ and $\size{S} \geq 2d+3$. If such a subgraph $S$ exists, then without loss of generality assume that the vertices of $S$ are $\{k,\ldots, N-1\}$ for some $k\leq N-(2d+3)$. Now, take $W$ as the set of vertices that are mapped by $F$ into $S_0$. Take a function $\Psi:W \rightarrow S_0$ to be the restriction of $F$ to $S_0$. Let $t=\max\limits_{v \in V(G)} F(v) - (2d+2)$. For $i \in [t]$, let $\Phi_i$ be $F$ restricted to the vertices that are mapped into $S_{2d+2+i}$.  Observe that $F=\Phi_{\Pi}$, where $\Pi=\{\Phi_i,i \in [t]\} \cup \{\Psi\}$. Now, it is easy to verify that $\Phi_1,\ldots,\Phi_t$ are all \emph{embeddable} with respect to $\Psi$.

Conversely, let $\{\Phi_1,\ldots,\Phi_t\}$ be a sequence of \emph{embeddable} feasible partial 
embeddings with respect to $\Psi: W \rightarrow S_0$. We make the following claims about the sequence of embeddable feasible partial embeddings, which we prove later.
\begin{cl}
\label{clm:cycle_vertex}
For each $v \in V(G)$, either $v \in W$ or there exists some $\Phi_i$ such that $v \in Dom(\Phi_i)$.
\end{cl}
\begin{cl}
\label{clm:cycle_inj}
 The following properties hold:
 \begin{itemize}
 \item[(a)] For any $v \in V(G) \setminus W$, let $A_v=\{\md(\Phi_i) \in V(C) \setminus S_0: v \in Dom_{\Phi_i}, 1 \leq i \leq t\}$. Then the subgraph of $C$ induced  by $A_v$, is connected.
 \item[(b)] $v \in Dom_{\Phi_i} \cap Dom_{\Phi_j}$ implies $\Phi_i(v)=\Phi_j(v)$. 
 \end{itemize} 
\end{cl}

Consider $F=\Phi_\Pi$ where $\Pi = \{\Phi_i, i \in [t]\} \cup \{ \Psi \}$. We  are done with
the proof of the converse part by the following claims, which we will prove later.
\begin{cl}
\label{clm:cycle_dist}
The function $F=\Phi_{\Pi}$ is a distortion $d$ embedding of $G$ into $C$.
\end{cl}
\begin{cl}
\label{clm:cycle_nct}
The function $F=\Phi_{\Pi}$ is a non-contracting embedding of $G$ into $C$.
\end{cl}
\end{proof}

\begin{proof}[Proof of Claim~\ref{clm:cycle_vertex}]
 If $v \in W$ or $v \in {\sf Dom}_{\Phi_1}$, then we are done. Assume that $v \notin W \cup \Phi_1$. 
This implies $v \in R(\Phi_1)$ as the graph $G$ is connected and $L(\Phi_1) = \emptyset$. Then by Lemma~\ref{lem:succeed}, $v \in R(\Phi_2) \cup {\sf Dom}_{\Phi_2}^{d+1}$. If $v \in  {\sf Dom}_{{\Phi_2}}^{d+1}$, then $v \in {\sf Dom}_{\Phi_2}$ and we are done. Otherwise, $v \in R(\Phi_2)$. Let $m-1$
be the maximum index such that $v \in R(\Phi_{m-1})$. Observe that $m \leq t$ as $R(\Phi_t)=\emptyset$ and $v \notin W$. By Lemma~\ref{lem:succeed}, we can conclude that $ v \in {\sf Dom}_{\Phi_m}^{d+1}$, i.e., $v \in {\sf Dom}_{\Phi_m}$.
\end{proof}

\begin{proof}[Proof of Claim~\ref{clm:cycle_inj}]
\remove{Let $k \in [t]$ be the smallest index such that $v \in Dom_{\Phi_k}$.
Let $k'=\min\{t,f_{k}+d+1\}$. Using the fact that
 $\Phi_{i+1}$ succeeds $\Phi_{i}$ for all $i \in [t-1]$, for all $k \leq a,b \leq k'$,
 $\Phi_{a}(v)=\Phi(v)$. $v \in Dom_{}$}
  \begin{itemize}
 \item[(a)] We give a proof by contradiction. Let $v \in {\sf Dom}_{\Phi_i}$, $v \in {\sf Dom}_{\Phi_j}$ such that $\md(\Phi_i)$ and $\md(\Phi_j)$
are not adjacent in the graph $C \setminus W$. W.l.o.g assume that $\md(\Phi_i) < \md(\Phi_j)$.
 Consider the path $P=\{x_1,\ldots,x_k\}, k \geq 3,$  from $\md(\Phi_i)$ to $\md(\Phi_j)$ in the graph 
 $C \setminus W$, where $x_1=\md(\Phi_i) $ and $x_k =\md(\Phi_j)$.
 
 Recall the definition of succession from Definition~\ref{defi:succeed_cycle}. The partial embedding $\Phi_j$ cannot succeed $\Phi_i$ as $\md(\Phi_i)$ and $\md(\Phi_j)$
are not adjacent. So, there must exist some $\Phi_{i'}$ such that $\mbox{mid}(\Phi_{i'})$ is a vertex in $P$ such that
 $\md(\Phi_i)< \md(\Phi_{i'})< \md(\Phi_j)$. This implies $v \in L(\Phi_{i'})$ and $v \in R(\Phi_{i'})$, i.e.,
 $L(\Phi_{i'}) \cap R(\Phi_{i'}) \neq \emptyset$. This is not possible as $\Phi_{i'}$ is a feasible partial embedding.
 \item[(b)] The statement follows from part (a) and by using the definition of succession.
 \end{itemize}
\end{proof}

\begin{proof}[Proof of Claim~\ref{clm:cycle_dist}]
By Claim~\ref{clm:cycle_vertex}, ${\sf Dom}_F=V(G)$. Consider any pair of vertices $u, v \in V(G)$. We need to show the following 
statement. $D_C(F(u),F(v)) \leq d. D_G(u,v)$. We use the method of induction on $D_G(u,v)$. For now, assume 
that the statement is true for $D_G(u,v)=1$. We will prove it after the inductive step. Let the statement be true for
all $D_G(u,v) < l$. Consider a pair of vertices $u,v$ such that $D_G(u,v)=l > 1$. Let $x$ be some vertex, other than
$u$ and $v$, in the shortest path from $u$ to $v$, i.e, $D_G(u,v) =D_G(u,x) + D_G(x,v)$. 
By triangle inequality, $D_C(F(u),F(v)) \leq D_C(F(u),F(x)) + D_C(F(x),F(v)) $ holds. By the induction hypothesis, $D_C(F(u),F(x)) \leq d. D_G(u,x)$  and $D_C(F(x),F(v)) \leq d. D_G(x,v)$ hold. Putting everything together, 
$D_C(F(u),F(v)) \leq d. D_G(u,v)$. Thus, we need to prove the base case of this induction hypothesis. Consider an edge $(u,v)$ in $G$. We break the analysis into the following cases.

\begin{itemize}
\item {\bf Case 1($v \in W$):}
If $u \in W$, then $D_C(F(u),F(v)) \leq d$ as $\Psi$ is a distortion $d$ embedding of $W$.
If $u \notin W$, then $u $ is in the domain of some $\Phi_i$.  The path between $\Psi(v)=F(v)$ and $\Phi_i(u)=F(u)$ does not expand by a factor of more than $d$, due to condition (iii) of Definition~\ref{defi:feasible_cycle}. This implies $D_C(F(u),F(v)) \leq d$.

\item {\bf Case 2 ($v \in {\sf Dom}_{\Phi_1}^L$):}
In this case, again we denote the vertex set $\{1,\ldots,d+1\} \in V(C)$ as $C_b$. $F(u) \in S_a \cup C_b$ by condition (ii) of Definition~\ref{defi:embed_cycle}. 

If $F(u) \in S_a$, both $u,v \in {\sf Dom}_{\Phi_1}$, $D_C(F(u),F(v)) \leq d$. This is because $\Phi_1$ is a feasible
partial embedding of ${\sf Dom}_{\Phi_1}$. 

If $F(u) \in  C_b$, then $u \in W$.  The path between $\Phi_1(v)=F(v)$ and $\Psi(u)=F(u)$ is not expanding by a factor of more than $d$ by condition (iii) of Definition~\ref{defi:feasible_cycle}. So, $D_C(F(u),F(v)) \leq d$.

\item {\bf Case 3 ($v \in {\sf Dom}_{\Phi_t}^R$):}
In this case, we denote the vertex set $\{-(d+1),\ldots,1\}\in V(C)$ as $C_e$. $F(u) \in S_b \cup C_e$ by  condition (ii) of Definition~\ref{defi:embed_cycle}. 

If $F(u) \in S_b$, both $u,v \in {\sf Dom}_{\Phi_t}$, $D_C(F(u),F(v)) \leq d$. This is because $\Phi_t$ is a feasible
partial embedding of ${\sf Dom}_{\Phi_t}$. 

If $F(u) \in  C_e$, then $u \in W$.  The path between $\Phi_t(v)=F(v)$ and $\Psi(u)=F(u)$ is not expanding by a factor of more than $d$ by condition (iii) of Definition~\ref{defi:feasible_cycle}. So, $D_C(F(u),F(v)) \leq d$.

\item {\bf Case 4 ($v \notin W \cup {\sf Dom}_{\Phi_1}^L \cup {\sf Dom}_{\Phi_t}^R$):}
It is easy to observe that there exists a $\Phi_i$ such 
that $v \in {\sf Dom}_{\Phi_i}^0$. By condition (iv) of 
Definition~\ref{defi:feasible_cycle}, we can say that $u \in {\sf Dom}_{\Phi_i}$. Note that
both $u$ and $v$ are in ${\sf Dom}_{\Phi_i}$ and $\Phi_i$ is a feasible partial embedding.
So, $D_C(F(u),F(v)) \leq d$.
\end{itemize}
\end{proof}

\begin{proof}[Proof of Claim~\ref{clm:cycle_nct}]
By Claim~\ref{clm:cycle_vertex}, ${\sf Dom}_F=V(G)$. Let us consider 
$u,v \in V(G)$. Let $P_C(u,v)=\{x \in V(G),~\mbox{$F(x)$ is in the shortest path from $F(u)$ to $F(v)$ in $C$} \}$. 
Note that $u,v \in P_C(u,v)$. We have to show that 
$D_G(u,v) \leq D_C(F(u),F(v))$ for any $u,v \in V(G)$.

 We divide the proof into the following cases.
\begin{itemize}
\item {\bf  Case 1($P_C(u,v) \cap W \neq  \phi$):}

For some $x \in W$, $d_C(F(u),F(v)) = d_C(F(u),F(x))+d_C(F(x),F(v)) $. Using the fact that
 $u(v)$ is in the domain of some feasible partial embedding and recalling condition (iii) of Definition~\ref{defi:feasible_cycle}, observe that $D_G(u,x) \leq D_C(F(u),F(x))$ and $D_G(x,v) \leq D_C(F(x),F(v))$. This implies $D_G(u,v) \leq D_C(F(u),F(v))$.
 
 \item {\bf Case 2($P_C(u,v) \cap W =  \phi$):}
 {\bf Fact 1:} If $D_C(F(u),F(v)) \leq 2d+2$, then there exists
 some $\Phi_i$ such that $u,v \in Dom(\Phi_i)$. 
 
 This implies the path between $F(u)$ and $F(v)$ is non-contracting and has expansion of at most $d$ due to condition (ii) of Definition~\ref{defi:feasible_cycle}.
 
  Now assume that $D_C(F(u),F(v)) \geq  2d +3 $. Let $P_C(u,v)=\{x_1,\ldots,x_l\}$ such that $D_C(F(u),F(v))= \sum\limits_{i=1}^{l-1}D_C(F(x_i),F(x_{i+1}))$ and
  $P_C(x_i,x_{i+1})=\{x_i,x_{i+1}\}$. Note that $u=x_1$, $v=x_l$, and $D_C(F(x_i),F(x_{i+1})) \leq 2d + 2$ .
   Observe that
   \begin{center}
   $D_G(x_i,x_{i+1}) \leq D_C(F(x_i),F(x_{i+1}))$ by Fact 1
\end{center}   
   Hence, $D_G(u,v) \leq D_C(F(u),F(v)) $.
     \end{itemize}
   Hence, we are done.
\end{proof}

Thus, we have seen that finding a sequence of feasible partial embeddings is enough to find a non-contracting distortion $d$ embedding of $G$ into $C$. What remains to be proven is that the number of such sequences is bounded by an FPT function. 
\remove{
\begin{lem}
\label{lem:local_cycle}
If $\Delta (G) > 2d$, then there does not exist a distortion $d$ embedding of $G$
into a cycle.
\end{lem}
}

In the following observation, we describe the structure of a partial embedding.

\begin{obs}
\label{obs:seq_cycle}
Let $f:U \rightarrow S_a$ be any feasible partial embedding with respect to $\Psi$.
 Then we can describe it by the following properties.
\begin{itemize}
\item[(i)] An integer $a$ such that $2d+3 \leq a \leq N-(2d+3)$.
\item[(ii)] A sequence of vertices $u_1,\ldots,u_p$.
\item[(iii)] A sequence of non-negative integers $x_0,\ldots,x_{p-1}$ such that 

$x_0+\sum\limits_{i=1}^{p-1}(D_G(u_i,u_{i+1})+x_i) \leq 2d+2$.
\end{itemize}
\end{obs}
\begin{proof}
Consider the vertex set $\{u_1,\ldots,u_p\}$ as ${\sf Dom}_{f}=U$, such that $f(u_1) < f(u_2) \ldots <f(u_p)$. Let $x_0 = d+1 - (a - f(u_1))$. For $i \in \{1,\ldots, p-1\}$ let $x_i = D_C(f(u_i),f(u_{i+1})) - D_G(u_i,u_{i+1})$. Then, $f(u_j)=a-(d+1)+ x_0 + \left(\sum\limits_{i=1}^{j-1}(D_G(u_i,u_{i+1}\right) +x_i)$. This implies the statement of the Observation.
\end{proof}

Now, we are ready to bound the number of feasible partial embeddings possible.

\begin{defi}
\label{defi:count}
Let $x$ be a non-negative integer and $\cN_{u_1}(x)$ be the number of sequences $u_1,\ldots,u_p$ such that $\sum\limits_{i=1}^{p-1}  D_C(f(u_i),f(u_{i+1}))=x$. We define $\cN(x)=\max\limits_{u_1 \in V(G)} \cN_{u_1}(x)$.
\end{defi}

\begin{lem}
\label{lem:cycle_N(x)}
For any non-negative integer $x$, $\cN(x)\leq (4d(2d+2))^{x}$.
\end{lem}
\begin{proof}
We prove the statement using induction on $x$.
For $x=0$, the statement is trivially true. Assume that the statement
 is true for every $y<x$. Let $\cS_{i}$ be the set of sequences such that
 $D_C(f(u_1),f(u_2))=i$, where $i \in [x]$. As $f$ is a non-contracting embedding, $D_G(u_1,u_2) \leq i$.
 Let $A_{u_1}(i)$ be the set of vertices $u$ such that $D_G(u_1,u)\leq i$. Observe that, 
 $\cN_{u_1}(x)\leq \sum\limits_{i=1}^x  \size{\cS_{i}} \leq \sum\limits_{i=1}^x \size{A_{u_1}(i)}  \cdot \cN(x-i) $. 
 
 By the induction hypothesis, 
 \begin{center}
$ \cN(x-i) \leq (4d(2d+2))^{x-i}$.
 \end{center}
 Note that by Observation~\ref{obs:degree-bd}, if there is a non-contracting distortion $d$ embedding of $G$ into $C$, it must be true that $\Delta(G)\leq 2d$. This implies that $\size{A_{u_1}(i)}\leq  2 \cdot (2d)^{i}$.  
 So,
 \begin{eqnarray*}
 \cN{u_1}(x) &\leq&  \sum\limits_{i=1}^x  2\cdot (2d)^{i}\cdot (4d(2d+2))^{x-i} \\
  &\leq& (4d(2d+2))^{x} \sum\limits_{i=1}^x  \frac{1}{(2d+2)^{i}} \\
  &\leq& (4d(2d+2))^{x} 
 \end{eqnarray*}
  Note that the above discussion is true for any $u_1 \in V(G)$. Hence, 
  \begin{center}
  $\cN(x) \leq (4d(2d+2))^{x}$.
  \end{center}
\end{proof}

\begin{lem}
\label{lem:num_par_cycle}
Let $\size{V(G)}=n, \size{V(H)}=N$. The total number of feasible partial
 embeddings with respect to a particular $\Psi$, is at most $\Oh \left(N\cdot n\cdot(4d(2d+2))^{2d+2} \right)$ and the total
 number of possible $\Psi:W \rightarrow S_0$ is $\Oh \left(n\cdot (4d(2d+2))^{2d+2}\right)$.
\end{lem}
\begin{proof}
By Observation~\ref{obs:seq_cycle} and Definition~\ref{defi:count}, we can have
 at most $\cN \left( 2d+2-x_1 \right)$ feasible partial embeddings with respect to $\Psi$ for a fixed integer 
 $a$ and fixed vertices $u_1 \in V(G)$, where $u_1$ is mapped to $a-(d+1)+x_1$. Hence, the total
 number of feasible partial embeddings is at most 
\begin{eqnarray*}
N\cdot n\cdot \sum\limits_{x_1=0}^{2d+2}\cN \left( 2d+2-x_1\right)
&\leq& N\cdot n\cdot \sum\limits_{x_1=0}^{2d+2}(4d(2d+2))^{2d+2-x_1} \\
&=& O\left(N\cdot n\cdot (4d(2d+2))^{2d+2} \right).
\end{eqnarray*} 

 Observe that a particular $\Psi$ can be described by a vertex sequence $u_1 \ldots u_p$ along with a sequence of non-negative integers $x_0,\ldots ,x_{p-1}$ such that $x_0+\sum\limits_{i=1}^{p-1}(D_G(u_i,u_{i+1})+x_i) \leq 2d+2$. Hence, the stated bound on the total number of possible $\Psi$ follows from Lemma~\ref{lem:cycle_N(x)}.
\end{proof}

Putting everything together, our algorithm tries to find a sequence of feasible partial embeddings in order to find a non-contracting distortion $d$ embedding of $G$ into $C$. The bound on feasible partial embeddings makes the algorithm run in FPT time.

\begin{theo}
\label{theo:main_cycle}
Given an undirected unweighted graph $G$ on $n$ vertices, a cycle $C$ and a distortion parameter $d$, there exists an algorithm that either finds a non-contracting distortion $d$ embedding of $G$ into $C$ or decides that there does not exist such an embedding in $\Oh\left(n^3 \cdot d^{2d+3} \cdot (4d(2d+2))^{4d+4}\right)$ time.
\end{theo}

\begin{proof}
Suppose there exists a desired embedding $F$ from $G$ to $C$. By Observation~\ref{obs:unwt-length_linecycle}, there exists a maximal subpath $P$ of length
$2dn$ of $C$ such that $F(x) \in V(P)$ for all $x \in V(G)$. If $\size{V(C)}=N > 4dn$, then 
consider an induced connected subgraph $C'$ of $C$ having $2dn$ vertices. Observe that
$C'$ is a line and  
$G$ can be embedded into $C$ if and only if $G$ can be embedded into $C'$. We run the algorithm 
of embedding an unweighted graph into a line~\cite{FellowsFLLRS13}, to find a possible embedding of $G$ into $C'$. Note that the time complexity of the mentioned algorithm in ~\cite{FellowsFLLRS13}, is $\Oh(nd^4(2d+1)^{2d})$.

If $N \leq 4dn$, then we do the following. Let $F$ be a distortion $d$ embedding of $G$ into $C$, if one exists. We find $\Delta(G)$ in $O(n)$ time.
By Observation~\ref{obs:degree-bd}, if $\Delta(G)> 2d$ then we report no $F$ exists. Otherwise, we try to find an embeddable sequence of partial embeddings, according to Lemma~\ref{lem:succeed_embed_cycle}. We first fix $\Psi: W \rightarrow S_0$. Let $
 \cF_{\Psi}$ be the set of all possible partial embeddings with respect to $\Psi$ 
 from $G$ to $C$. By Observation~\ref{obs:seq_cycle} and Definition~\ref{defi:count}, $\size{\cF_{\Psi}} \leq O\left(\cbound \right)$. Then we make the following construction:
 \begin{itemize}
 \item[(i)] Construct a directed graph $\cG$, where $V(\cG)=\cF_\Psi \cup \{\cS,\cD\}$ and $E(\cG)=E_1 \cup E_2 \cup E_3$,
  where 
  \begin{eqnarray*}
  E_1 &=& \{(\Phi_a,\Phi_b) : \Phi_b ~\mbox{succeeds}~\Phi_a ~\mbox{w.r.t.}~ \Psi\}\\
   E_2 &=& \{(\cS,\Phi_1):L(\Phi_1)=\emptyset, R(\Phi_1)=V(G)\setminus ({\sf Dom}_{\Phi_1} \cup W)\} \\
    E_3 &=& \{(\Phi_t,\cD): L(\Phi_t)=V(G) \setminus ({\sf Dom}_{\Phi_t} \cup W), R(\Phi_t)=\emptyset\}.
  \end{eqnarray*}
   
 \item[(ii)] Then we check for the existence of a directed path from $\cS$ to $\cD$. This implies that there is a directed path from some $\Phi_1 \in \cF_{\Psi}$ to some $\Phi_t \in \cF_{\Psi}$ in $\cG$ 
 such that conditions (ii) and (iii) of Definition~\ref{defi:embed_cycle} are satisfied. If such a path exists, then there exists an embeddable sequence, say $\Phi_1,\ldots,\Phi_t$, of feasible partial embeddings with respect to $\Psi$. Report $\Phi_{\Pi}$ as $F$, where $\Pi = \{\Phi_i:i \in [t]\}$.
 
 Otherwise, there does not exist an \emph{embeddable} sequence of feasible partal embeddings with respect to $\Psi$.
  \end{itemize}  
 If there does not exist an \emph{embeddable} sequence of feasible partal embeddings with respect to $\Psi$,
 then we conclude that we cannot have a non-contracting and distortion $d$ embedding of $G$ into $C$.
Observe that we execute the above procedure for all possible $\Psi$ in the worst case. But, the total number of possible $\Psi$ is at most $O\left(n\cdot(4d(2d+2))^{2d+2}\right)$ by Lemma~\ref{lem:num_par_cycle}. The number of vertices in graph $\cG$ is at most $O\left(\cbound\right)$. Every vertex of $\cG$ has outdegree at most $O(d^{2d})$. This is because, if $f_b$ succeeds $f_a$, then $f_b$ is determined by $f_a$ along with the condition (i) of Definition~\ref{defi:succeed_cycle}. Similar to~\cite{FellowsFLLRS13}, we can test whether a
partial embedding succeeds another using a prefix tree like data structure in $\Oh(d^2)$ time. So, the total running time of Step-(i)
 is $\Oh\left( d^2\cdot d^{2d}\cdot \cbound \right)$. We can test the existence of a path of the mentioned type in Step-(ii), by running the DFS algorithm in $\Oh\left(\cbound \right)$ time. Hence, our algorithm takes $\Oh\left( d^2\cdot d^{2d}\cdot \cbound \right)$ time for each $\Psi$ and $\Oh\left(N\cdot n^2\cdot d^{2d+2}\cdot(4d(2d+2))^{4d+4}\right)$ in total. As $N \leq 2dn$, the stated bound of the time complexity follows.
\end{proof}

\paragraph*{A Note on Embedding into lines}

Note that we can design a similar algorithm for embedding into lines. An FPT algorithm for embedding a graph metric $(G,D_G)$ 
into a graph metric $(H,D_H)$, where $H$ is a line, already exists in~\cite{FellowsFLLRS13}. A key observation made for embedding into lines 
was that it is enough to search for a pushing embedding of $G$ into $H$. For a non-contracting distortion $d$ embedding $F$, 
let $v_1, v_2,\ldots, v_n$ be an ordering of the vertices such that $F(v_1) < F(v_2) < \ldots < F(v_n)$. Then $F$ is called a 
pushing embedding if $D_H(f(v_i),f(v_{i+1})=D_G(f(v_i),f(v_{i+1}))$ for all $1 \leq i \leq n-1$. Our algorithm does not 
assume pushing embeddings and therefore will work even for embedding into lines, where the maps of certain vertices are 
already fixed. Specifically in this part, we mention two results which will be useful for us in Section~\ref{sec:theta}.

A line with $N$ vertices can be thought to have vertices named $\{1,2,\ldots,N\}$. 

\begin{lem}\label{lem:path-st-algo}
Let $(G,D_G)$ and $(H,D_H)$ be two graph metrics on $n$ and $N$ vertices,respectively. Let $a_1,a_2$ be two positive integers such that $a_1 +a_2 \leq N$. Let $Z_1 = \{1,2,\ldots,a_1\}$ and $Z_2 =\{N-a_2+1, N-a_2+2\ldots,N\}$. Let $\Psi_1:U_1\rightarrow Z_1$ and $\Psi_2:U_2\rightarrow Z_2$ be two fixed non-contracting distortion $d$ embeddings of two disjoint sets $U_1,U_2\subseteq V(G)$. Then there is an FPT algorithm to determine whether there exists a non-contracting distortion $d$ embedding of $G$ into $H$ that extends $\Psi_1$ and $\Psi_2$. The running time of the algorithm is $\Oh\left(\linethetabound\right)$.
\end{lem}

\begin{proof}
Without loss of generality, assume that there exists a vertex of $G$ mapped to $1$ and $N$. If $N > 2dn$, then there exists an induced maximal subpath $P$ of $H$ such that all vertices of $G$ are mapped into it and $\size{V(P)} >2dn$.
This is impossible by Observation~\ref{obs:unwt-length_linecycle}. Hence, there
does not exist a desired embedding if $N > 2dn$.

Now assume that $N \leq 2dn$. The algorithm is very similar to the algorithm for embedding into a cycle. As before, for any $\ell \in \{1,2,\ldots,N\}$, $S_{\ell} = \{\ell-(d+1),\ldots, \ell+(d+1)\}$. We explicitly define a feasible partial function.

\begin{defi}
\label{defi:partial_line}
\begin{itemize}
\item[(i)] For an $a \in V(H)$, a \emph{partial embedding} of $U \subseteq V(G)$, with respect to $\Psi_1$ and $\Psi_2$, is a function $f_a : U \rightarrow S_a$. As before, $U$ is referred to as ${\sf Dom}_{f_a}$ and $a$ is referred to as $\md(f_a)$.
\item[(ii)] ${\sf Dom}_{f_a}^{[p,q]}$ is the set of vertices of $U$ that are mapped into
 $\{a+p,\ldots a+q\}$, where $-(d+1) \leq p \leq q \leq (d+1)$. ${\sf Dom}_{f_a}^x $ is the set of vertices of $U$ that are mapped into $\{a+x\}$.
 \item[(iii)] ${\sf Dom}_{f_a}^L= Dom_{f_a}^{[-(d+1), -1]}$ and ${\sf Dom}_{f_a}^R= Dom_{f_a}^{[1, (d+1)]}$. $L(f_a)$ ($R(f_a)$) is the union of the vertex sets of connected components of $V(G) \setminus (U_1 \cup U_2\cup U)$ that have neighbours in ${\sf Dom}_{f_a}^L$ ($Dom_{f_a}^R$).
\end{itemize}   
\end{defi}

\begin{defi}
\label{defi:feasible_line}
Let $f_a:U \rightarrow S_a$ be a \emph{partial embedding} with respect to $\Psi_1$ and $\Psi_2$, where $U \subseteq V(G)$. Then $f$ is said to be \emph{feasible with respect to $\Psi_1$ and $\Psi_2$}, if it satisfies the following conditions. 
\begin{itemize}
\item[(i)] $U \neq \emptyset$, $U \cap U_1 = U \cap U_2 = \emptyset$ and $S_a \cap Z_1 = S_a \cap Z_2 = \emptyset$.
\item[(ii)] $f_a$ is a non-contracting distortion $d$ embedding of $U$.
\item[(iii)] For every pair $u \in U$ and $w \in U_1$, $D_H(f_a(u),\Psi_1(w))$ is neither contracting nor expanding by a factor of more than $d$. A similar condition holds for every pair $u \in U$ and $w \in U_1$, 
\item[(iv)] If $u$ is a neighbour of a vertex in ${\sf Dom}_{f_a}^0$, then $f_a(u) \in S_a$.
\item[(v)] $L(f_a) \cap R(f_a) = \emptyset$.
\end{itemize}
\end{defi}

The definition of succession of feasible partial embeddings and embeddability of a sequence of partial embeddings are analogous to Definitions~\ref{defi:succeed_cycle} and \ref{defi:embed_cycle}. We
 can also state results analogous to Lemma~\ref{lem:succeed} and \ref{lem:succeed_embed_cycle}. 

What remains is to bound the number of partial embeddings. By Observation~\ref{obs:degree-bd}, $\Delta(G) \leq 2d$ if $G$ is embeddable into $H$. The structure of a partial embedding for embedding into lines is same as that described in Observation~\ref{obs:seq_cycle}. Therefore, Lemma~\ref{lem:cycle_N(x)} is true in this case as well. Given $\Psi_1$ and $\Psi_2$, analogous to Lemma~\ref{lem:num_par_cycle}, the total number of feasible partial embeddings with respect to $\Psi_1$ and $\Psi_2$ is $\Oh(N\cdot n(4d(2d+2))^{2d+2})$. Now, we design an algorithm similar to that given in Theorem~\ref{theo:main_cycle}. The correctness of the algorithm can be argued in a similar way as that in Theorem~\ref{theo:main_cycle}. The running time of the algorithm turns out to be $\Oh\left(N\cdot n\cdot d^{2d+2}\cdot (4d(2d+2))^{2d+2}\right)$. As $N \leq 2dn$, the 
claimed bound of the time complexity holds. 
follows.
\end{proof}

We also obtain the following corollary using similar analysis. 
\begin{coro}\label{cor:path-st-algo}
Let $(G,D_G)$ and $(H,D_H)$ be two graph metrics on $n$ and $N$ vertices,respectively. Let $a_1$ be a positive integer such that $a_1 \leq N$. Let $Z_1 = \{1,2,\ldots,a_1\}$. Let $\Psi_1:U_1\rightarrow Z_1$ be a fixed non-contracting distortion $d$ embedding of $U_1\subseteq V(G)$. Also consider a fixed vertex $v \in V(G)$. Then there is an FPT algorithm to determine whether there exists a non-contracting distortion $d$ embedding $F$ of $G$ into $H$ that extends $\Psi_1$ and where for all $u\in V(G)$, $F(u) \leq F(v)$. The running time of the algorithm is $\Oh\left(\linethetabound\right)$.
\end{coro}

\begin{proof}
Without loss of generality, assume that there exists a vertex of $G$ that is mapped to $1$.
Then, by Observation~\ref{obs:unwt-length_linecycle}, there does not exist any vertex of $G$ that is mapped to some $j > 2dn$. So, we delete 
all vertices of $H$ that are at a distance more than $2dn$ from $1$. Call the new graph $H'$.
 Observe that finding a required embedding of $G$ into $H$ is equivalent to finding a required embedding of 
 $G$ into $H'$. Let $N'$ denote the number of vertices in $H'$. Note that $N' \leq 2dn$.
 
The definition of feasible partial embeddings, succession of feasible partial embeddings, embeddability of 
a sequence of partial embeddings are similar to the definitions in Lemma~\ref{lem:path-st-algo}. Similarly, 
we can also argue the correctness of finding an embeddable sequence of partial embeddings and a bound on 
the number of feasible partial embeddings. The FPT algorithm will be very similar to that of Lemma~
\ref{lem:path-st-algo}. The only change is that now there is an added condition that $v$ is the last vertex 
that gets mapped in the line $H$. However, this can be ensured very easily, by making modifications in the directed graph $\mathcal{G}$ associated with partial embeddings, that was built for the algorithms of Theorem~
\ref{theo:main_cycle} and of Lemma~\ref{lem:path-st-algo}. The running time of the algorithm is 
$\Oh\left(N'\cdot n\cdot d^{2d+2}\cdot(4d(2d+2))^{2d+2}\right)$. As $ N' \leq 2dn$, the claimed bound of the time complexity follows. 
\end{proof}

As mentioned earlier, these algorithms on embedding into the line will be useful to us in Section~\ref{sec:theta}.

\subsection{Embedding a weighted graph into a cycle}

In this part, we consider embedding weighted graph metrics into cycles. Let $G$ be a weighted graph and $w:E(G) \rightarrow \mathbb{R} $ be the weight function. The objective is to either find a non-contracting distortion $d$ embedding of $G$ into $C$ or decide that no such embedding exists. We will call this problem {\sc Weighted Graph Metric Embedding}.
 In general this problem is NP-complete, which we show shortly. However, we first give an FPT algorithm by taking $M=\max\limits_{e \in E(G)} w(e)$ as a parameter along with $d$. The terminologies and the algorithm discussed in Section~\ref{ssec:cycle_unwt}, can be extended to the an FPT algorithm for embedding of a weighted graph metric $G$ into a cycle, parameterized by $M$ and $d$.\remove{ Note that the weight function $w$ is some. Along with distortion paramater $d$, here  is
also given as a parameter to our algorithm. }

\subsubsection*{FPT algorithm taking $d$ and $M$ as parameter}
We have an observation analogous to Observation~\ref{obs:specialtype_cycle1}, 
with the change that $\size{V(S)}\geq (2d+4)M$. $S_\ell=\{\ell-(dM+1),\ldots,\ell+(dM+1)\}$. 
We first fix a function $\Psi:W \rightarrow S_0$. We define a partial embedding as $f_a: U \rightarrow S_a$, where $U \subseteq V(G)$ and other terminologies in Definition~\ref{defi:partial_cycle} can be defined in a similar manner. We can define the notion of \emph{feasible} partial embedding, how a feasible partial 
embedding \emph{succeeds} another and what an \emph{embeddable} sequence of feasible partial embeddings is similarly. We can also state results analogous to Lemma~\ref{lem:succeed} and \ref{lem:succeed_embed_cycle}. 

Analogous to Observation~\ref{obs:degree-bd}, we check whether $\Delta(G) \leq 2dM$. If this is false, we conclude that we cannot embed the given 
 graph into the cycle $C$ with distortion $d$. If the condition is true, we execute an algorithm similar to that explained
 in the proof of Theorem~\ref{theo:main_cycle}. By following similar analysis done in 
Lemma~\ref{lem:cycle_N(x)} and Lemma~\ref{lem:num_par_cycle}, the total number of feasible partial embeddings with respect to some $\Psi$ can be bounded by $\Oh \left( n\cdot N\cdot (4dM(2dM+2))^{2dM +2} \right)$ and the total
number of possible $\Psi$ can be bounded by $\Oh \left(  n\cdot (4dM(2dM+2))^{2dM +2} \right)$. One can easily go through the analogous steps explained in Theorem~\ref{theo:main_cycle} and verify that the time complexity
 of our algorithm to embed a weighted graph into a cycle $C$ is $\Oh \left(n^2\cdot N\cdot (dM)^{2dM+2}(4dM(2dM+2))^{4dM+4}\right)$. Using Observation~\ref{obs:unwt-length_linecycle} and an arguement similar to that given in the 
 first paragraph of the proof of Theorem~\ref{theo:main_cycle}, we can assume that $N \leq 4dMn$. Hence, the time complexity of our algorithm to embed a weighted graph into a cycle $C$ is $\Oh \left(n^3(dM)^{2dM+3}(4dM(2dM+2))^{4dM+4}\right)$.  
 \begin{theo}
 \label{theo:cycle_wt}
 Let $G$ be the given weighted graph on $n$ vertices such that $M=\max\limits_{e \in E(G)} w(e)$, and $C$ be a cycle. Then there
  exists an algorithm that either finds a non-contracting distortion $d$ embedding of $G$ into $C$ or
  concludes that there does not exist such an embedding. The running time of the algorithm is 
  $\Oh \left(n^3(dM)^{2dM+3}(4dM(2dM+2))^{4dM+4}\right)$.
 \end{theo}

\subsubsection*{Hardness proof}
Consider the {\sc Weighted Graph Metric Embedding} problem where the input metric $(G,D_G)$ has an associated weight function $w:E(G) \rightarrow \mathbb{R}^{\geq 0}$ and the distortion is $d >2$. We use the notation~$W(G)$ to denote maximum weight over all edges of the graph $G$.
\remove{
\paragraph*{{\bf Embedding a weighted graph into another given graph: \embed($G,H,d$)}}
\begin{description}
\item[Instance:] Given a wighted graph $G$ and a unweighted graph $H$ along with distortion parameter $d >2$.
\item[Question:] Does there exist a noncontracting distortion $d$ embedding of $G$ into $H$.
\end{description}
}
In this section, we prove the following Theorem.
 \begin{theo}
 \label{theo:cycle_hard}
 {\sc Weighted Graph Metric Embedding} is NP-complete for any rational $d >2$. The problem remains NP-Complete even if $W(G) \leq \size{V(G)}$.
 \end{theo}

To prove the theorem, we need the following observation, which is the weighted analogue of Observation~\ref{obs:unwt-length_linecycle}, and a known result of Fellows et al.~\cite{FellowsFLLRS13}.

\begin{obs}
\label{obs:length_linecycle}
Let $F$ be a non-contracting distortion $d$ embedding of a weighted graph $G$, into $H$ such that $H$ is either a line or a cycle. Let $P'$ be the maximal subpath of $H$ such that $F(x) \in V(P')$ for all $x \in V(G)$. Then $\size{V(P')} \leq 2dMn$, where $\size{V(G)}=n$ and $W(G)  = M$.
\end{obs}
\begin{proof}
For any two vertices $x,y \in V(G)$, $D_G(x,y) \leq M(n-1)$. This implies $D_H(F(x),F(y)) \leq dM(n-1)$.
Let $u \in V(H)$ be such that some vertex of $G$ is mapped to it. Observe that any other vertex of $H$ that has a vertex of $G$ as its pre-image, must be within a distance of $dM(n-1)$ from $u$ in $H$. As the degree of $u$ is at most 2, the result follows.
\end{proof}
 
 \begin{pre}[\cite{FellowsFLLRS13}]
\label{prop:line_hard}
 Let $G$ be the given graph with weight function $w$, and $L$ be a given line having sufficiently large number of vertices. 
 {\sc Weighted Graph Metric Embedding} on the instance ($G,w,L,d$) is NP-complete for any $d >2$. The problem 
 remains NP-Complete even if $ w(G) \leq \size{V(G)}$.
 \end{pre}

 \begin{proof}[Proof of Theorem~\ref{theo:cycle_hard}] 
 Clearly the problem is in NP. To show NP-hardness, we reduce the {\sc Weighted Graph Metric Embedding} problem for cycles from the {\sc Weighted Graph Metric Embedding} problem for lines. 
 We give the reduction for a special case where $w(G) \leq \size{V(G)}=n$ and $\size{V(C)}= 4dn^2+1$. We need to prove that there exists an embedding of $G$ into $L$ if and only if there exists an embedding of $G$ into $C$.
 
Let $F$ be a required embedding from $G$ to $L$. By Observation~\ref{obs:length_linecycle}, there exists a maximal subpath $P'$ of $2dn^2$ vertices of $L$ such that $F(x) \in V(P')$ for all $x \in V(G)$. 
   Let $Q'$ be a maximal subpath of $2dn^2$ vertices of $C$ and $\Phi$ be a bijection from $P'$ to $Q'$ such that 
  $D_L(u,v)=D_C(\Phi(u),\Phi(v))$ for any $u,v \in V(P')$. Such a bijection exists as $\size{V(C)} =4dn^2+1$. Now consider the  function
  $F'$ from $V(G)$ to $V(C)$ defined as follows. For any $x \in V(G)$, $F'(x)=\Phi(F(x))$. Since $\size{V(C)} > 2 \size{V(P')}=2 \size{V(Q')}$, $F'$ is a non-contracting and distortion $d$ embedding of $G$ into $C$. 
  
  For the converse part, let $F$ be a non-contracting distortion $d$ embedding from $G$ to $C$. By Observation~\ref{obs:length_linecycle}, there exists a maximal subpath $Q'$ of $2dn^2$ vertices of $C$ such that $F(x) \in V(Q')$ for all $x \in V(G)$. 
   Let $P'$ be a subpath of $2dn^2$ vertices of $L$ and  $\Psi$ be a bijection from $Q'$ to $P'$ such that 
  $D_C(u,v)=D_L(\Psi(u),\Psi(v))$ for any $u,v \in V(Q')$. Such a bijection exists as $\size{V(C)} =4dn^2+1$. Now consider the function
  $F'$ from $V(G)$ to $V(C)$ defined as follows. For any $x \in V(G)$, $F'(x)=\Psi(F(x))$. Since $\size{V(C)} > 2 \size{V(Q')} = 2\size{V(P')}$, it follows that $F'$ is a non-contracting and distortion $d$ embedding of $G$ into $L$. 
 \end{proof}
\section{Bijective {\sc Graph Metric Embedding} for bounded treewidth graphs}
\label{sec:bij}

In this section, we consider the bijective {\sc Graph Metric Embedding} problem for the graph class $\mathcal{G}$, which consists of graphs with treewidth at most $\alpha$ for a given constant $\alpha$. Let $(G,D_G)$ be the input connected graph metric to be embedded into the output graph metric $(H,D_H)$, where $H\in \mathcal{G}$. Also, $\size{V(G)}=\size{V(H)}$. We design an FPT algorithm for the problem parameterized by the distortion $d$ and the maximum degree $\Delta(H)$. We will also refer to $\Delta(H)$ as $\Delta$. Again, the strategy is to define feasible partial embeddings locally, and a notion of succession amongst partial embeddings such that we can build towards a bijective non-contracting distortion $d$ metric embedding of $G$ into $H$. We use a tree decomposition of $H$ to define the notions of partial embeddings and successions. We also use the tree decomposition to design a dynamic programming algorithm to derive a non-contracting distortion $d$ embedding of $G$ into $H$ from a set of partial embeddings that is computed in FPT time. The ideas used to design this algorithm is essentially similar to those used for designing an FPT algorithm for embedding into trees with bounded degree~\cite{FellowsFLLRS13}. However, we need to modify the definitions and the algorithm according to the need of our problem.

Let $\cT=(T,\{X_\bu \}_{\bu \in V(T) })$ be a \emph{nice tree decomposition}~\cite{saketbook15} of $H$. Recall that $X_\br=\emptyset$ and $X_\bl=\emptyset$, where $\br$ is the root of the tree $T$ and $\bl$ is any leaf of $T$. We use $N_T(\bu), C_T(\bu)$ to denote the set of neighbours and children of $\bu$ in $T$, respectively. Let $T_\bu$ denote the subtree rooted
  at the node $\bu \in V(T)$ and $H_\bu$ denote the subgraph induced by $\underset{\bv \in V(T_\bu)}{\bigcup} X_\bv$, i.e, the vertices present in the bags corresponding to the nodes present in $T_\bu$. Note that $H_\br = H$. For a $\bv \in N_T(\bu)$, we denote $T_{\bu}(\bv)$ as the subtree containing $\bv$ in the graph $T\setminus \{\bu\}$ and $H_{\bu \bv}$ as the subgraph induced on all vertices in $\underset{\bw \in V(T_\bu(\bv))}{\bigcup} X_\bw$. 
    For $x \in V(H)$, $B(x,d+1)$ is the set of vertices of $H$ that are at distance of at most $d+1$ from $x$. For $\bu \in V(T)$, $\cB (\bu,d+1)=\underset{x \in X_\bu}{\bigcup} B(x,d+1)$. 
    
    Next, we define partial embeddings with respect to the sets  $\cB (\bu,d+1)$, $\bu \in V(T)$. For ease of notations, we abuse the usual set theoretic terminology by using $A \setminus a$ to denote $A \setminus \{ a\}$.

  \begin{defi}
  \label{defi:partial_tw}
 \begin{itemize}
 \item[(i)] A \emph{$\bu$-partial embedding} of $U  \subseteq V(G)$ is a function $f_\bu: U \rightarrow \cB(\bu,d+1) $, where $\bu \in V(T)$ and $U \subseteq V(G)$. We refer to $U$ as ${\sf Dom}_{f_\bu}$.
 \item[(ii)] Let $\bv \in N_T(\bu)$. Then ${\sf Dom}_{f_\bu}({\bv})=\{x \in U : f_\bu(x) \in V(H_{\bu \bv})\}$, 
 ${\sf Dom}_{f_\bu}^{[p,q]}=\{x \in U : p \leq D_H(y,f_\bu(x)) \leq q ~\mbox{for some } y\in X_\bu\}$, ${\sf Dom}_{f_\bu}^{[p,q]}(\bv)={\sf Dom}_{f_\bu}^{[p,q]} \cap {\sf Dom}_{f_\bu}(\bv)$,
  ${\sf Dom}_{f_\bu}^{k}(\bv)={\sf Dom}_{f_\bu}^{[k,k]}(\bv)$ for $k \geq 1$  and ${\sf Dom}_{f_\bu}^0={\sf Dom}_{f_\bu}^{[0,0]}=X_\bu$.
  \item[(iii)] $M[f_\bu,\bv]$ is the union of vertex sets of all connected components of $G \setminus {\sf Dom}_{f_\bu}$ that have neighbours in ${\sf Dom}_{f_\bu}(\bv)$.
\end{itemize}  
    \end{defi}
  
  The feasibility of partial embeddings in order for them to build up to a metric embedding is defined as below.
  \begin{defi}
  \label{defi:feasible_tw}
  A $\bu$-partial embedding $f_\bu: U \rightarrow \cB(\bu,d+1)$ is \emph{feasible} if it satisfies the following conditions.
  \begin{itemize}
  \item[(i)] $f_\bu$ is a non-contracting and distortion $d$ embedding of $U$ into $\cB(\bu,d+1)$.
  \item[(ii)] $M[f_\bu,\bv] \cap M[f_\bu,\bw] =\emptyset$ for any $\bv,\bw \in N_T(\bu)$ and $\bv \neq \bw$.
  \item[(iii)] All neighbours of ${\sf Dom}_{f_\bu}^0$ are in $U$. 
  \end{itemize}
  \end{defi}
  
  Similarly, we define succession of partial embeddings, keeping in mind that we want to find a set of partial embeddings and put them together to form a non-contracting distortion $d$ metric embedding.
  
  \begin{defi}
  \label{defi:succeed_tw}
  Let $f_\bu$ and $f_\bv$ be two feasible partial embeddings and $\bv \in C_T(\bu)$. Then $f_\bv$ succeeds $f_\bu$ if the following conditions are satisfied.
  \begin{itemize}
  \item[(i)] ${\sf Dom}_{f_\bu} \cap {\sf Dom}_{f_\bv}= {\sf Dom}_{f_\bv}$ if $\bu$ is an \emph{introduce} node and ${\sf Dom}_{f_\bu} = {\sf Dom}_{f_\bv}$, otherwise.
 \item[(ii)] $\forall x \in {\sf Dom}_{f_\bu} \cap {\sf Dom}_{f_\bv}$, $f_\bu (x)=f_\bv (x)$.
 \item[(iii)] $M[f_\bu,\bv]=\underset{\bw\in N_T(\bv) \setminus \bu}{\bigcup} (M[f_\bv,\bw] \cup {\sf Dom}_{f_\bv}^{d+1}(\bw))$.
 \item[(iv)] $M[f_\bv,\bu]=\underset{\bw \in N_T(\bu) \setminus \bv}{\bigcup} (M[f_\bu,\bw] \cup {\sf Dom}_{f_\bu}^{d+1}(\bw))$.
  \end{itemize}
  \end{defi}
  
  Next, we describe how we put together feasible partial embeddings. In the case of cycles, we used the structure of the cycle. Here, we use the structure of the tree $T$.
  
  \begin{defi}
  \label{defi:embed_tw}
 Let $\Pi=\{f_\bu : \bu \in V(T) \}$. $\Pi$ is said to be an \emph{embeddable} set of partial embeddings if the following are satisfied.
 \begin{itemize}
 \item[(i)] Each $f_\bu$ is feasible, $\bu \in V(T)$.
 \item[(ii)] Let us take any $\bu \in V(T)$. Then for every $\bv \in C_T(\bu)$, $f_\bv$ succeeds $f_\bu$. 
 \end{itemize}
  \end{defi}
  
 \remove{ \begin{obs}
  Let $F$ be a non-contracting distortion $d$ embedding of $G$ into $H$ and $(x,y) \in E(G)$. Then
  $D_G(x,y) \leq D_H(F(x),F(y)) \leq d.D_$
  \end{obs}}
  
  The next Lemma shows that it is enough to design an algorithm to find an embeddable set of partial embeddings.
  \begin{lem}
  \label{lem:iff_tw}
  Let $G$ be the given graph to be embedded into $H$ such that $\size{V(G)}=\size{V(H)}$. $G$ has a
   bijective, non-contracting and distortion $d$ embedding into $H$ if and only if  there exists an embeddable set of partial embeddings.
  \end{lem}
  \begin{proof}
  Let $F$ be a non-contracting distortion $d$ embedding of $G$ into $H$. For $\bu \in V(T)$, consider $f_\bu$
   to be $F$ restricted to the preimage of $\cB(\bu,d+1)$. Let $\Pi=\{f_\bu : \bu \in V(T) \}$. It is easy to verify that $\Pi$ is an embeddable set of partial embeddings.
   
   For the converse part, Let $\Pi=\{f_\bu : \bu \in V(T) \}$ be an embeddable set of partial embeddings. 
  Let us construct $F=\Phi_\Pi$, as in Definition~\ref{defi:union_function}. \remove{$F=f_{\Pi}$.}First consider the following claims, which we prove later.
   \begin{cl}
  \label{cl:dom_tw}
 For every $x \in V(G)$, there exists some $f_\bu$ such that $x \in {\sf Dom}_{f_\bu}$.
  \end{cl}
  \begin{cl}
 \label{cl:inj_tw}
 \begin{itemize}
 \item[(a)] The subgraph of $T$ induced by $A_x=\{\bu \in V(T): x \in {\sf Dom}_{f_\bu}\}$ is connected. 
 \item[(b)] If $x \in  {\sf Dom}_{f_\bu} \cap {\sf Dom}_{f_\bv}$, then $f_\bu(x) = f_\bv (x)$.
 \end{itemize} 
 \end{cl}

  By Claim~\ref{cl:dom_tw} and Claim~\ref{cl:inj_tw}, $F$ is a bijective function 
  from $V(G)$ to $V(H)$ as $\size{V(G)}=\size{V(H)}$. We are done with the proof of the converse part by the following claims, which we prove later.
 
  \begin{cl}
  \label{cl:dist_tw}
  $F$, as an embedding of $G$ into $H$, has expansion at most $d$.
  \end{cl}
  
  \begin{cl}
  \label{cl:nct_tw}
  $F$ is a non-contracting embedding of $G$ into $H$.
  \end{cl}
  \end{proof}
    
  \begin{proof}[Proof of Claim~\ref{cl:dom_tw}]
If $x \in {\sf Dom}_{f_\br}$, $\br$ is the root of the tree $T$, we are done. Assume that $x \notin {\sf Dom}_{f_\br}$. This implies $x \in \underset{\bv \in C_T(\br)}{\bigcup} {M [f_\br, \bv ]}$. By condition (ii) of Definition~\ref{defi:feasible_tw}, we can say that there exists exactly 
one $\bv _1 \in C_T(\br)$ such that $x \in M [f_\br, \bv _1 ]$. The choice of $\bv _1$ implies that $x \notin M[f_{\bv _1}, \br]$. Now, by condition (iii) of Definition~\ref{defi:succeed_tw}, 
$x \in \underset{\bw \in N_T(\bv _1) \setminus \br}{\bigcup} (M[f_{\bv _1},\bw] \cup {\sf Dom}_{f_{\bv _1}}^{d+1}
(w))$. If $x \in {\sf Dom}_{f_{\bv _1}}$, then we are done. Otherwise, $x \in M [f_{\bv _1}, \bv _2 ]$ for 
exactly one $\bv _2 \in N_T(\bv _1) \setminus \br$, due to condition (ii) of Definition~\ref{defi:feasible_tw}. The choice of $\bv _2$ implies that $x \notin M[f_{\bv 
_2}, \bv _1]$. Again by condition (iii) of Definition~\ref{defi:succeed_tw} and condition (ii) of Definition~\ref{defi:feasible_tw}, we can say that either $x \in 
{\sf Dom}_{f_{\bv _2}} $, or $x \in M [f_{\bv _2}, \bv _3 ]
$ for exactly one $\bv _3 \in N_T(\bv _2) \setminus \bv _1$.

 In both cases, we proceed as before. Either we get a non-leaf node $\bu \in V(T)$ such that $x \in {\sf Dom}_{f_\bu}$, or 
a leaf $\bv_ i$ such that $x \in {\sf Dom}_{f_{\bv _i}} \cup \underset{\bl \in C_T(\bv _i)}{\bigcup}{ M[f_{\bv _i}, \bl]}$.
However, $\underset{\bl  \in C_T(\bv _i)}{\bigcup}{ M[f_{\bv _i}, \bl]}= \emptyset$ as $\bv _i$ is a leaf. So, $x \in {\sf Dom}_{f_{\bv _i}}$.
 \end{proof}
  
   \begin{proof}[Proof of Claim~\ref{cl:inj_tw}]
 \begin{itemize}
 \item[(a)]  We give a proof by contradiction. Let us consider $\bu,\bv \in V(T)$ such that $x \in {\sf Dom}_{f_\bu} \cap  {\sf Dom}_{f_\bv}$ 
 and $(\bu,\bv) \notin E(T)$. Let $P=\bu_1\ldots \bu_k,~k \geq 3$, be the path 
 from $\bu=\bu_1$ to $\bv=\bu_k$ in $T$ such that for some $i,~1 < i <k,~x \notin {\sf Dom}_{f_{\bu_i}}$. 
 By Definition~\ref{defi:partial_tw}, $x \in M[f_{\bu_i},\bu_{i+1}]$ and 
 $x \in M[f_{\bu_i},\bu_{i-1}]$. In other words, both $\bu_{i-1},\bu_{i+1} \in N_T(\bu_i)$ and $x \in    M[f_{\bu_i},\bu_{i+1}] \cap M[f_{\bu_i},\bu_{i-1}]$. This contradicts the fact that $f_{\bu_i}$ is a feasible partial embedding.
 \item[(b)] The statement follows from Part (a) and by using the definition of succession, i.e., Definition~\ref{defi:succeed_tw}.
 \end{itemize}
 \end{proof}
  
   \begin{proof}[Proof of Claim~\ref{cl:dist_tw}]
   Consider an edge $(x,y)$. Let $F(x) \in X_{\bu}$ for some $\bu \in V(T)$. Observe that in this case 
    $x \in {\sf Dom}_{f_\bu}^0$. By condition (iii) of Definition~\ref{defi:feasible_tw},
    $y \in {\sf Dom}_{f_\bu}$. Also, note that $f_{\bu}$ is a feasible partial embedding.
    This implies $D_H(F(x),F(y)) \leq d$.  
  
     Therefore, for any edge $(x,y) \in E(G)$, $D_H(F(x),F(y))\leq d$. Now, we apply induction similar to that in the proof of Claim~\ref{clm:cycle_dist}, and get the required result.
    \end{proof}
  
   \begin{proof}[Proof of Claim~\ref{cl:nct_tw}]
   We need to show that $D_G(x,y)\leq D_H(F(x),F(y))$ for any $x,y \in V(G)$. We prove it by the method of induction on $D_H(F(x),F(y))$. For the base case consider $D_G(F(x),F(y))=1$, or $(F(x),F(y))\in E(H)$. By the definition of tree decomposition, there exists $\bu \in V(T)$ such that $F(x), F(y) \in X_\bu$. Now using the fact that both $x,y \in {\sf Dom}_{f_\bu}$ and $f_\bu$ is a feasible partial embedding of ${\sf Dom}_{f_\bu}$, $D_G(x,y)\leq D_H(F(x),F(y)) $. Hence, the base case holds.
      
 For the inductive step assume that $D_G(x,y)\leq D_H(F(x),F(y))$ holds for any $x,y$
 such that $D_H(F(x),F(y)) < \ell$. Now, we have to show for the case $D_H(F(x),F(y)) =\ell > 1$. There exists
 $\bu \in V(T)$ such that $F(x) \in X_\bu$. Let
 $f_\bu:U \rightarrow \cB(\bu,d+1)$. The shortest $F(x)$ to $F(y)$ path must contain a vertex $a \in V(H)$ 
 such that $a \in \cB(\bu,d+1)$ and $D_H(F(x),F(y)) =D_H(F(x),a)+D_H(a,F(y))$. Let $a=F(w)$ for some $w \in V(G)$. Such a $w$ exists as the embedding $F$ is a bijection. By the induction hypothesis, $D_G(x,w) \leq D_H(F(x),F(w))$ and $D_G(w,y) \leq D_H(F(w),F(y))$. Hence, $D_G(x,y)\leq D_H(F(x),F(y))$.

\remove{
     $X_u$. The partial embedding $f_u$ is a non-contracting distortion $d$ embedding of ${\sf Dom}_{f_u}$ Let us consider two vertices $x,y \in V(G)$. 
   Let $x~\mbox{-}a_1~\mbox{-}\ldots~\mbox{-}a_l~\mbox{y}$ be the shortest path from $x$ to $y$ in 
   $G$. Then let $P=F(x)\mbox{-}x_1\mbox{-}x_2\mbox{-}\ldots\mbox{-}x_l\mbox{-}y$ some path from $F(x)$ to $F(y)$ in $H$.}
\end{proof}

Putting everything together, we have an FPT algorithm for the problem.

  \begin{theo}
  \label{thm:main_bintw}
  Let $G,H$ be two given graphs such that $\size{V(G)}=\size{V(H)}=n$, the maximum degree of H is
  $\Delta$ and $tw(H) \leq \alpha$. Then there exists an algorithm that either finds a bijective non-contracting distortion $d$ embedding
    of $G$ into $H$ or decides no such embedding exists in $\bijbound$ time.
  \end{theo}
  
\begin{proof}
  Due to Observation~\ref{obs:degree-bd}, if $\Delta(G) > {\Delta}^d$, then there does not exist an embedding from $G$ to $H$. So, assume that $\Delta(G) \leq \Delta ^d$. 
   
   Let $\cT=(T,\{X_\bu\}_{\bu \in V(T)})$ be a nice tree decomposition. Note that $\size{X_\bu} \leq \alpha$, $X_\br = \emptyset$
   and $X_\bl = \emptyset$, where $\br$ is the root of $T$ and $\bl$ is any leaf of $T$. We do a
   dynamic programming. For each $\bu \in V(T)$, we create a binary list of all possible feasible 
   $\bu$-partial embeddings from some
    subset of $V(G)$ to $\cB(\bu,d+1)$. For any leaf $\bl \in V(T)$, make all entries to be true. For the root $\br$, just create one
     entry in the list, whose value is to be assigned. Let $\bu$ be a non-leaf node. Let $f_\bu$ be a feasible partial embedding.
      Suppose for every $\bv \in C_T(\bu)$, there exists a feasible partial embedding $f_\bv$ and $f_\bv$ succeeds $f_\bu$. Then assign
       the corresponding entry of $f_\bu$ to be true. Assign true to the corresponding entry of $\br$  if there exists a feasible partial embeddings for each child of $\br$. By Lemma~\ref{lem:iff_tw}, one can observe that the algorithm is correct.
       
       For the running time, we have to bound the total number of feasible partial embeddings.
       \begin{obs}
       The total number of $\bu$-partial embedding is at most $\ubound$.
       \end{obs}
       \begin{proof}
        Let us consider a $\bu$-partial embedding $f_\bu$ from some subset of $V(G)$ to $\cB(\bu,d+1)$.
        Observe that $\size{\cB(\bu,d+1)} \leq \alpha.\Delta ^{d+1}$. As we are considering
        non-contracting embeddings, the domain of $f_\bu$ must be contained in the union of 
        at most $\alpha$ balls of radius $2d+2$, where each ball is centered at some vertex of $G$. As $\Delta(G) \leq \Delta^{d}$, the domain of $f_\bu$ can contain at most
        $\Delta^{\Oh(\alpha.d^2)}$. So, the total number of feasible $\bu$-partial embeddings is 
        at most $\ubound$.
        \end{proof}
            
  Let $f_\bu$ and $f_\bv$ be given  such that $\bv \in C_T(\bu)$. From Definition~\ref{defi:succeed_tw}, 
  one can decide whether $f_\bv$ succeeds $f_\bu$ in time $\Oh\left(n.\alpha.\Delta^{d+1}\right)$. 

Recall that $\cT$ is a nice tree decomposition, which implies $\size{V(T)} = \Oh(\alpha.n)$ and $
\Delta(T) = \Oh(1)$. So, observe that there are at most $\alpha.n.\ubound$ partial embeddings. Note that the bags corresponding to two adjacent vertices of $T$, differ by at most one vertex of $H$. So, 
each feasible partial embedding can take part in at most $n \cdot \left( \alpha \cdot \Delta ^{d+1} \right)^{\Delta^{\Oh(\alpha.d^2)}}$ number of successions. Hence, the total running time is $\bijbound$.
\end{proof}

  \paragraph*{A note on bijective {\sc Gen-Graph Metric Embedding} for bounded treewidth graphs}
  
  Assume that the input of the {\sc Red Blue Graph Metric Embedding} problem is such that the output metric $(H,D_H)$ has the following property: given $V(H)=R\uplus B$, any path of length $d+1$ in $H$ has at least one internal vertex in $R$. Also, we want to find a non-contracting distortion $d$ embedding that is a bijection between $V(G)$ and $R$. For this variant of the {\sc Red Blue Graph Metric Embedding}, the above algorithm can be easily modified to give a solution. Most details of the above algorithm remain the same. Note that for a $\bu \in V(T)$, a $\bu$-partial embeddings of $U \in V(G)$ is a function $f_{\bu}:U \rightarrow \mathcal{B}(\bu,d+1)\cap R$. We explicitly prove the equivalent of Claim~\ref{cl:nct_tw}, since that is the only major place of difference.
  
  We restate the claim for a map $F$ obtained from an embeddable set of partial embeddings, as in Lemma~\ref{lem:iff_tw}. 
  
  \begin{cl}
  \label{cl:nct_tw_rb}
  $F$ is a non-contracting embedding of $G$ into $H$.
  \end{cl}
  
   \begin{proof}
   We need to show that $D_G(x,y)\leq D_H(F(x),F(y))$ for any $x,y \in V(G)$. We prove it by the method of induction on $D_H(F(x),F(y))$. For the base case consider $D_G(F(x),F(y))\leq d$. By the definition of tree decomposition, there exists $\bu \in V(T)$ such that $F(x), F(y) \in \mathcal{B}(\bu,d+1)$. Now using the fact that both $x,y \in {\sf Dom}_{f_\bu}$ and $f_\bu$ is a feasible partial embedding of ${\sf Dom}_{f_\bu}$, $D_G(x,y)\leq D_H(F(x),F(y)) $. Hence, the base case holds.
      
 For the inductive step assume that $D_G(x,y)\leq D_H(F(x),F(y))$ holds for any $x,y$
 such that $D_H(F(x),F(y)) < \ell$. Now, we have to show for the case $D_H(F(x),F(y)) =\ell > 1$. There exists
 $\bu \in V(T)$ such that $F(x) \in X_\bu$. Let
 $f_\bu:U \rightarrow \cB(\bu,d+1)$. Since $\ell \geq d+1$, the shortest $F(x)$ to $F(y)$ path must contain a vertex $a \in R$ 
 such that $a \in \cB(\bu,d+1)$ and $D_H(F(x),F(y)) =D_H(F(x),a)+D_H(a,F(y))$. Let $a=F(w)$ for some $w \in V(G)$. Such a $w$ exists as the embedding $F$ is a bijection between $V(G)$ and $R$. By the induction hypothesis, $D_G(x,w) \leq D_H(F(x),F(w))$ and $D_G(w,y) \leq D_H(F(w),F(y))$. Hence, $D_G(x,y)\leq D_H(F(x),F(y))$.
\end{proof}

Thus, we obtain an FPT algorithm for this variant of {\sc Red-Blue Graph Metric Embedding}. Due to Lemma~\ref{lem:bij-prelim} and Proposition~\ref{prop:scaling}, we obtain the following result.

\begin{coro}\label{cor:bij-tw-gen}
  Assume that the bijective {\sc Gen-Graph Metric Embedding} problem takes as input two graph metrics $(G,D_G)$ and $(H,D_H)$ such that $\size{V(G)}=\size{V(H)}=n$. Let the maximum degree of H be
  $\Delta$ and $tw(H) \leq \alpha$. Then there exists an algorithm that either finds a bijective distortion $d$ embedding
    for this instance of bijective {\sc Gen-Graph Metric Embedding} or decides that no such embedding exists in $\genbijbound$ time.
\end{coro}

\section{{\sc Graph Metric Embedding} and connected treewidth}
\label{sec:ctw}

In this Section, we will look at the {\sc Graph Metric Embedding} problem with respect to the added parameters of treewidth and longest geodesic cycle of the output graph metric. Let $(G,D_G)$ be the input connected graph metric to be embedded into $(H,D_H)$. We show that this problem is FPT, when parameterized by the distortion $d$, the treewidth $tw(H)=\alpha$, the length $\ell_g$ of the longest geodesic cycle of $H$, and the maximum degree $\Delta(H) = \Delta$. From~\cite{Diestel2017} it can be shown that for a graph with longest geodesic cycle $\ell_g$, a tree decomposition of treewidth $\alpha'$ can be converted into a connected tree decomposition of width $\alpha' + {\alpha' \choose 2}(\ell_g(\alpha'-2)-1)$ in polynomial time. Since trees have constant connected treewidth, our algorithm is a generalization of the FPT algorithm for {\sc Graph Metric Embedding} for trees, parameterized by distortion $d$ and maximum degree $\Delta$~\cite{FellowsFLLRS13}. As before, we employ a dynamic programming to build a non-contracting distortion $d$ metric embedding using a set of partial embeddings that are computed in FPT time.

Let $(G,D_G)$ be a graph metric to be embedded into $(H,D_H)$. Here the parameters are the treewidth $\alpha$ of $H$, the length of the longest geodesic cycle $\ell_g$ in $H$, the distortion $d$ and the maximum degree $\Delta$ of $H$. Let $\mathcal{T}$ be a nice tree decomposition of $H$ with width $\mu$. Since from~\cite{Diestel2017} $H$ has a connected tree decomposition of width $\mu$, we may assume that the nice tree decomposition is derived from the connected tree decomposition~\cite{saketbook15} and therefore the maximum distance between any two vertices inside a bag in $\mathcal{T}$ is $\Gamma\leq \mu$.

We borrow the definition of feasible partial embeddings and succession of feasible partial embeddings from Definitions~\ref{defi:partial_tw}, \ref{defi:feasible_tw}, and \ref{defi:succeed_tw}.

Unlike the arguments of Section~\ref{sec:bij}, now ensuring non-contraction for a non-contracting distortion $d$ metric embedding $F$ is more elaborate. Local non-contraction no longer implies global non-contraction. This problem was dealt with in~\cite{FellowsFLLRS13} by introducing the notion of \emph{types}. For our algorithm too, for a vertex $\bu \in V(T)$ we need to define a \emph{type} for every vertex of $V(G)$ that is mapped into the subgraph $H_{\bu}$, to indicate how it behaves with the rest of the graph. More specifically, let $x\in V(G)$ be a vertex that is mapped to $F(x) \in V(H_{\bu})$. Consider any other vertex $y \in V(G)$ such that $F(y)$ is a vertex from $V(H) \setminus V(H_{\bu})$. Then, a type relates $D_H(F(x),F(y))$ with $D_G(x,y)$. The crucial observation here is that (i) the shortest path between $F(x)$ and $F(y)$ must pass through a vertex in $X_{\bu}$, and (ii) the shortest path between $x$ and $y$ in $G$ is such that there is at least one internal vertex $x'$ such that $F(x')\in \mathcal{B}(\bu, d+1)$, as otherwise $F$ does not have distortion $d$. Thus, the types store information of the interaction of vertices of the graph seen so far with the boundary vertices, and this is enough to ensure global non-contraction.

\begin{defi}
\label{defi:types_ctw}
Let $\bu \in V(T)$, $f_\bu$ be a feasible partial embedding and $X_\bu =\{ u_1,\ldots,u_k\}$, $1 \leq k \leq \alpha_c$.
Then 
\begin{itemize}
\item[(i)] For $\bv \in N_T(\bu)$ and $u_i \in X_\bu$, $[f_\bu, \bv, u_i]$ \emph{type} is a function $t^{u_i} : {\sf Dom}_{f_\bu}  (\bv)  \rightarrow \{\infty, 2\Gamma+3d+3,\Gamma+d+1,\ldots,-(\Gamma+d+1)\}$.
\item[(ii)] A $[f_\bu, \bv]$ \emph{type} $\bt$ is  a tuple $(t^{u_i},\ldots,t^{u_k})$, where $t^{u_i}$ is a 
$[f_\bu, \bv, u_i]$ type.
\item[(iii)] A $[f_\bu, \bv]$ \emph{type-list} is a set of $[f_\bu, \bv]$ types.
\end{itemize} 
\end{defi}

Intuitively, we want to define a type corresponding to each vertex mapped into $H_{\bu}$. However, this blows up the number of types. In order to handle this, it can be shown that we do not need to remember the type of each vertex, and that it is enough to only remember the type of vertices "close to" the vertices in $X_{\bu}$. Now we present the formal arguments. To bound the total number of possible types, we define a function $\beta$ as follows: $\beta(k)=k$ if $k < 2\Gamma + 3d +3$, and $\beta(k)=\infty$ otherwise. In the following definitions, treat $\beta(k)=k$ and the definition of $\beta$ will be clear while we prove our claims.

\begin{defi}
\label{defi:comp_ctw}
Let us consider $\bu \in V(T)$, $\bv \in N_T(\bu)$. Let $f_\bu$ be a feasible partial embedding and $\cL$ be a $[f_\bu,\bv]$ type-list. Then $\cL$ is said to be \emph{compatible} with ${\sf Dom}_{f_\bu} (\bv)$ if the following condition is satisfied. 

\begin{itemize}
\item For each $x \in {\sf Dom}_{f_\bu}(\bv)$ there exists a type $\bt \in \cL $,  such that for each $y \in {\sf Dom}_{f_\bu}(\bv)$, for all $u_i \in X_\bu$ $D_H(f_\bu (x), u_i) - D_G(x,y) =t^{u_i}(y)$ .
\end{itemize}
\end{defi}

\begin{defi}
\label{defi:agree_ctw}
Let $\bu \in V(T)$ and $f_\bu$ be a feasible partial embedding. Also consider $\bv , \bw \in N_T(\bu)$ along with
 a $[f_\bu,\bv]$ type-list $\cL_1$ and a $[f_\bu,\bw]$ type-list $\cL_2$ such that $\bv \neq \bw$. Then 
 $\cL_1$ and $\cL_2$ \emph{agree} if the following condition is satisfied for all $u_i \in X_\bu$.
 
 \begin{itemize}
  \item For every $\bt_1 \in \cL_1$ and $\bt_2 \in \cL_2$, there exists $x \in {\sf Dom}_{f_\bu}(\bv)$ and $y \in {\sf Dom}_{f_\bu}(\bw)$ such that 
 $t_1^{u_i}(x) + t_2^{u_i}(y) \geq D_G(x,y)$ for all $u_i \in X_\bu$.
 \end{itemize}
\end{defi}

Next, we define a state with respect to a vertex in $T$.

\begin{defi}
\label{defi:state_ctw}
Let $\bu \in V(T)$. A \emph{$\bu$-state} constitutes of a feasible partial embedding $f_\bu$, a $[f_\bu,\bv]$ type-list
$\cL[f_\bu,v]$
for each $\bv \in N_T(\bu)$.
\end{defi}

Notice that it is no longer enough to consider feasibility and succession of partial embeddings. We also need to take care of the types of vertices. Therefore, we define feasibility and succession of states.

\begin{defi}
\label{defi:ustate_ctw}
A $\bu$-state is said to be \emph{feasible} if 
 the following conditions are satisfied.
\begin{itemize}
\item[(i)] $\cL[f_\bu,\bv]$ is compatible with ${\sf Dom}_{f_\bu}(\bv)$, for each $\bv \in N_T(\bu)$.
\item[(ii)] $\cL[f_\bu,\bv]$ agrees with $\cL[f_\bu, \bw]$, for any $\bv, \bw \in N_T(\bu)$ and $\bv \neq \bw$.
\end{itemize}
\end{defi}

\begin{defi}
\label{defi:succ_ctw}
Let $\bu \in V(T)$ and $\bv \in C_T(\bu)$. Let $\cS_\bu, \cS_\bv$ be feasible
$\bu$-state and $\bv$-state, respectively. $\cS_\bv$ is said to \emph{succeed}
$\cS_\bu$ if the following properties hold.
\begin{itemize}
\item[(i)] $f_\bv$ succeeds $f_\bu$.
\item[(ii)]\remove{If $\bu$ is not a \emph{forget} node, then} For every $\bw \in N_T(\bv) \setminus \bu$ and a type $\bt_1 \in \cL[f_\bv,\bw]$ 
	there exists  a type $\bt_2 \in \cL[f_\bu,\bv]$ satisfying the following conditions.

\begin{itemize}
 			
\item[(a)] For all $x \in {\sf Dom}_{f_\bu}(\bv) \cap {\sf Dom}_{f_\bv}(\bw)$ and $a \in X_\bu \cap X_\bv$, $t_2^{a}(x)= t_1^{a}(x)$.
 			
\item[(b)] For all $x \in {\sf
                            Dom}_{f_\bu}(\bv) \cap {\sf
                            Dom}_{f_\bv}(\bw)$ and $a \in X_\bu
                          \setminus X_\bv$, 
$$
t_2^{a}(x)=\beta \left( \min\limits_{b \in X_\bv}( D_H( a,b) + t_1^{b}
  (x)) \right) .
$$
 				
\item[(c)] For all $x \in {\sf Dom}_{f_\bu}(\bv) \setminus {\sf
    Dom}_{f_\bv}(\bw)$ and $a \in X_\bu \cap X_\bv$, 
$$
t_2^{a}(x)= \beta \left( \max\limits_{y \in {\sf Dom}_{f_\bv}(\bw)} \left(   t_1^{a} (y)) -D_G(x,y) \right) \right).
$$		

\item[(d)] For all $x \in {\sf Dom}_{f_\bu}(\bv) \setminus {\sf
    Dom}_{f_\bv}(\bw)$ and $a \in X_\bu \setminus X_\bv$, 
$$
 		 t_2^{a}(x)= \beta \left( \max\limits_{y \in {\sf
                       Dom}_{f_\bv}(\bw)} \left( \min\limits_{b \in
                       X_\bv}( D_H( a,b) + t_1^{b} (y)) -D_G(x,y)
                   \right) \right).
$$
\end{itemize} 	
		 
\item[(iii)]\remove{  If $\bu$ is a \emph{forget} node, then} For every $\bw \in N_T(\bu) \setminus \bv$ and a type $\bt_1 \in \cL[f_\bu,\bw]$
 			there exists  a type $\bt_2 \in \cL[f_\bv,\bu]$ satisfying the following conditions.
 		\begin{itemize}
 			
\item[(a)] For all $ x \in {\sf Dom}_{f_\bv}(\bu) \cap {\sf
    Dom}_{f_\bu}(\bw)$ and  $a \in X_\bu \cap X_\bv$, 
$t_2^{a}(x)= t_1^{a}(x)$.
 			
\item[(b)] For all $x \in {\sf Dom}_{f_\bv}(\bu) \cap {\sf Dom}_{f_\bu}(\bw)$ and $a \in X_\bv \setminus X_\bu$, 
$$ 							
t_2^{a}(x)=\beta\left( \min\limits_{b \in X_\bu}( D_H( a,b) + t_1^{b}
  (x)) \right).
$$			

\item[(c)] For all $x \in {\sf Dom}_{f_\bv}(\bu) \setminus {\sf Dom}_{f_\bu}(\bw)$ and  $a \in X_\bu \cap X_\bv$, $t_2^{a}(x)= t_1^{a}(x)$.
 		
\item[(d)] For all $x \in {\sf Dom}_{f_\bv}(\bu)
                  \setminus {\sf Dom}_{f_\bu}(\bw)$ and  $a \in X_\bv
                  \setminus X_\bu$, 
\remove{$t_2^{a}(x)=\beta\left(\max\limits_{y \in {\sf Dom}_{f_\bu}(\bw)} \left(  t_1^{a} (y) -D_G(x,y) \right)\right)$}
$$
 		 t_2^{a}(x)= \beta\left(\max\limits_{y \in {\sf
                       Dom}_{f_\bu}(\bw)} \left( \min\limits_{b \in
                       X_\bu}( D_H( a,b) + t_1^{b} (y)) -D_G(x,y)
                   \right)\right).
$$
\end{itemize}
\end{itemize}
\end{defi}

Now, we define the embeddability of a set of feasible states.
\begin{defi}
\label{defi:embed_ctw}

For $u \in V(T)$, let $\cS_u$ denote a $u$-state. The set $\{\cS_\bu:\bu \in V(T)\}$ is said to be an \emph{embeddable} set of feasible states if the following conditions are satisfied.
\begin{itemize}
\item[(i)] For each $\bu \in V(T)$, $\cS_\bu$ is a feasible state.
\item[(ii)] For $\bu \in V(T)$ and $\bv \in C_T(\bu)$, $\cS_\bv$ succeeds $\cS_\bu$.
\end{itemize} 
\end{defi}

The above definitions are enough to show the relation between the existence of a non-contracting distortion $d$ embedding of $G$ into $H$ and the existence of an embeddable set of feasible states. This is proved over the following two Lemmas.

\begin{lem}
\label{lem:ifthen_ctw}
Let $F$ be a non-contracting and distortion $d$ embedding of $G$ into $H$. Then 
there exists an embeddable set of feasible states.
\end{lem}
\begin{proof}
For $\bu \in V(T)$, consider $f_\bu$ to be the function $F$ restricted to $\cB(\bu,d+1)$. 
Observe that each $f_\bu$ is feasible. It is also easy to verify that for $\bu \in V(T)$ and $\bv \in C_T(\bu)$,
 $f_\bv$ succeeds $f_\bu$.
 
 To show the existence of feasible $\bu$-states, we have to define types. 
 Let $\bu =\{u_1,\ldots,u_k\} \in V(T)$ and $\bv \in C(\bu)$. We create a $[f_\bu,\bv]$ type-list $\cL[f_\bu,\bv]$.
 For each $x \in {\sf Dom}_{f_\bu}(\bv) \cup M[f_\bu,\bv] $ and $u_i \in X_\bu$, we define $t^{u_i}_x (y)= \beta \left( D_H(F(x),u_i) - D_G(x,y)\right) $ for all $y \in {\sf Dom}_{f_\bu} (\bv)$. We add $\bt_x[f_\bu,\bv]=(t_x^{u_1},\ldots,t_x^{u_k})$ to $\cL[f_\bu,\bv]$. $\bt_x$ is a $[f_\bu,\bv]$ type by the following observation, which we prove later.
 \begin{obs}
 \label{obs:type_ctw}
 For any $ u_i \in X_\bu$ and $y \in {\sf Dom}_{f_\bu} (\bv)$,  $t_x^{u_i}(y) \geq -(\Gamma +d+1)$.
 \end{obs}

 For any $x,y \in {\sf Dom}_{f_\bu}(\bv)$, we have the type $\bt_x[f_\bu,\bv]$ such that $t_x^{u_i}[f_\bu,\bv](y)=D_H(F(x),u_i) - D_G(x,y)$ for all $u_i \in X_\bu$. From Definition~\ref{defi:comp_ctw}, we can say that say each $\cL[f_\bu,\bv]$ is compatible with ${\sf Dom}_{f_\bu}(\bv)$.
 
 Let $\cS_\bu$ constitute the feasible $\bu$-partial embedding $f_\bu$ along with $\cL(f_\bu,\bv)$ for each $\bv \in N_T(\bu)$. From Definition~\ref{defi:ustate_ctw}, $\cS_\bu$ is a feasible 
 $\bu$-state by the the following claim, which we prove later.
 
 \begin{cl}
 \label{clm:agree_ctw}
 For two different $\bv,\bw \in N_T(\bu)$ such that $\bv \neq \bw$, $\cL[f_\bu, \bv]$ agrees with $\cL[f_\bu,\bw]$.
 \end{cl}

To complete the proof of existence of a set of embeddable feasible states, we have to prove that conditions (ii) and (iii) of Definition~\ref{defi:succ_ctw} hold for the above description of types. Condition (ii) holds by the following claim, and Condition (iii) can 
be proved similarly. The following claim will be proved later.

\begin{cl}
\label{clm:succ_ctw}
Let $\bu \in V(T)$ and $\bv \in N_T(\bu)$.\remove{ If $\bu$ is not a
  forget node,} Then for every $\bw \in N_T(\bv) \setminus \bu$ and a
type $\bt_1 \in \cL[f_\bv,\bw]$ there exists  a type $\bt_2 \in
\cL[f_\bu, \bv]$ satisfying the following conditions.
\begin{itemize}
 		
\item[(a)] For all $x \in {\sf Dom}_{f_\bu}(\bv) \cap {\sf Dom}_{f_\bv}(\bw)$ and $a \in X_\bu \cap X_\bv$, $t_2^{a}(x)= t_1^{a}(x)$

\item[(b)] For all $x \in {\sf Dom}_{f_\bu}(\bv) \cap {\sf Dom}_{f_\bv}(\bw)$ and $a \in X_\bu \setminus X_\bv$, 
 							$t_2^{a}(x)=\beta \left( \min\limits_{b \in X_\bv}( D_H( a,b) + t_1^{b} (x)) \right) $.

\item[(c)] For all $x \in {\sf Dom}_{f_\bu}(\bv) \setminus {\sf
    Dom}_{f_\bv}(\bw)$ and $a \in X_\bu \cap X_\bv$, 
$$
t_2^{a}(x)= \beta \left( \max\limits_{y \in {\sf Dom}_{f_\bv}(\bw)}
  \left(   t_1^{a} (y)) -D_G(x,y) \right) \right).
$$		

\item[(d)] For all $x \in {\sf Dom}_{f_\bu}(\bv) \setminus {\sf
    Dom}_{f_\bv}(\bw)$ and $a \in X_\bu \setminus X_\bv$, 
\begin{center}
 		 $t_2^{a}(x)= \beta \left( \max\limits_{y \in {\sf Dom}_{f_\bv}(\bw)} \left( \min\limits_{b \in X_\bv}( D_H( a,b) + t_1^{b} (y)) -D_G(x,y) \right) \right)$.
\end{center}

\end{itemize} 	
\end{cl}
 \end{proof}
 
 \begin{lem}
 \label{lem:converse_ctw}
 Let $\Pi=\{f_\bu : \bu \in V(T)\}$ be an embeddable set of feasible states. Then there exists a non-contracting
 and distortion $d$ embedding of $G$ into $H$.
 \end{lem}
 \begin{proof}
 This Lemma will be proved by a series of claims. The Claims~\ref{clm:dom_ctw}, \ref{clm:inj_ctw} and 
 ~\ref{clm:dist_ctw} can be proved in the similar way that we proved Claims~\ref{cl:dom_tw}, \ref{cl:inj_tw}
  and \ref{cl:dist_tw}, respectively. We prove Claim~\ref{clm:prop_ctw} and \ref{clm:nct_ctw} later.
 \begin{cl}
 \label{clm:dom_ctw}
 For every $x \in V(G)$, there exists a feasible $u$-state such that $x \in {\sf Dom}_{f_\bu}$.
 \end{cl}
 \remove{\begin{proof}
  This is similar to the Proof of Claim~\ref{cl:bij_tw}.
 \end{proof}}
 \begin{cl}
 \label{clm:inj_ctw}
 The subgraph of $T$ induced by $A_x=\{\bu \in V(T): x \in {\sf Dom}_{f_\bu}\}$ is connected. Moreover, 
 $x \in  {\sf Dom}_{f_\bu} \cap {\sf Dom}_{f_\bv}$ implies $f_\bu(x) = f_\bv (x)$.
 \end{cl}
 \remove{
 \begin{proof}
 Let us consider $\bu,\bv \in V(T)$ such that $x \in {\sf Dom}_{f_\bu} \cap  {\sf Dom}_{f_\bu}$ 
 and $(\bu,\bv) \notin E(T)$. Let $P=\bu_1\ldots \bu_k,k \geq 3$, be the path 
 from $\bu=\bu_1$ to $\bv=\bu_k$ in $T$ such that for some $i,1 < i <k, x \notin {\sf Dom}_{f_{\bu_i}}$. 
 By Definition~\ref{defi:partial_tw}, $x \in M[f_{\bu_i},\bu_{i+1}]$ and 
 $x \in M[f_{\bu_i},\bu_{i-1}]$. That is $\bu_{i-1},\bu_{i+1} \in N_T(\bu_i)$ and $x \in    M[f_{\bu_i},\bu_{i+1}] \cap M[f_{\bu_i},\bu_{i-1}]$. It contradicts the fact that $f_{\bu_i}$ is a feasible partial embedding.
 \end{proof}
 }
Now by Definition~\ref{defi:union_function}, we construct $F=\Phi_{\Pi}$, where $\Pi=\{f_\bu: \bu \in V(T)\}$.

 \begin{cl}
 \label{clm:dist_ctw}
 $F$ is a metric embedding with expansion at most $d$.
 \end{cl}
 
 \begin{cl}
 \label{clm:prop_ctw}
Consider a path $P=\bu_1 \bu_2\ldots \bu_k$ from $\bu=\bu_1$ to $\bv=\bv_k$ in $T$.\remove{ such that
either $\bu_2 \in C(\bu_1)$ and $\bu_1$ is not a forget node, or $\bu_1 \in C(\bu_2)$ and $\bu_2$ is a forget node} Then for every $x \in {\sf Dom}_{f_\bv}$, at least one of the following properties hold.
 \begin{description}
 \item[ Prop-1:] There exists a $\bu_j \in P$ and $y \in {\sf Dom}_{f_{\bu_j}}$ such that $D_H(F(x), u')- D_G(x,y)\geq 2\Gamma + 3d + 3 $  for all $u' \in X_{\bu_j}$.
 \item[Prop-2:] There exists a type $\bt_x \in \cL[f_\bu, \bu_2]$ such that   
 $\bt_x^{u'} (y)=D_H(F(x), u') - D_G(x,y)$ for all $y \in {\sf Dom}_{f_{\bu}}(\bu_2)$ and  $u' \in X_\bu$.
 \end{description} 
\end{cl}

\begin{cl}
\label{clm:nct_ctw}
Consider a path $P=\bu_1 \bu_2\ldots \bu_k$ from $\bu$ to $\bv$ in $T$, where $\bu =\bu_1$ and $\bv = \bu_k$. Then
$F$ restricted to $\underset{\bu_i \in P}{\bigcup}{ {\sf Dom}_{f_{\bu_i}}}$ is non-contracting.
\end{cl}

Now we will be done if we prove that $F$ is a non-contracting embedding.
Let us consider two vertices $x,y \in V(G)$. Note that each of $F(x)$ and $F(y)$ is in some bag. Fix $\bu,\bv \in V(T)$ such that $F(x) \in X_\bu$ and $F(y) \in X_\bv$. Consider the path $P=\bu_1 \bu_2\ldots \bu_k$ from $\bu$ to $\bv$ in $T$, where $\bu =\bu_1$ and $\bv = \bu_k$. By Claim~\ref{clm:nct_ctw}, the shortest
path between $F(x)$ to $F(y)$ is non-contracting as $x,y \in \underset{\bu_i \in P}{\bigcup} {\sf Dom}_{f_{\bu_i}}$.
 \end{proof}

 \begin{proof}[Proof of Observation~\ref{obs:type_ctw}] Observe that
\begin{eqnarray*}
D_G(x,y) \leq D_H(F(x),F(y)) &\leq& D_H(F(x),u_i)+D_H(u_i,F(y)) \\
&\leq& D_H(F(x),u_i)+\Gamma + (d+1).
\end{eqnarray*}
 The first inequality follows from the fact that $F$ is a non-contracting embedding, while the second inequality follows from
 triangle inequality. The third inequality follows from the fact that $y \in \cB(\bu,d+1)$ along with the fact that $H$ has bounded 
 connected treewidth.
 
 So, $D_H(F(x),u_i) - D_G(x,y) \geq -(\Gamma + d+1)$. This implies 
 \begin{center}
  $t^{u_i}_x (y)= \beta\left(D_H(F(x),u_i) - D_G(x,y) \right) \geq -(\Gamma + d+1)$.
  \end{center}
 \end{proof}

\begin{proof}[Proof of Claim~\ref{clm:agree_ctw}]
 For contradiction, let there exist $\bt_p \in \cL[f_\bu,\bv]$ and $\bt_q \in \cL[f_\bu,\bw]$
 such that for all $x \in {\sf Dom}_{f_\bu}(\bv)$ and $y \in {\sf Dom}_{f_\bu}(\bw)$ there is some $u_i \in X_\bu$, $t_p^{u_i}(x) + t_q^{u_i}(y) < D_G(x,y)$.
 Let $x' \in {\sf Dom}_{f_\bu}(\bv)$ and $y' \in {\sf Dom}_{f_\bu}(\bw)$ be two vertices such that there is a $u_i$ with $t_p^{u_i}(x') + t_q^{u_i}(y') < D_G(x',y')$ and for all $x \in {\sf Dom}_{f_\bu}(\bv)$, $y \in {\sf Dom}_{f_\bu}(\bw)$ and $u_j$ with $t_p^{u_j}(x) + t_q^{u_j}(y) < D_G(x,y)$, $t_p^{u_i}(x') + t_q^{u_i}(y') - D_G(x',y') \leq t_p^{u_j}(x) + t_q^{u_j}(y) - D_G(x,y)$. This implies
 \begin{eqnarray*}
  t_p^{u_i}(x') + t_q^{u_i}(y') &=& \beta \left( D_H(F(p),u_i) -
                                    D_G(p,x')\right) + \beta
                                    \left(D_H(F(q),u_i) -
                                    D_G(q,y')\right) \\
&<& D_G(x',y')\\
   &=& D_H(F(p),u_i) - D_G(p,x')+ D_H(F(q),u_i) - D_G(q,y') \\
&<& D_G(x',y')
 \end{eqnarray*}

Now, we can say
 \begin{center}
 $D_H(F(p),F(q)) \leq D_H(F(p),u_i) + D_H(F(q),u_i) < D_G(p,x') +  D_G(q,y') + D_G(x',y')$.
 \end{center}
 By the choice of $x'$ and $y'$, $D_G(p,x') +  D_G(q,y') + D_G(x',y')=D_G(p,q)$ as any $p$ to $q$ path in $G$ must pass through
 some vertex in ${\sf Dom}_{f_\bu}(\bv)$ and ${\sf Dom}_{f_\bu}(\bw)$. Putting everything together, $D_H(F(p),F(q)) < D_G(p,q)$. It contradicts the fact that $F$ is non-contracting. 
\end{proof} 
\begin{proof}[Proof of Claim~\ref{clm:succ_ctw}]
Let $\bt_p =(t_p^{v_1},\ldots, t_p^{v_l}) \in \cL[f_\bv,\bw] $, where $v_j \in X_\bv$ for all $j \in [l]$ and $p$ is some vertex in ${\sf Dom}_{f_\bv}(\bw) \cup M[f_\bv, \bw]$. Note that $t_p^{v_j} (x) =\beta( D_H(F(p),v_j) - D_G(p,x))$ for all $x \in {\sf Dom}_{f_\bv}(\bw)$. As $\bu$ is not a forget node, ${\sf Dom}_{f_\bv}(\bw) \cup M[f_\bv, \bw] \subseteq {\sf Dom}_{f_\bu}(\bv) \cap M[f_\bu, \bv]$. As we are assuming that we are working with a nice tree decomposition and that the graph has bounded connected treewidth, $p \in {\sf Dom}_{f_\bu}(\bv) \cup M[f_\bu, \bv]$. 
Consider the type $\bz_p =(z_p^{u_1},\ldots, z_p^{u_k}) \in \cL[f_\bu, \bv] $, where $u_i \in X_\bu$ for all $i \in [k]$.
$z_p(x)=\beta\left( D_H(F(p),u_i) - D_G(p,x) \right)$ for all $x \in {\sf Dom}_{f_\bu}(\bv)$. 

\begin{itemize}
\item  For all $x \in {\sf Dom}_{f_\bu}(\bv) \cap {\sf Dom}_{f_\bv}(\bw)$ and $ a\in X_\bu \cap X_\bv$, it is easy to see that  $z_p^a(x) = t_p^a(x)$ .
\item For all $x \in {\sf Dom}_{f_\bu}(\bv) \cap {\sf Dom}_{f_\bv}(\bw)$ and $ a\in X_\bu \setminus  X_\bv$, we can say the following.
 $D_H(F(p),a) = \min\limits_{b \in X_\bv} \left( D_H(F(p),b) + D_H(b,a ) \right)$. It 
is because we are assuming each bag is connected and any path from $F(p)$ to $a$ in $H$, must 
pass through some vertex in $X_\bv$. Hence, 
\begin{eqnarray*}
z_p^a(x) &=& \beta \left( D_H(F(p),a) - D_G(p,x) \right)\\
&=&\beta \left(\min\limits_{b \in X_\bv} \left(D_H(F(p),b) + D_H(b,a ) \right)  - D_G(p,x) \right)\\
&=&\beta \left(\min\limits_{b \in X_\bv} \left( D_H(b,a ) + \beta(
    D_H(F(p),b)   - D_G(p,x)) \right)\right) \\
&&\quad\quad\quad\quad\quad\quad \quad\quad\quad\quad\quad\quad \quad
\quad\quad\quad\quad\quad \quad\quad\quad\quad\because\mbox{By the definition of }\beta\\ 
 &=& \beta \left( \min\limits_{b \in X_\bv} \left( D_H(a,b) + t_p^b (x) \right) \right).
\end{eqnarray*}

  \item   For all $x \in {\sf Dom}_{f_\bu}(\bv) \setminus {\sf Dom}_{f_\bv}(\bw)$ for any $ a\in X_\bu \cap  X_\bv$, we can say the
  following. Any path from $p$ to $x$ in $G$ must pass through a vertex in ${\sf Dom}_{f_\bv}(\bw)$, i.e.,
 $D_G(p,x) = \min\limits_{y \in {\sf Dom}_{f_\bv}(\bw)} \left( D_G(p,y) + D_G(y,x) \right)$. Recall that $z_p^a(x) = \beta \left(D_H(F(p),a) - D_G(p,x)\right)$. This implies
 
 \begin{eqnarray*}
 z_p^a(x) &=& \beta \left(D_H(F(p),a) -  \min\limits_{y \in {\sf Dom}_{f_\bv}(\bw)} \left( D_G(p,y) + D_G(y,x)
   \right) \right)\\
    &=& \beta \left( \max\limits_{y \in {\sf Dom}_{f_\bv}(\bw)} \left( D_H(F(p),a) - D_G(p,y) - D_G(y,x) \right) \right) \\
     &=& \beta \left( \max\limits_{y \in {\sf Dom}_{f_\bv}(\bw)}
         \left(\beta \left( D_H(F(p),a) - D_G(p,y) \right) - D_G(y,x)
         \right) \right) \\
&&\quad\quad\quad\quad\quad\quad \quad\quad\quad\quad\quad\quad \quad
\quad\quad\quad\quad\quad \quad\quad\quad\quad
\because D_G(x,y)\leq \Gamma +2d+2\\
    &=&  \beta \left(\max\limits_{y \in {\sf Dom}_{f_\bv}(\bw)} \left( t_p^a(y) - D_G(x,y) \right)\right)
 \end{eqnarray*}

 \item   For all $x \in {\sf Dom}_{f_\bu}(\bv) \setminus {\sf Dom}_{f_\bv}(\bw)$ for any $ a\in X_\bu \setminus  X_\bv$, we can say the
  following. Any path from $p$ to $x$ in $G$ must pass through a vertex in ${\sf Dom}_{f_\bv}(\bw)$, i.e.,
 $D_G(p,x) = \min\limits_{y \in {\sf Dom}_{f_\bv}(\bw)} \left( D_G(p,y) + D_G(y,x) \right)$. Recall that $z_p^a(x) = D_H(F(p),a) - D_G(p,x)$. This implies
 
 \begin{eqnarray*}
 z_p^a(x) &=&\beta\left( D_H(F(p),a) -  \min\limits_{y \in {\sf Dom}_{f_\bv}(\bw)} \left( D_G(p,y) + D_G(y,x)
   \right)\right)\\
    &=&  \beta\left(\max\limits_{y \in {\sf Dom}_{f_\bv}(\bw)} \left( D_H(F(p),a) - D_G(p,y) - D_G(x,y) \right)\right) \\
      &=&  \beta\left(\max\limits_{y \in {\sf Dom}_{f_\bv}(\bw)}
          \left(\beta( D_H(F(p),a) - D_G(p,y)) - D_G(x,y)
          \right)\right)\\
&&\quad\quad\quad\quad\quad\quad \quad\quad\quad\quad\quad\quad \quad
\quad\quad\quad\quad\quad \quad\quad\quad\quad
 \because D_G(x,y)\leq \Gamma +2d+2\\
\end{eqnarray*}

Observe that $D_H(F(p),a) = \min\limits_{b \in X_\bv} \left( D_H(F(p),b) + D_H(b,a ) \right)$. Hence, 

\begin{eqnarray*}
z_p^a(x) &=& \beta\left(\max\limits_{y \in {\sf Dom}_{f_\bv}(\bw)} \left( \beta \left( \min\limits_{b \in X_\bv} \left( D_H(F(p),b) + D_H(b,a ) \right)- D_G(p,y) \right) - D_G(x,y) \right) \right) \\
&=&\beta\left(\max\limits_{y \in {\sf Dom}_{f_\bv}(\bw)} \left( \min\limits_{b \in X_\bv} \left( D_H(a,b ) + \beta\left(D_H(F(p),b) - D_G(p,y)\right)   \right) - D_G(x,y) \right) \right)\\
&=& \beta\left(\max\limits_{y \in {\sf Dom}_{f_\bv}(\bw)} \left( \min\limits_{b \in X_\bv} \left( D_H(a,b ) +t_p^b(y)   \right) - D_G(x,y) \right)\right)
\end{eqnarray*}
\end{itemize}
 \end{proof}

\begin{proof}[Proof of Claim~\ref{clm:prop_ctw}]
We prove the claim by induction on $k$. First consider $k=1$. Note that $\cS_\bu$ is a feasible
 state. Observe that Prop-2 holds by Definition~\ref{defi:ustate_ctw} and \ref{defi:comp_ctw}. Therefore, the base case holds.
 
 Let the statement be true for all $k' < k$. Now we show for $k$.
 
\remove{First assume the following holds. If $\bu_2 \in C(\bu_1)$, then $\bu_1$
 is a not a forget node and if $\bu_1 \in C(\bu_2)$, then $\bu_2$
 is a forget node. Otherwise, we are done by induction hypothesis.}
   If Prop-1 holds for some $x \in {\sf Dom}_{f_v}$, we are done for that $x$. So, consider some
    $x \in {\sf Dom}_{f_\bv}$ such that for every
  $\bu_j \in P$ and $y \in {\sf Dom}_{f_{\bu_j}}$ there exists $u' \in X_{\bu_j}$
 such that $D_H(F(x), u')- D_G(x,y) \leq \Gamma + d + 1$. 
 
  Consider the sub-path $P'=\bu_2\ldots\bu_k$. Recall that Prop-1 does not hold for the $x \in {\sf Dom}_{f_\bv}$ with respect to $P'$. By the induction 
 hypothesis, there exists a type $\bt_{x} \in \cL[f_{\bu_2},\bu_3]$ such that $\bt_x^{u'} (y)=D_H(F(x), u') - 
 D_G(x,y)$ for all $y \in {\sf Dom}_{f_{\bu_2}}(\bu_3)$ and  $u' \in X_{\bu_2}$. Note that either $\cS_{\bu_2}$ 
 succeeds $\cS_{\bu_1}$ or $\cS_{\bu_1}$ succeeds $\cS_{\bu_2}$ depending on $\bu_2 \in C(\bu_1)$ or $\bu_1 \in 
 C(\bu_2)$.\remove{ Also, it is given that either $\bu_2 \in C(\bu_1)$ and $\bu_1$ is not a forget node, or $\bu_1 \in C(\bu_2)$ and $\bu_2$ is a forget node.} Now by Definition~\ref{defi:succ_ctw}, we will have a type $\bz_x \in \cL[f_\bu, \bu_2]$ such that the following properties hold.
 \begin{itemize}
 \item For all $y \in {\sf Dom}_{f_\bu}(\bu_2) \cap {\sf Dom}_{f_{\bu_2}}(\bu_3) $ and $a \in X_\bu \cap X_{\bu_2}$,
       \begin{eqnarray*}
       z_x^a(y) = t_x^a(y) = D_H(F(x), a) -  D_G(x,y)
       \end{eqnarray*}
 \item For all $y \in {\sf Dom}_{f_\bu}(\bu_2) \cap {\sf Dom}_{f_{\bu_2}}(\bu_3) $ and $a \in X_\bu \setminus X_{\bu_2}$,
  \begin{eqnarray*}
       z_x^a(y) &=& \beta \left(\min\limits_{b \in X_{\bu_2}} \left( D_H(a,b) + t_x^b(y) \right)\right)\\
       &=&\beta \left(\min\limits_{b \in X_{\bu_2}} \left( D_H(a,b) +
           D_H(F(x), b) -  D_G(x,y)\right)\right) \\
&&\quad\quad\quad\quad\quad\quad \quad\quad\quad\quad\quad\quad \quad
\quad\quad\quad\quad\quad
\because \mbox{By Induction Hypothesis(I.H)}\\
        &=&  \beta \left( D_H(F(x),a)-D_G(x,y)\right)\\
       &=&  D_H(F(x),a)-D_G(x,y)\quad\quad\quad\quad\quad\quad\quad \because \mbox{Prop-1 does not hold}
       \end{eqnarray*}
       
        \item For all $y \in {\sf Dom}_{f_\bu}(\bu_2) \setminus {\sf Dom}_{f_{\bu_2}}(\bu_3) $ and $a \in X_\bu \cap X_{\bu_2}$,
        \begin{eqnarray*}
       z_x^a(y) &=& \beta \left(\max\limits_{y' \in {\sf Dom}_{f_{\bu_2}}(\bu_3)} \left( t_x^{a} (y') -D_G(y,y') \right)\right)\\
       &=& \beta \left( \max\limits_{y' \in {\sf Dom}_{f_{\bu_2}}(\bu_3)} \left( D_H(F(x), a) -  D_G(x,y') -D_G(y,y') \right)\right) ~~ (\because \mbox{By I.H})\\
       &=& \beta \left( D_H(F(x),a) - \min\limits_{y' \in {\sf Dom}_{f_{\bu_2}}(\bu_3)}  \left( D_G(x,y') +D_G(y',y) \right)\right)\\
       &=& \beta \left(D_H(F(x),a)-D_G(x,y)\right) \\
       &=& D_H(F(x),a)-D_G(x,y) ~~(\because \mbox{Prop-1 does not hold})
       \end{eqnarray*}
       
        \item For all $y \in {\sf Dom}_{f_\bu}(\bu_2) \setminus {\sf Dom}_{f_{\bu_2}}(\bu_3) $ and $a \in X_\bu \setminus X_{\bu_2}$,
        \begin{eqnarray*}
       z_x^a(y) &=& \beta \left(\max\limits_{y' \in {\sf Dom}_{f_{\bu_2}}(\bu_3)} \left( \min\limits_{b \in X_{\bu_1}}( D_H( a,b) + t_x^{b} (y')) -D_G(y,y') \right)\right)\\
       &=& \beta \left( \max\limits_{y' \in {\sf Dom}_{f_{\bu_2}}(\bu_3)} \left( \min\limits_{b \in X_{\bu_1}}( D_H( a,b) + D_H(F(x), b) -  D_G(x,y')) -D_G(y,y') \right)\right)~~ (\because \mbox{By I.H})\\
       &=& \beta \left( \max\limits_{y' \in {\sf Dom}_{f_{\bu_2}}(\bu_3)} \left( D_H(F(x), a) -  D_G(x,y') -D_G(y,y') \right)\right)\\
       &=& \beta \left( D_H(F(x),a) - \min\limits_{y' \in {\sf Dom}_{f_{\bu_2}}(\bu_3)}  \left( D_G(x,y') +D_G(y',y) \right)\right)\\
       &=& \beta \left(D_H(F(x),a)-D_G(x,y)\right) \\
       &=& D_H(F(x),a)-D_G(x,y) ~~(\because \mbox{Prop-1 does not hold})
       \end{eqnarray*}
\end{itemize}
\end{proof}

\begin{proof}[Proof of Claim~\ref{clm:nct_ctw}]
We prove by the method of induction on $k$. If $k=1$, the statement is valid as $\cS_\bu$ is a feasible $u$-partial embedding.
Now assume that the statement is true for all $k' < k$. Now, we show for $k$. 

\remove{{\bf Fact 2:} Assume the following holds. If $\bu_2 \in C(\bu_1)$, then $\bu_1$
 is a not a forget node and if $\bu_1 \in C(\bu_2)$, then $\bu_2$
 is a forget node. Otherwise, we are done by induction hypothesis.}
 Observe that it is enough to show the following. For any two vertices $x \in {\sf Dom}_{f_{\bu_1}} \setminus {\sf Dom}_{f_{\bu_2}}$ and $y \in {\sf Dom}_{f_{\bu_k}} \setminus {\sf Dom}_{f_{\bu_{k-1}}}$, $D_H(F(x), F(y)) \geq D_G(F(x),F(y))$. All other cases are taken care of by the Induction Hypothesis on shorter length paths. Note that $x$ satisfies either Prop-1 or Prop-2 in Claim~\ref{clm:prop_ctw}.
  \begin{itemize}
\item {\bf Case 1(Prop-1 holds)}
  In this case there exists a $\bu_j$ and $z \in {\sf Dom}_{f_{\bu_j}}$, in the path $P$ such that $D_H(F(y), u')- D_G(y,z) \geq 2\Gamma + 3d+3$  for all $u' \in X_{\bu_j}$\remove{and $y \in {\sf Dom}_{f_{\bu_j}}$}. 
  
  Observe that any path from $F(x)$ and $F(y)$ must pass through some vertex $u'' \in X_{\bu_j}$. So, 
  \begin{eqnarray*}
  D_H(F(x),F(y)) &=&  D_H(F(x),u'') + D_H(u'',F(y)) \\
  &\geq&  D_H(F(x),F(z)) - D_H(u'',F(z)) + D_H(u'',F(y))~\mbox{(By triangle inequality)}
    \end{eqnarray*}
    By the induction hypothesis, $D_H(F(x),F(z)) \geq D_G(x,z)$ and by Prop-1  $D_H(u'',F(y) \geq D_G(z,y) + 2\Gamma +3d+3$. Hence,
    \begin{eqnarray*}
     D_H(F(x),F(y)) &\geq&  D_G(x,z) - D_H(F(z),u') + D_G(z,y)+ 2\Gamma +3d+3 \\
  &\geq& D_G(x, z) -  D_H(u'',F(z)) + D_G(y,z) + 2\Gamma +3d + 3\\
  &\geq& D_G(x,y) + 2\Gamma +3d + 3  - D_H(u'',F(z)) \\
  &\geq& D_G(x,y) + \Gamma +2d +2 
    \end{eqnarray*}
 The last inequality holds due to the following. $z \in {\sf Dom}_{f_{\bu_j}}$ and $u'' \in X_{\bu_j}$. Recalling the fact that 
 we are considering bounded connected treewidth, we can say that $D_H(u'',F(z)) \leq \Gamma + d+1$.

\item {\bf Case 2(Prop-2 holds)}
In this case there exists a type $\bt_y \in \cL[f_\bu,\bu_2]$ such that $t_y^{u'} (y)=D_H(F(y), u') - D_G(y,z)$ for all $z \in {\sf Dom}_{f_{\bu_2}(\bu)}$ and  $u' \in X_\bu$. Let $x \in {\sf Dom}_{f_\bu}(\bw)$ for some $\bw \in N_T(\bu) \setminus \bu_1$. Note that $\cS_\bu$ is a feasible partial embedding and hence $\cL[f_\bu,\bw]$ is compatible with 
$ {\sf Dom}_{f_\bu}(\bw)$. So, there exists a type $\bt_x \in \cL[f_\bu, \bw]$ such that $t_x^{u'} (z)= D_H(F(x),u') - D_G(x,z) $ for all $z \in {\sf Dom}_{f_\bu}(\bw)$ and $u' \in X_\bu$.

Again considering the fact that $\cS_\bu$ is a feasible partial embedding, $\cL[f_\bu,\bu_2]$ and $L[f_\bu,\bw]$ must agree. Now by Definition~\ref{defi:agree_ctw}, there exists $y' \in {\sf Dom}_{f_\bu}(\bu_2)$ and $x' \in {\sf Dom}_{f_\bu}(\bw)$ such that
$t_x^{u'}(x')+t_y^{u'}(y') \geq D_G(x',y')$. 

 As the shortest path from $F(x)$ to $F(y)$ passes through some vertex $u'' \in X_\bu$, 
 we can deduce the following.

\begin{eqnarray*}
D_H(F(x),F(y)) &=& D_H(F(x),u'') +D_H(u'' + F(y)) \\
& \geq & t_x^{u''} (x') + D_G(x,x') + t_y^{u''} (y')+D_G(y,y')\\
& \geq & D_G(x',y')+ D_G(x,x')+ D_G(y,y') \geq D_G(x,y)
\end{eqnarray*} 
 \end{itemize}
\end{proof}

Now, we are ready to design the FPT algorithm. 
\begin{theo}
\label{thm:main_ctw}
 Let $G,H$ be two given graphs with $n$ and $N$ vertices, respectively, such that the maximum degree of H is
  $\Delta$, $tw(H) \leq \alpha_c$ and the length of the longest geodesic cycle in $H$ is $\ell_g$. Then there exists an algorithm that either finds a non-contracting distortion $d$ embedding of $G$ into $H$ or decides no such embedding exists in running time $\ctwbound$, where $\mu = \ctwconsbound$.
\end{theo}
\begin{proof}
Due to Observation~\ref{obs:degree-bd}, if $\Delta(G) > {\Delta}^d$, then there does not exist an embedding from $G$ to $H$. So, assume that $\Delta(G) \leq \Delta ^d$. First, by~\cite{FlumGrohe}, we find a tree decomposition of width $4\alpha$ in time $2^{\Oh(\alpha)}\cdot n^2$. Then using the arguments in~\cite{Diestel2017}, we can convert this tree decomposition in polynomial time to a connected tree decomposition whose width is $\mu = \ctwconsbound$.
   
   Let $\cT=(T,\{X_\bu\}_{\bu \in V(T)})$ be a nice tree decomposition derived from it is constant time. Note that $\size{X_\bu} \leq \mu$, $X_\br = \emptyset$
   and $X_\bl = \emptyset$, where $\br$ is the root of $T$ and $\bl$ is any leaf of $T$. We do a
   dynamic programming. For each $\bu \in V(T)$, we create a binary list of all possible feasible 
   $\bu$-state. For any leaf $\bl \in V(T)$, make all entries to be true. 
   Let $\bu$ be a non-leaf node. Let $\cS_\bu$ be a feasible $\bu$-state.
      Suppose for every $\bv \in C_T(\bu)$, there exists a feasible partial embedding $\cS_\bv$ and $\cS_\bv$ succeeds $\cS_\bu$. Then assign
       the corresponding entry of $\cS_\bu$ to be true. Assign true to the corresponding entry of $\br$  if there exists a feasible $\bu$-state  for each child of $\br$. By Lemmas~\ref{lem:ifthen_ctw} and \ref{lem:converse_ctw}, one can observe that the algorithm is correct.
       
       For the running time, we have to bound the total number of feasible partial embeddings and total number of feasible states. Before proceeding further,  let $n_\bu$ denote the number of feasible $\bu$-states, where $\bu \in V(T)$, and let $t_s$ denote the time required
to test whether a feasible state succeeds another. 

       \begin{obs}
       The total number of $\bu$-partial embeddings is at most $\ufbound$ and the total number of feasible $\bu$-states 
       is at most $\ustate$.
             \end{obs}
       \begin{proof}
        Let us consider a $\bu$-partial embedding $f_\bu$ from some subset of $V(G)$ to $\cB(\bu,d+1)$.
        Observe that $\size{\cB(\bu,d+1)} \leq \mu.\Delta ^{d+1}$. Note that the distance between two vertices in $\cB(\bu,d+1)$, is at most $\Gamma$. As we are considering
        non-contracting embedding, the domain of $f_\bu$ must be contained in a ball of radius $\Gamma + 2d +2$ centered at some vertex of $G$. As $\Delta(G) \leq \Delta^{d}$, the domain of $f_\bu$ can contain 
        at most $\Delta^{\Oh(\Gamma .d + d^2)}$ vertices. So, the total number of feasible $\bu$-partial embeddings is 
        at most $\ufbound$.
        
        Now consider
  some $\bu \in V(T)$ and $\bv \in N_T(\bu)$. The total number of $[f_\bu,\bv]$ types is at most $\fuvtype$ and hence the total number of  $[f_\bu,\bv]$ typelists is at most $2^{\fuvtype}$. Note that the maximum degree of $T$ is $\Oh(1)$. So, the total number of feasible $\bu$-states, i.e., $n_\bu$  is at most $\ustate$.
        \end{proof}

 Recalling Definition~\ref{defi:succeed_tw} and \ref{defi:succ_ctw}, we can say the following. Let $f_\bu$ and $f_\bv$ correspond to $\bu \in V(T)$ and $\bv \in C_T(\bu)$. One can decide whether $f_\bv$ succeeds $f_\bu$ in time $\Oh\left(n\cdot \mu \cdot \Delta^{d+1}\right)$ and hence the time required 
  to test whether a feasible state succeeds another $t_s=n\cdot \mu \cdot \Delta^{d+1}+ \mu \cdot \Delta^{d+1}\cdot 2^{\Oh(\fuvtype)}$.

First, a tree decomposition has to be constructed and converted into a connected tree decomposition in $2^{\Oh(\alpha)}\cdot n^2 + n^{\Oh(1)}$ time. This connected tree decomposition is converted into a nice tree decomposition in polynomial time, with the guarantee that the distance between two vertices in a bag of the nice tree decomposition is at most $\Gamma$. Recall that $\cT$ is a nice tree decomposition, which implies $\size{V(T)} = \Oh(\mu.N)$ and $
\Delta(T) = \Oh(1)$.  So, there are at most $\Oh\left( \mu \cdot N n_\bu \right)$ feasible states.
Note that the bags corresponding to two adjacent vertices of $T$, differ by at most one vertex of $H$. So, 
each feasible partial embedding can take part in at most $\left( \mu \cdot \Delta ^{d+1} \right)^{\Delta^{\Oh(\mu.d^2)}}$ number of successions. Thus the total running time of the algorithm is 

$= 2^{\Oh(\alpha)}\cdot n^2 + n^{\Oh(1)} + \mu \cdot N\cdot n_\bu \cdot  \left( \mu \cdot \Delta ^{d+1} \right)^{\Delta^{\Oh(\mu.d^2)}} \cdot t_s $. 
Since $\Gamma \leq \mu$, the running time is $\ctwbound$, where $\mu = \ctwconsbound$.      
\end{proof}
\section{{\sc Graph Metric Embedding} for Generalized Theta graphs}
\label{sec:theta}
In this section, we design an FPT algorithm for embedding unweighted graphs into generalized theta graphs. Our FPT algorithm is parameterized by the distortion $d$ and the number $k$ of paths in the generalized theta graph. The strategy for the algorithm is still the same: that of putting together partial embeddings to obtain a non-contracting distortion $d$ metric embedding. For this algorithm, we also observe structural properties of graphs that are embeddable into generalized theta graphs. We exploit these properties to obtain an FPT algorithm to compute a set of partial embeddings, and then use a dynamic programming algorithm to put together partial embeddings from the set to obtain the solution metric embedding. Our notion of partial embeddings will be more involved in this algorithm.
 
Let $(G,D_G)$ be the graph metric that we want to embed into the graph metric $(H,D_H)$. Here $H$ is a generalized theta graph defined at $s$,$t$ and let $\mathcal{P}$ be the family of $s-t$ paths that define $H$. To begin with, we try to guess the non-contracting distortion $d$ embedding of $(G,D_G)$ into $(H,D_H)$, when restricted to a $f(d)$-ball, for some suitable $f$, around $s$ and around $t$.

\begin{defi}\label{def:s-t-balls}
Let $F$ be a non-contracting distortion $d$ embedding of $G$ into $H$. Define 
$B_s = \{v \in V(H) ~\vert~ D_H(v,s)\leq d\}$ and $B_t = \{v \in V(H) ~\vert~ D_H(v,t)\leq d\}$;
$ B_{s'} = \{v \in V(H) ~\vert~ D_H(v,s)\leq 2d^2\}$ and $B_{t'} = \{v \in V(H) ~\vert~ D_H(v,t)\leq 2d^2\}$. For an embedding $F:V(G)\rightarrow V(H)$, define ${\sf Dom}_s^F = \{u \in V(G) ~\vert~ F(u) \in B_s\}$ and ${\sf Dom}_t^F = \{u \in V(G) ~\vert~ F(u) \in B_t\}$; ${\sf Dom}_{s'}^F = \{u \in V(G) ~\vert~ F(u) \in B'_{s}\}$ and ${\sf Dom}_{t'}^F = \{u \in V(G) ~\vert~ F(u) \in B_{t'}\}$.
\end{defi}

The following observation talks about the degree bound on the vertices of a graph that is embeddable into a generalized theta graph.

\begin{obs}\label{obs:degree}
 If there exists a non-contracting distortion $d$ embedding $F$ of $G$ into $H$, then:
 \begin{itemize}
\item Each vertex in ${\sf Dom}_{s}^F$ can have degree at most $(k+1)d$. Similarly, each vertex in ${\sf Dom}_t^F$ can have degree at most $(k+1)d$. 
\item All other vertices of $G$ can have degree at most $2d$.
\end{itemize}
\end{obs}
Hence, Each vertex in ${\sf Dom}_{s'}^F \cup {\sf Dom}_{t'}^F$ can have degree at most $(k+1)d$.
\begin{obs}\label{obs:number-fixed-window}
The number of possible non-contracting distortion $d$ embeddings of some $U \subseteq V(G)$ into $B_{s'} \cup B_{t'}$ is
at most  $n^2 \cdot (2kd)^{(4d)^{\Oh(kd)}}$.
\end{obs}

\begin{proof}
Observe that $\size{B_{s'}},\size{ B_{t'}} \leq 2kd^2+1$. As we are considering non-contracting embeddings, the preimage of $B_{s'}$ ($B_{t'}$) lies 
 in a ball of radius $4d^2$ centered at some vertex of $G$. By Observation~\ref{obs:degree}, the size of the pre-image of $B_{s'}$ ($B_{t'}$) is at most $(4d^2)^{\Oh(kd)}$. Hence, the number of required embeddings can be at most $n^2 \cdot (2kd^2+1)^{2.(4d^2)^{\Oh(kd)}}=n^2 \cdot (2kd)^{(4d)^{\Oh(kd)}}$.
\end{proof}
\remove{
\begin{obs}\label{obs:number-fixed-window}
The number of possible non-contracting distortion $d$ embeddings of some $U \subseteq V(G)$ into $B_s \cup B_t$ is
at most  $n^2 \cdot (kd+1)^{(2d)^{\Oh(kd)}}$.
\end{obs}

\begin{proof}
Observe that $\size{B_s},\size{ B_t} \leq kd+1$. As we are considering non-contracting embeddings, the preimage of $B_s$ ($B_t$) lies 
 in a ball of radius $2d$ centered at some vertex of $G$. By Observation~\ref{obs:degree}, the size of the pre-image of $B_s$ ($B_t$) is at most $(2d)^{\Oh(kd)}$. Hence, the number of required embeddings can be at most $n^2 \cdot (kd+1)^{2.(2d)^{\Oh(kd)}}$.
\end{proof}}
\remove{
\begin{obs}\label{obs:number-fixed-window}
The number of possible ${\sf Dom}_s^F \cup {\sf Dom}_t^F$, over all distortion $d$ embeddings into the generalized theta graph $H$ is $\Oh\left(n^2 \cdot 4^{((k+1)d)^d} \right)$
\end{obs}

\begin{proof}
We bound the number of possible ${\sf Dom}_s^F$ and the bound for ${\sf Dom}_t^F$ will follow in a similar manner. If ${\sf Dom}_s^F \neq \emptyset$, we guess one vertex $u\in {\sf Dom}_s^F$. Look at the set $S_u$ of vertices that are at most distance $d$ away from $u$. By Observation~\ref{obs:degree}, the number of such vertices is at most $((k+1)d)^d$. Once $u$ has been fixed, the vertex set ${\sf Dom}_s^F$ is a subset of $S_u$. Therefore, there are $n \cdot 2^{((k+1)d)^d}  $ choices for ${\sf Dom}_s^F$ and the claim follows.  
\end{proof}
}
We prove several properties of graphs that are embeddable into generalized theta graphs. For the given input graph $G$, let $F$ be a non-contracting distortion $d$ embedding and $\Psi:\domstd \rightarrow B_{s'} \cup B_{t'}$ be the restriction of $F$ to $B_{s'} \cup B_{t'}$. Let $C_1,C_2,\ldots C_a$ be the components of $\gdash$. We will derive certain properties of $G$ with the help of the embedding $F$. Let $P_1,\ldots,P_k$ be the $k$ edge disjoint paths from $s$ to $t$. For each $i \in [k]$, let $P_i' = P_i \setminus (B_s \cup B_t)$ and $P_i^{''} = P_i \setminus (B_{s'} \cup B_{t'})$. If $P_i^{''}$ is a non-empty path, let $s_i$ be the endpoint of $P_i^{''}$ that has an edge to $B_{s'}$ while $t_i$ be the endpoint of $P_i^{''}$ that has an edge to $B_{t'}$. Let $S_i$ ($T_i$) denote
the set of vertices of ${\sf Dom}^F_{s'}$ (${\sf Dom}^F_{t'}$)  that are mapped into $P_i$.

\begin{rem}
\label{rem:path_length}
We are assuming each $P_i$ has length at least $4d^2 + 2d$. Otherwise, we guess the set of vertices that are mapped into the path. Note that, for a fixed $\Psi$, there are at most $(2d)^{2d}$ many guesses per each path of length less than $4d^2+2d$. In total there will be at most $(2d)^{2dk}$ guesses for mapping of vertices into all paths of length less than $4d^2+2d$. For simplicity of presentation, we assume that there are no $s-t$ path of length less than $4d^2+2d$.
\end{rem}

\begin{obs}\label{obs:comp-path}
 Let $F$ be a non-contracting distortion $d$ embedding, $\Psi:\domstd \rightarrow B_{s'} \cup B_{t'}$ be the restriction of $F$ to $B_{s'} \cup B_{t'}$, and $C_1,C_2,\ldots C_a$ be the components of $\gdash$. Then each component of $\gdash$ can have it's vertices mapped into exactly one $P_i'$, $i \in [k]$. 
\end{obs}
\begin{proof}
Since a non-contracting distortion $d$ embedding requires an edge to have expansion at most $d$, all vertices of a
 component of $\gdash$ must be mapped into exactly one
  $P_i'$, $i\in [k]$. 
\end{proof}

Also, there cannot be many components of $\gdash$ that has some vertex mapped into a $P_i^{''}$, $i\in[k]$.

\begin{obs}\label{obs:path-comp}
Given a non-contracting distortion $d$ embedding $F$, each $P_i^{''}$, $i\in [k]$, can have at most $2$ connected components of $\gdash$ having some vertex mapped into $P_i^{''}$, in the non-contracting distortion $d$ embedding $F$.
\end{obs}

\begin{proof}
Let $F$ be a non-contracting distortion $d$ embedding. Assume that there are three connected components $C_{i_1},C_{i_2},C_{i_3}$ of $\gdash$ having some vertex mapped into path $P_i^{''}$ for some $i \in [k]$.  By Observation~\ref{obs:comp-path}, $C_{i_1},C_{i_2},C_{i_3}$ are are mapped into $P_i'$. So, there exist two components such that either both of them have neighbours, in $G$, mapped into $B_{s'}$ or both of them have neighbours, in $G$, mapped into  $B_{t'}$. With out loss of generality, assume that 
\begin{itemize}
\item $C_{i_1}$ and $C_{i_2}$ have some vertex mapped into $B_{s'}$;
\item There exist vertices $u$ and $v$ in $C_{i_1}$ and $C_{i_2}$, respectively, such that $u~(v)$ is mapped to some vertex in $P_i^{''}$, i.e., $u~(v)$ is not mapped into any vertex in $B_{s'}$;
\item $u~(v)$ has some neighbours mapped to some vertex in $B_{s'}$;
\item $D_H(s,F(u)) < D_H(s,F(v))$.
\end{itemize} 
Let $v'$ be some neighbour of $v$ such that it is mapped to some vertex 
 in $B_{s'}$, Observe that $D_H(F(v'),F(v)) \leq d$. Recall that we $P_i$ is of length at least $4d^2+2d$ and $F$ is distortion $d$ embedding. This implies $D_H(F(u),F(v)) < d$. As $F$ is a non-contracting, $D_G(u,v) < d$. Note that $u$ and $v$ belong to different components in $\gdash$. So, the shortest path between $u$ and $v$ in $G$ contains a vertex $w$ that is mapped to some vertex in $B_s$. Note that 
 $D_G(u,w),D_G(v,w) < d$. Considering the fact that $F$ is distortion $d$ embedding, 
  $D_H(F(u),F(w))$ and $D_H(F(v),F(w)) < d^2$. Using the fact that $w$ is mapped to some vertex in $B_s$, 
  $D_H(s,F(u))$ and $D_H(s,F(v)) < d^2$. This is impossible as $u$ and $v$ are not mapped into any vertex in $B_{s'}$.
\end{proof}
Let $C_1,\ldots,C_a$ be the components of $ \gdash$. We say a component $C_i$ is \emph{residual} 
if  $C_i \setminus \domstd$ is non-empty. By Observation~\ref{obs:path-comp}, there can be at most $2k$ 
components of $\gdash$ such that there exists a vertex in any component which will be mapped to some vertex 
not in $B_{s'}$ or $B_{t'}$. Equivalently, there are at most $2k$ residual componets of $ \gdash$.

\begin{defi}\label{def:empty-path}
Let $F$ be a non-contracting distortion $d$ embedding. An empty subpath of $F$ is a subpath of the generalized theta graph where none of the vertices  have any preimage. If a path $P_i^{''}$, $i\in [k]$, has an empty subpath with one endpoint at $t_i$, then such a subpath is called a \emph{$t$-empty subpath}. Similarly, if a path $P_i^{''}$ has an empty subpath with one endpoint at $s_i$, then such a subpath is called a \emph{$s$-empty subpath}. If a path $P_i^{''}$ contains an empty subpath that coincides with neither $s_i$ nor $t_i$, then such a subpath is called an \emph{internal-empty subpath}. Finally, it is possible that the path $P_i^{''}$ itself is an empty subpath and then $P_i^{''}$ is called a \emph{fully-empty subpath}.
\end{defi}
Note that a path $P_i'$ can have at most one empty subpath with respect to $F$. Similarly, we classify the components of $\gdash$.

\begin{defi}\label{def:components}
Let $F$ be a non-contracting distortion $d$ embedding. A residual component in $\gdash$ is called an \emph{$s$-component} if it has neighbours to ${\sf Dom}_{s'}^F$ and not to ${\sf Dom}_{t'}^F$. Similarly, we define a $t$-component. A \emph{full component} is a component that has neighbours to both ${\sf Dom}_{s'}^F$ and ${\sf Dom}_{t'}^F$.
\end{defi}

Since $F$ is a non-contracting distortion $d$ embedding, the following observation is true.

\begin{obs}\label{obs:config}
Let $F$ be a non-contracting distortion $d$ embedding. Any path $P_i$, $P_i^{'} \neq \emptyset$, can be one of the following \emph{form}s.
\begin{itemize}
\item[(i)]\emph{form-1:} It has an $s$-component mapped into it by $F$, and a $t$-empty subpath, 
\item[(ii)]\emph{form-2:} It has a $t$-component mapped into it by $F$, and an $s$-empty subpath,
\item[(iii)]\emph{form-3:} It has an $s$-component and a $t$-component mapped into it by $F$, and an internal-empty subpath,
\item[(iv)]\emph{form-4:} It has a full component mapped into it by $F$,
\item[(iv)]\emph{form-5:} It contains a fully-empty subpath.
\end{itemize}
If we refer $P_i$ to be of \emph{form-\tst}, then $P_i$ is of form-1 or form-2 or form-3.
\end{obs}

\remove{
\begin{defi}
\label{defi:last}
 Let $F$ be a non-contracting distortion $d$ embedding. For a $j$-component $C$, $j \in \{s,t\}$, 
\begin{itemize}
\item  The \emph{first} vertex of $C$, with respect to embedding $F$ is the vertex  $a \in C$ such that $F(a)$ is the nearest vertex from $j$ in $H$.
\item The \emph{last} vertex of $C$ with respect to embedding $F$ is the vertex $\ell \in C$ such that $F(\ell)$ is the furthest vertex from $j$ in $H$.
\end{itemize} 
\end{defi}

}

The objective is to find a non-contracting distortion $d$ embedding $F$, if it exists. Although we do not know about $F$, we want to store a snapshot of $F$.

\begin{defi}
\label{defi:config}
A \emph{configuration} $\cX$ is a tuple $(\Psi,\mathcal{P}',\hat{\mathcal{P}})$ where:
\begin{itemize}
\item Let $U \subseteq U'\subseteq V(G)$ be such that $G\setminus U$ creates a set of residual components $\{C_1,C_2,\ldots,C_a\}$, $a \leq 2k$. $\Psi:U' \rightarrow B_{s'} \cup B_{t'}$ is a non-contracting distortion $d$ embedding of $U'$ and $U$ is the set of vertices that are mapped into $B_s \cup B_t$.  
\item $\mathcal{P}' \subseteq \mathcal{P}$, where $\cP$ is the set of all $s\mbox{-}t$ paths.
\item $\hat{\mathcal{P}}$ is a family of $\vert \mathcal{P} \setminus \mathcal{P}' \vert$ tuples such that for each path $P_i \in \mathcal{P} \setminus \mathcal{P}'$, there is a tuple $({\sf form}_i,\cC_{P_i}, {\sf comp}_{i})$ with the following information:
\begin{itemize}
\item[(i)] ${\sf form}_i$ assigns the name of a form to $P_i$.
\item[(ii)] The set $\cC_{P_i}$ is a set of at most $2$ residual components of $G\setminus U$ that are assigned to $P_i'$ and to no other $P_j',j\neq i$. 
\item[(iii)] The function ${\sf comp}_i$ indicates for each $C \in \cC_{P_i}$ whether it is an $s$-component or a $t$-component or full-component, with respect to $\Psi$.
\end{itemize}
\item $\bigcup_{P_i \in \mathcal{P} \setminus \mathcal{P}'} \cC_{P_i}$ has all the residual components of $G \setminus U$. 
\end{itemize}
\end{defi}

The number of configurations is bounded for a fixed $\Psi$.
\begin{obs}\label{obs:pathformfix}
For any fixed $\Psi$, the total number of configurations is $\Oh(k^{2k})$.
\end{obs}
\begin{proof}
 When $\Psi:U \rightarrow B_{s'}\cup B_{t'}$ is fixed, the residual components $\{C_1,C_2,\ldots,C_a\}$ of $G \setminus U$ are fixed. Note that $a \leq 2k$. Also, whether a residual component is an $s$-component, a $t$-component or a full-component gets fixed. The total number of ways in which the residual components can be assigned to paths of $\mathcal{P}$ is $\Oh(k^{2k})$. Once an assignment of the residual components to the paths is fixed, we can find out $\mathcal{P}'$, and the tuple $({\sf form}_i,\cC_{P_i}, {\sf comp}_{i})$ for each $P_i \in \cP \setminus \cP'$. Therefore, the total number of configurations for a fixed $\Psi$ is $\Oh(k^{2k})$. 
\end{proof}

Next, we define feasible configurations that can be associated with metric embeddings.

\begin{defi}
\label{defi:feasible-config}
A configuration $\cX = (\Psi,\mathcal{P}',\hat{\mathcal{P}})$ is said to be \emph{feasible} with respect to a non-contracting distortion $d$ embedding $F$ of $G$ into $H$ if the followings hold:
\begin{itemize}
\item $\Psi:\domstd \rightarrow B_{s'} \cup B_{t'}$ is the restriction of $F$ to $\domstd$.  
\item $P_i^{''} = P_i \setminus (B_{s'} \cup B_{t'})$ is empty for each $P_i \in \cP' \subseteq \cP$. \remove{$\mathcal{P}' \subseteq \mathcal{P}$ are the empty paths with respect to $F$},
\item For each $P_i \in \hat{\mathcal{P}}$,there is a tuple $({\sf form}_i,\cC_{P_i}, {\sf comp}_{i})$ with the following information:
\begin{itemize}
\item[(i)] ${\sf form}_i$ is the form of $P_i$ in $F$.
\item[(ii)] The set $\cC_{P_i}$ is the set of at most $2$ residual components of $G\setminus U$ that are embedded into $P_i'$ by $F$. 
\item[(iii)] The function ${\sf comp}_i$ indicates for each $C \in \cC_{P_i}$ whether it is an $s$-component or a $t$-component or full-component, with respect to $F$.
\end{itemize}
\item $\bigcup_{P_i \in \mathcal{P} \setminus \mathcal{P}'} \cC_{P_i}$ has all the residual components of $\gdash$. 
\end{itemize}
\end{defi}

By Observation~\ref{obs:comp-path}, in a feasible configuration $\size{C_{P_i}}\leq 2$. Note that if $\size{C_{P_i}}=0$, then $P_i$ can only be of form-5. If $\size{C_{P_i}}=1$, then $P_i$ can be of form-1, form-2 or form-4. If $\size{C_{P_i}}=2$, then $P_i$ can only be of form-3 where one component of $C_{P_i}$ is a $s$-component and the other one is a $t$-component. Note that each non-contracting distortion $d$ embedding $F$ induces a feasible configuration. We denote it by $\cX(F)$. Also, the total number of feasible configurations is bounded by the total number of configurations.

 Next, we define the notion of a last vertex for a residual component of $\gdash$ with respect to the embedding $F$.
\begin{defi}
\label{defi:last}
 Let $F$ be a non-contracting distortion $d$ embedding. Let  $C$ be a $j$-component, $j \in \{s,t\}$. A vertex $\ell$ in $C $ is the \emph{last} vertex of $C$ with respect to embedding $F$ if $D_H(j,F(\ell)) \geq D_H(j,F(x))$ for all $x \in C$.
\end{defi}

The following Lemma gives a bound on the potential last vertices of a component of $\gdash$ if $G$ is embeddable into $H$.

\begin{lem}\label{lem:last-vertex-comp}
Let $\cF$ be a family of non-contracting distortion $d$ embedding of $G$ into $H$ such that $\cX(F_1)=\cX(F_2)$ for any $F_1, F_2 \in \cF$. Then for any form-\tst path $P_i$ and any $s$($t$)-component $C \in \cC_{P_i}$,  there are $d^{\Oh(d^2)}$ vertices that are candidates for being the last vertex of $C$ with respect to some $F \in \cF$.
\end{lem}

\begin{proof}
Without loss of generality, assume that $C$ is an $s$-component.
Let $v\in C $ be a candidate for the last vertex of $C$ with respect to some  $F\in \cF$. Note that 
$v \in C \setminus {\sc Dom}_{s'}^F$. Let $u \in S_i $ be some vertex in ${\sf Dom}_{s'}^F$ that is mapped to a vertex in $P_i$. Since $G$ is a connected graph, such a vertex always exists. We show that, in $G$, $v$ must be either a vertex furthest away from $u$ or at most $d^2$ distance away from a furthest vertex. Let $\hat{Q}$ be the subpath of $P_i$ between $F(u)$ and $F(v)$. Consider the subpath $Q\subseteq \hat{Q}$ that has length $d^2$ and has an endpoint at $F(v)$. Before proceeding further, we need the following Claim.
\begin{cl}
For any vertex $x$ that is mapped into some vertex of $\hat{Q} \setminus Q$, $D_G(u,x) < D_G(u,v)$.
\end{cl}
\begin{proof}
If the shortest path between $u$ and $v$ passes through $x$ in $G$, then we are done.
 Therefore, assume that the shortest path between $u$ and $v$ passes through two vertices of an edge $(y,z)$ such that $D_H(F(u),F(y)) < D_H(F(u),F(x))$ and $D_H(F(u),F(z)) > D_H(F(u),F(x))$. For ease of notation, when $D_H(F(u),F(y)) < D_H(F(u),F(x))$ we say that $y$ is \emph{mapped before} $x$. Similarly, when $D_H(F(u),F(z)) > D_H(F(u),F(x))$ we say that $z$ is \emph{mapped after} $x$. Note that such an edge $(y,z)$ exists on the shortest path between $u$ and $v$. Also, since the expansion is at most $d$, $D_H(F(y),F(z)) \leq d$. Now, we give an upper bound for $D_G(u,x)$
 \begin{eqnarray*}
 D_G(u,x) &\leq& D_G(u,y)+D_G(y,x) \\
 &\leq& D_G(u,y) + D_H(F(y),F(x))\\
 &<& D_G(u,y) + D_H(F(y),F(z)) \\
 &\leq& D_G(u,y) + d
 \end{eqnarray*}
 For a lower bound of $D_G(u,v)$,
 \begin{eqnarray*}
D_G(u,v) &=&D_G(u,y)+1+D_G(z,v)\\
&\geq& D_G(u,y)+1+\frac{D_H(F(z),F(v))}{d}\\
&>& D_G(u,y)+1 +\frac{d^2+1-d}{d}\\
&>& D_G(u,y)+d.
 \end{eqnarray*}
 \end{proof}

 Therefore, all vertices that are furthest away from $u$ must be mapped to vertices of $Q$. As we are constructing a non-contracting embedding, all vertices that are mapped to vertices of $Q$ must be within distance $d^2$ of the furthest vertices from $u$. 
 
 There can be at most $d^2$ furthest vertices from $u$. By Observation~\ref{obs:degree}, each vertex is of degree at most $2d$. Hence, the total number of candidate vertices for $v$ is bounded by $d^2.(2d)^{d^2}$.
\end{proof}

Next, we define the notion of a shortest embedding in the context of residual component(s) of $\gdash$ in a path of $\mathcal{P}$.
\begin{defi}
\label{defi:shortest}
Let $\cY$ be a feasible configuration such that $\cY=\cX(F)$ for a non-contracting distortion $d$ embedding $F$. Let $P_i$ be a form-\tst path, $C \in \cC_{P_i}$ be a $s$-component of $\gdash$ and $\ell \in C$ be a candidate to be the last vertex of $C$ with respect $F$. 

Recall that $S_i $ is the set of vertices of ${\sf Dom}_{s'}^F$ that are mapped into $P_i$. Let $\cA$ be a family of non-contracting and distortion $d$ embedding of $C \cup S_i$ into $P_i$ such that the 
 following conditions hold, 
\begin{itemize}
 \item[(i)] $f_1\vert _{S_i}=f_2\vert_{S_i}$ for any $f_1,f_2 \in \cA$.
 \item[(ii)] For each $f \in \cA$, $f(x)$ is a vertex of $P_i'$ for any $x \in C$.
 \item[(iii)] For each $f \in \cA$, $F\vert_{C \cup S_i}=f$ and $\ell$ is the last vertex of $C$ with respect to $F$. 
 \item[(iv)] For each $f \in \cA$, for any $x \in \domstd$, the path between $f(\ell)$ and $f(x)$ is non-contracting with expansion at most $d$.
\end{itemize}

Then the \emph{shortest} embedding of $C \cup S_i$ into $P_i$ with respect to $\cY$ and $\ell$, is an embedding $f \in \cA$ such that $D_H\left(s,f(\ell)\right) \leq D_H\left(s,f'(\ell)\right) $ for all $f' \in \cA$.
If $C$ is a $t$-component, $T_i $ is taken to be the set of vertices of ${\sf Dom}_{t'}^F$ that are mapped into $P_i$ and we can define the shortest embedding of $C \cup T_i$ with respect to $\cY$ and $\ell$ in a similar way.
\end{defi}

We can extend the notion of shortest embedding of a component into a path of $\mathcal{P}$ to that of a non-contracting distortion $d$ embedding of $G$ into $H$ that has shortest embeddings for all $s$-components and $t$-components.

\begin{defi}
Let us consider a non-contracting distortion $d$ embedding $F$ of $G$ into $H$. We say $F$ is a \emph{special embedding with respect to feasible configuration $\cX(F)$}
if for every path $P_i$ of form-\tst and $s$ ($t$)-component $C \in \cC_P$, the following holds: $F\vert_{C \cup S_i}$ ($F\vert_{C \cup T_i}$) is the shortest embedding of $C \cup S_i$ ($C \cup T_i)$ into $P_i$ with respect to  the feasible configuration $\cX(F)$ and the last vertex of $C$ with respect to $F$.
\end{defi}

The next lemma shows that it is enough to look for a special embedding of $G$ into $H$.  

\begin{lem}
\label{lem:main_theta}
If there exists a non-contracting distortion d embedding of $G$ into $H$, then there exists
a special embedding of $G$ into $H$ with respect to some configuration. 
\end{lem}
\begin{proof}
Let $F$ be a desired embedding of $G$ into $H$. Consider $\Psi$ to be $F$ restricted to $B_{s'} \cup B_{t'}$ and let the feasible configuration with respect to $F$ be $\cX(F)$. Let $F'$ be a function from $V(G)$ to $V(H)$ satisfying the following conditions.
\begin{itemize}
\item[(i)] $\cX(F)=\cX(F')$.
\item[(ii)] For every $x \in \domstd$, $F'(x)=F(x)$. That is $\domstd =\domstddash$ and $\domst =\domstdash$
\item[(iii)] For every path $P_i$ of form-4, component $C \in \cC_{P_i}$ and  vertex $x \in C$, $F'(x)=F(x)$. In other words if $x$ is a vertex of any full component $C$, then $F'(x)=F(x)$.
\item[(iv)] For every path $P_i$ of form-\tst and  $s(t)$-component $C \in \cC_{P_i},  F'\vert_{C \cup S_i}$ ($F'\vert_{C \cup T_i}$) is the shortest embedding of $C \cup S_i(C \cup T_i)$ into $P_i$ with respect to $\cX(F')$ and the last vertex of $C$ with respect to $F$.  
\end{itemize}

It is easy to see that $F'$ is an injection from $V(G)$ to $V(H)$. Note that the last vertices of residual components in $F'$ remain the same as the last vertices of residual components in $F$. Let $L$ denote the set of last vertices of the residual components of $\gdash = \gdashdash$. To conclude that $F'$ is a special embedding from $G$ to $H$, we have to show that $F'$ is a non-contracting distortion $d$ embedding of $G$ into $H$. We will be done by the following claims.
 \begin{cl}
 $F'$ has expansion at most $d$.
 \end{cl} 
 \begin{proof}
We show that $D_H(F'(x),F'(y)) \leq d$ holds for any edge $(x,y) \in E(G)$. Then we can apply induction similar to that in the proof of Claim~\ref{clm:cycle_dist}.
 
 If both $x,y \in \domstddash$, then $D_H(F'(x),F'(y))=D_H(F(x),F(y)) \leq d$. This is because $\Psi$ is a distortion $d$ embedding of $\domstddash$.  If both $x$ and $y$ are mapped
 into the same residual component $C$, then also $D_H(F'(x),F'(y))\leq d$.  Note that both $x$ and $y$ cannot belong to different residual components. So, the only case that remains is when
 $x \in \domstddash$ and $y$ is in some residual component $C$.
 
  If $C$ is a full component, then $D_H(F'(x),F'(y))=D_H(F(x),F(y)) \leq d$. Let $P_i$ be the path 
  that contains $C$.
 If $C$ is an (a) $s$ ($t$)-component, then $x$ has to be a vertex in $S_i$ ($T_i$). However, $F'\vert_{C \cup S_i}$($F'\vert_{C \cup T_i}$) is a shortest embedding of $C \cup S_i$ ($C \cup T_i$) into $P_i$. Note that every shortest embedding is a non-contracting distortion $d$ embedding. Hence, $D_H(F'(x),F'(y)) \leq d$.
 \end{proof}

  \begin{cl}\label{cl:nct_theta0}
 Let $C$ be any residual component of $\gdashdash$.
  Then  for any $x \in C$ and $y \in \domstddash$, the shortest path between $F'(x)$ and $F'(y)$ is non-contracting.
 \end{cl}
 
 \begin{proof}
 
 Note that $F'(y)=F(y)$. If $C$ is a full component, then $F'(x)=F(x)$. So, $D_H(F'(x),F'(y))=D_H(F(x),F(y))\geq D_G(x,y)$. Now consider the case if $C$ is an (a) $s$ ($t$)-component.
  The shortest path from $F'(x)$ and $F'(y)$ must pass through $F'(z)$, where $z$ is either a vertex in $\domstdash$ or the last vertex of $C$ with respect to $F'$. 

In either case,
\begin{eqnarray*}
D_H(F'(x),F'(y)) &=& D_H(F'(x),F'(z))+D_H(F'(z),F'(y)) \\
 &\geq& D_G(x,z)+D_G(z,y) ~~(\mbox{By the construction of $F'$})\\
 &\geq& D_G(x,y). 
\end{eqnarray*} 
 \end{proof}
 
 Recall that $L$ is the set of last vertices of residual components in $\gdashdash$ with respect to $F$, and by construction $F'$.

 \begin{cl}
 \label{cl:nct_theta}
 $F'$ is a non-contracting embedding.
 \end{cl}
   \begin{proof}
Let $x$ and $y$ be two vertices of $G$ such that either both $x$ and $y$ are in $\domstddash$ or both are in the same residual component $C$. Then by the construction of $F'$, $F'(x)=F(x)$ and $F'(y)=F(y)$. Hence, the shortest path between $F'(x)$ and $F'(y)$ is non-contracting.

If one of $x$ and $y$ is in $\domstddash$ and the other vertex is in some residual component $C$, then also we are done 
by Claim~\ref{cl:nct_theta0}.

So, assume that $x$ and $y$ are in different residual components. We look at the following cases:
\begin{itemize}
\item Let $x \in C_1$ and $y \in C_2$. Assume that one of $C_1$ and $C_2$ is a full component. Without loss of generality, let $C_1$ be the full component. Then the shortest path from $F'(x)$ to $F'(y)$ passes through $F'(u)$, where $u$ is some vertex in $\domstddash~(\domstdash)$. So,
\begin{eqnarray*}
D_H(F'(x),F'(y)) &=& D_H(F'(x),F'(u))+D_H(F'(u),F'(y))\\
&\geq& D_G(x,u)+D_G(u,y) ~~(\mbox{By Claim~\ref{cl:nct_theta0}})\\
&\geq& D_G(x,y)
\end{eqnarray*}

\item Let $x \in C_1$ and $y \in C_2$ such that neither is a full component. Let $\ell_1\in V(G)$ be the last vertex of $C_1$ with respect to $F$, and $\ell_2 \in V(G)$ be the last vertex of $C_2$ with respect to $F$. By construction they remain the last vertices of the respective residual components with respect to $F'$ as well. Suppose the shortest path from $F'(x)$ to $F'(y)$ passes through $F'(u)$ where $u$ is some vertex in $\domstddash~(\domstdash)$. Then 
\begin{eqnarray*}
D_H(F'(x),F'(y)) &=& D_H(F'(x),F'(u))+D_H(F'(u),F'(y))\\
&\geq& D_G(x,u)+D_G(u,y) ~~(\mbox{By Claim~\ref{cl:nct_theta0}})\\
&\geq& D_G(x,y)
\end{eqnarray*}

Otherwise, the shortest path from $F'(x)$ to $F'(y)$ contains $F'(\ell_1)$ and $F'(\ell_2)$. By construction of $F'$, $D_H(F'(\ell_1),F'(\ell_2)) \geq D_H(F(\ell_1),F(\ell_2))$. Since, $F$ was a non-contracting distortion $d$ embedding, this implies that $D_H(F'(\ell_1),F'(\ell_2)) \geq D_G(\ell_1,\ell_2)$. Then 
\begin{eqnarray*}
D_H(F'(x),F'(y)) &=& D_H(F'(x),F'(\ell_1))+D_H(F'(\ell_1),F'(\ell_2))+D_H(F'(\ell_1),F'(y))\\
&\geq& D_G(x,\ell_1)+D_G(\ell_1,\ell_2)+D_G(\ell_2,y) ~~(\mbox{Due to shortest embeddings})\\
&\geq& D_G(x,y)
\end{eqnarray*}
\end{itemize}
\end{proof}
\end{proof}

Therefore, we have shown that if $G$ is embeddable into $H$ then it is enough to find a special embedding. We design an FPT algorithm for finding a special embedding. 

\begin{theo}
\label{theo:main_theta}
 {\sc Metric Embedding} into generalized theta graphs is FPT parameterized by distortion $d$ and number $k$ of $s-t$ paths. The algorithm runs in time $ \Oh(N)+ n^5 \cdot k^{2k+1}\cdot (kd+1)^{(2d)^{\Oh(kd)}}\cdot d^{\Oh(d^2)} $, where $n$ and $N$ are the number of vertices in the input and output graph metrics, respectively.
\end{theo}
\begin{proof}
By Lemma~\ref{lem:main_theta}, it is sufficient to look for special embedding with respect to some configuration.
 We find $\Delta(G)$ and if $\Delta(G) > (k+1)d$, then we report NO. This is correct by Observation~\ref{obs:degree}. Note that $\Delta(G)$ can be found in linear time. 
 
 We first compute $D_H(s,u)$ and $D_H(t,u)$ for all $u \in V(H)$. We store this distance information in a matrix $\cD_{st}$, such that the look-up time for the distance from any $u \in V(H)$ to $s$ or $t$ is $\Oh(1)$. Note that we can compute $\cD_{st}$ in $\Oh(N+k)$ time as the number of edges in $H$ is $\Oh(N+k)$. $\cD_{st}$ will be required for checking whether an obtained function from $V(G)$ to $V(H)$ is a noncontracting distortion $d$ embedding.
 
 Let us fix a non-contracting distortion $d$ embedding $\Psi$ of $U' \subseteq V(G)$ into $B_{s'} \cup B_{t'}$ and a configuration $\cY$ containing $\Psi$. Let $U$ be the set of vertices that are mapped into 
 $B_s \cup B_t$. If the degree of any vertex in $G \setminus U$ is more than $2d$, then we decide that there does not exist any desired embedding with respect to $\cY$. Otherwise, we proceed as follows. Let $F$ be the special embedding of $G$ into $H$ with respect to $\cY$ that we want to find, if one exists. Note that $U = \domst$.

\begin{itemize}
\item[(i)] If a path $P_i$ is of form-5, we don't have to do anything for that.

\item[(ii)] Let a path $P_i$ be of form-4, and suppose $C \in \cC_{P_i}$ is the only full component mapping into $P_i$. The vertices in $S_i \cup T_i \subset U'$ are the points that are mapped into $P_i \setminus P_i^{''}$. Then we find a non-contracting distortion $d$ embedding $f_C$, if possible, of $C \cup 
S_i\cup T_i$ into $P_i$ such that $f_C\vert_{S_i}=\Psi\vert_{S_i}$ and $f_C\vert_{T_i}=\Psi\vert_{T_i} $. Note that 
$S_i$ ($T_i$) is the set of vertices of ${\sf Dom}_{s'}^F$ (${\sf Dom}_{t'}^F$) that are mapped into $P_i$. Such an embedding $f_C$ can be found by running the algorithm described in Lemma~\ref{lem:path-st-algo}. This algorithm admits a time complexity
of $\Oh\left(\linethetabound \right)=\Oh\left( n^2 \cdot d^{\Oh(d)}\right)$. If we cannot find such an embedding, then there does not exist any special embedding of $G$ into $H$ with respect to $\cY$. 

\item[(iii)] Let $P_i$ be a form-\tst path and $C \in \cC_{P_i}$ be an (a) $s$ ($t$)-component. Without loss of generality, assume that
$C$ is an $s$-component. Here, our objective is to find the
 shortest embedding $f$ of $C \cup S_i$ into $P_i$ with respect to $\cY$ and some $\ell$, where $\ell$ is the last vertex of $C$ with respect to $F$. We guess a vertex $\ell \in C$, as the last vertex. This vertex is either a furthest vertex from a fixed vertex $a\in S_i$ or lies within distance of $d^2$ from a furthest vertex from $a$ in $G$. By Lemma~\ref{lem:last-vertex-comp}, the total number of candidates for the last vertex of $C$ with respect to $F$ is $d^{\Oh(d^2)}$.

It is easy to see that $\size{C \cup S_i} \leq D_{H}(f(\ell),f(a)) \leq 2d.\size{C \cup S_i}$ by Observation~\ref{obs:unwt-length_linecycle}. Thus, the length of the shortest embedding of $C\cup S_i$, where $\ell$ is the last vertex, is also in this range. For each possible length $\size{C \cup S_i} \leq {\sf len} \leq 2d.\size{C \cup S_i}$, we try to find a non-contracting distortion $d$ embedding $f_{\sf len}$ of $C\cup S_i$ into a path $P_{\sf len} = \{1,2,\ldots,{\sf len}\}$ such that $f_{\sf len}$ restricted to the first $\size{S_i}$ vertices is same as the mapping by $\Psi$, and for each $u \in C\cup S_i$, $D_{P_{\sf len}}(1,f_{\sf len}(u)) \leq D_{P_{\sf len}}(1,f_{\sf len}(\ell))$. Observe that we have to run the algorithm described in Corollary~\ref{cor:path-st-algo}. If the algorithm returns no for all lengths, for every candidate $\ell$ for the last vertex, then there does not exist any special embedding of $G$ into $H$ with respect to $\cY$. Otherwise, assume that for the current guess $\ell$ the shortest length for which the algorithm returns a required non-contracting distortion $d$ embedding is ${\sf len}'$. Let this embedding be $f_{C}$. In the worst case this step will incur a time complexity of $d^{\Oh(d^2)}\cdot \Oh \left(2nd \cdot\linethetabound \right) = \Oh\left(n^3 \cdot d^{\Oh(d^2)}\right)$.
\end{itemize}

 Let $F=\Phi_\Pi$ be the function such that $\Pi=\{\Psi\} \cup \{f_C~\vert~ C~\mbox{is a residual component of }G \setminus U\}$.\remove{It is easy to see that $F$ is an injection.}
We verify whether the obtained $F$ is a non-contracting distortion $d$ embedding from $G$ to $H$. 
If yes, we are done. If not, then there does not exist any special embedding with respect to $\cY$.
Observe that the distance between two given points in $H$, can be computed in $\Oh(1)$ time using $\cD_{st}$.
 Since we only need to check for non-contraction and distortion $d$ for every pairs of mapped vertices in $H$, the verification of $F$ for a desired embedding can be done by spending $\Oh(n^2)$ time.
\remove{ If $\cX(F)=\cY$, then $F$ is a (special) embedding of $G$ into $H$ and we are done.} \remove{ If $\cX(F)=\cY$, then $F$ is a (special) embedding of $G$ into $H$ and we are done. If not, then
 there does not exist any special embedding with respect to $\cX$.}

Note that in the worst case, we have to run the above steps for all possible configurations. If we 
decide that there does not exist a special embedding with respect to all configurations, then we report that
$G$ does not admit the desired embedding of $G$ into $H$. The correctness of the algorithm
follows from Lemma~\ref{lem:main_theta}.

Now we analyze the running time of the algorithm. Recall that there can be $2k$ residual components of $G \setminus U$.
In the execution of our algorithm for a fixed configuration, we spent $2k\cdot n^3 \cdot d^{\Oh(d^2)} +\Oh(n^2)$ time. By Observation~\ref{obs:number-fixed-window} and \ref{obs:config}, the total number of
configuration is at most $k^{2k}\cdot n^2 \cdot (2kd)^{(4d)^{\Oh(kd)}} $. Note that we have spent
$\Oh(N + k)$ time to compute $\cD_{st}$. Putting everything together, 
the time complexity of our algorithm is bounded by 
$$ \Oh\left( N\right)+ n^5 \cdot k^{2k+1}    \cdot (2kd)^{(4d)^{\Oh(kd)}}\cdot d^{\Oh(d^2)}$$
\end{proof}

\section{Conclusion}\label{sec:conclusion}
In this paper, we presented several FPT algorithms for embedding into different graph classes. Note that for all the considered graph classes, our results can be modified to answer the {\sc Weighted Graph Metric Embedding} problem for the graph classes. Similar to the results in Section~\ref{sec:cycle}, for a particular graph class considered in this paper, when we take the maximum edge weight $M$ to be a parameter along with the set of parameters considered for the graph class, then the problem is still FPT. On the other hand, without $M$ as a parameter, the problem is NP-Complete for any distortion $d>2$. 
The question of the parameterized complexity of embedding into trees of unbounded degree, asked in~\cite{FellowsFLLRS13}, still remains open. Another important question is to determine the parameterized complexity of {\sc Graph Metric Embedding} for bounded treewidth graphs, even when the treewidth is taken to be some constant. Since trees with unbounded degree have treewidth $1$, the latter open problem is a generalization of the former.

\bibliographystyle{alpha}
\bibliography{reference}
\end{document}